\input harvmac.tex
\input epsf.tex

\def\figin{\epsfcheck\figin}\def\figins{\epsfcheck\figins}
\def\epsfcheck{\ifx\epsfbox\UnDeFiNeD
\message{(NO epsf.tex, FIGURES WILL BE IGNORED)}
\gdef\figin##1{\vskip2in}\gdef\figins##1{\hskip.5in}% blank space instead
\else\message{(FIGURES WILL BE INCLUDED)}%
\gdef\figin##1{##1}\gdef\figins##1{##1}\fi}
\def\DefWarn#1{}
\def\figinsert{\goodbreak\midinsert}
\def\ifig#1#2#3{\DefWarn#1\xdef#1{fig.~\the\figno}
\writedef{#1\leftbracket fig.\noexpand~\the\figno}%
\figinsert\figin{\centerline{#3}}\medskip\centerline{\vbox{\baselineskip12pt
\advance\hsize by -1truein\noindent\footnotefont{\bf
Fig.~\the\figno:} #2}}
\bigskip\endinsert\global\advance\figno by1}

 % \draftmode

%%%%%%%%%%%%%%%%%%%%%%%%%%%%%%%%%%%%%%%%5

\lref\shockthv{
  T.~Dray and G.~'t Hooft,
  %``The Gravitational Shock Wave Of A Massless Particle,''
  Nucl.\ Phys.\  B {\bf 253}, 173 (1985).
  %%CITATION = NUPHA,B253,173;%%
  G.~'t Hooft,
  %``Graviton Dominance in Ultrahigh-Energy Scattering,''
  Phys.\ Lett.\  B {\bf 198}, 61 (1987).
  %%CITATION = PHLTA,B198,61;%%
  H.~L.~Verlinde and E.~P.~Verlinde,
  %``Scattering at Planckian energies,''
  Nucl.\ Phys.\  B {\bf 371}, 246 (1992)
  [arXiv:hep-th/9110017].
  %%CITATION = NUPHA,B371,246;%%
}

\lref\nastaconf{
  K.~Kang and H.~Nastase,
  %``High energy QCD from Planckian scattering in AdS and the Froissart
  %bound,''
  Phys.\ Rev.\  D {\bf 72}, 106003 (2005)
  [arXiv:hep-th/0410173].
  %%CITATION = PHRVA,D72,106003;%%
}

%\cite{Erdmenger:1996yc}
\lref\erdmengerb{
  J.~Erdmenger and H.~Osborn,
  %``Conserved currents and the energy-momentum tensor in conformally  invariant
  %theories for general dimensions,''
  Nucl.\ Phys.\  B {\bf 483}, 431 (1997)
  [arXiv:hep-th/9605009].
  %%CITATION = NUPHA,B483,431;%%
}

%\cite{Banks:1998nr}
\lref\banksgreen{
  T.~Banks and M.~B.~Green,
  %``Non-perturbative effects in AdS(5) x S**5 string theory and d = 4 SUSY
  %Yang-Mills,''
  JHEP {\bf 9805}, 002 (1998)
  [arXiv:hep-th/9804170].
  %%CITATION = JHEPA,9805,002;%%
}

\lref\polscale{
  A.~M.~Polyakov, ``Ultraviolet And Infrared Interactions Of The Quantized Field,''
%\href{http://www.slac.stanford.edu/spires/find/hep/www?irn=444944}{SPIRES entry}
{\it  In *Stanford 1975, Symposium On Lepton and Photon Interactions At High Energies*, Stanford 1975, 855-867}.
}

%\cite{Kunszt:1989km}
\lref\kunsztqcd{
  Z.~Kunszt, P.~Nason, G.~Marchesini and B.~R.~Webber,
  %``QCD At Lep,''
  %%CITATION = C89-02-20.1;%%
}
%\cite{Glover:1994vz}
\lref\gloverqcd{
  E.~W.~N.~Glover and M.~R.~Sutton,
  %``The Energy-energy correlation function revisited,''
  Phys.\ Lett.\  B {\bf 342}, 375 (1995)
  [arXiv:hep-ph/9410234].
  %%CITATION = PHLTA,B342,375;%%
}
%\cite{Catani:1996jh}
\lref\cataniqcd{
  S.~Catani and M.~H.~Seymour,
  %``The Dipole Formalism for the Calculation of QCD Jet Cross Sections at
  %Next-to-Leading Order,''
  Phys.\ Lett.\  B {\bf 378}, 287 (1996)
  [arXiv:hep-ph/9602277].
  %%CITATION = PHLTA,B378,287;%%
}
%\cite{Nason:1996nu}
\lref\nasonqcd{
  P.~Nason {\it et al.},
  %``QCD,''
  arXiv:hep-ph/9602288.
  %%CITATION = HEP-PH/9602288;%%
}

%\StermanPD
\lref\stermanreview{
  G.~Sterman,
  %``QCD and jets,''
  arXiv:hep-ph/0412013.
  %%CITATION = HEP-PH/0412013;%%
}

%\cite{Catani:1998sf}
\lref\catani{
  S.~Catani and B.~R.~Webber,
  %``Resummed C-parameter distribution in e+ e- annihilation,''
  Phys.\ Lett.\  B {\bf 427}, 377 (1998)
  [arXiv:hep-ph/9801350].
  %%CITATION = PHLTA,B427,377;%%
}

%\DolanTT
\lref\dolanosborn{
  F.~A.~Dolan and H.~Osborn,
  %``Superconformal symmetry, correlation functions and the operator product
  %expansion,''
  Nucl.\ Phys.\  B {\bf 629}, 3 (2002)
  [arXiv:hep-th/0112251].
  %%CITATION = NUPHA,B629,3;%%
}

% References on lightcone OPE

\lref\lcopeearly{
  Y.~Frishman,
  %``Scale invariance and current commutators near the light cone,''
  Phys.\ Rev.\ Lett.\  {\bf 25}, 966 (1970).
  R.~A.~Brandt and G.~Preparata,
  %``Operator Product Expansions Near The Light Cone,''
  Nucl.\ Phys.\  B {\bf 27}, 541 (1972).
  %%CITATION = NUPHA,B27,541;%%
  N.~H.~Christ, B.~Hasslacher and A.~H.~Mueller,
  %``Light cone behavior of perturbation theory,''
  Phys.\ Rev.\  D {\bf 6}, 3543 (1972).
  %%CITATION = PHRVA,D6,3543;%%
}

\lref\ccps{
  L.~Cornalba, M.~S.~Costa, J.~Penedones and R.~Schiappa,
  %``Eikonal approximation in AdS/CFT: From shock waves to four-point
  %functions,''
  JHEP {\bf 0708}, 019 (2007)
  [arXiv:hep-th/0611122].
  %%CITATION = JHEPA,0708,019;%%
}

\lref\lcstring{ S. A. Anikin and O. I. Zavialov,
% `` Short-distance and light-cone expansions for products of currents,''
 Ann. Phys. 116, 135 (1978) ;
  I.~I.~Balitsky,
  %``String Operator Expansion Of The T Product Of Two Currents Near The Light
  %Cone,''
  Phys.\ Lett.\  B {\bf 124}, 230 (1983).
  %%CITATION = PHLTA,B124,230;%%
}

% end lightcone ope references

\lref\fallingmass{
 S.~Lin and E.~Shuryak,
  %``Toward the AdS/CFT Gravity Dual for High Energy Collisions: II. The Stress
  %Tensor on the Boundary,''
  arXiv:0711.0736 [hep-th].
  %%CITATION = ARXIV:0711.0736;%%
}
 \lref\fallingblackhole{
 J.~J.~Friess, S.~S.~Gubser, G.~Michalogiorgakis and S.~S.~Pufu,
  %``Expanding plasmas and quasinormal modes of anti-de Sitter black holes,''
  JHEP {\bf 0704}, 080 (2007)
  [arXiv:hep-th/0611005].
  %%CITATION = JHEPA,0704,080;%%
}

%\ClaySD
\lref\ellissecond{
  K.~A.~Clay and S.~D.~Ellis,
  %``A Precision calculation of the next-to-leading order energy-energy
  %correlation function,''
  Phys.\ Rev.\ Lett.\  {\bf 74}, 4392 (1995)
  [arXiv:hep-ph/9502223].
  %%CITATION = PRLTA,74,4392;%%
    G.~Kramer and H.~Spiesberger,
  %``A new calculation of the NLO energy-energy correlation function,''
  Z.\ Phys.\  C {\bf 73}, 495 (1997)
  [arXiv:hep-ph/9603385].
  %%CITATION = ZEPYA,C73,495;%%
}

\lref\takshap{
  A.~D.~Shapere and Y.~Tachikawa,
  %``Central charges of N=2 superconformal field theories in four dimensions,''
  arXiv:0804.1957 [hep-th].
  %%CITATION = ARXIV:0804.1957;%%
}

\lref\gkpstring{
  S.~S.~Gubser, I.~R.~Klebanov and A.~M.~Polyakov,
  %``A semi-classical limit of the gauge/string correspondence,''
  Nucl.\ Phys.\  B {\bf 636}, 99 (2002)
  [arXiv:hep-th/0204051].
  %%CITATION = NUPHA,B636,99;%%
}

\lref\parisipetr{
  G.~Parisi and R.~Petronzio,
  %``Small Transverse Momentum Distributions In Hard Processes,''
  Nucl.\ Phys.\  B {\bf 154}, 427 (1979).
  %%CITATION = NUPHA,B154,427;%%
}

%\MuellerFV
\lref\muellerlightray{
  B.~Geyer, D.~Robaschik and D.~Mueller,
  % ``Light Ray Operators And Their Application In QCD,''
  arXiv:hep-ph/9406259.
  %% CITATION = HEP-PH/9406259;%%
  D. Mueller, D. Robaschik, B. Geyer, F. M. Dittes and J. Horejsi,
  % ``Wave functions, evolution equations and evolution kernels from light-ray
  % operators of {QCD},''
  Fortsch.\ Phys.\  {\bf 42}, 101 (1994)
  [arXiv:hep-ph/9812448].
  %% CITATION = FPYKA,42,101; %%
}

\lref\gkp{
  S.~S.~Gubser, I.~R.~Klebanov and A.~M.~Polyakov,
  %``Gauge theory correlators from non-critical string theory,''
  Phys.\ Lett.\  B {\bf 428}, 105 (1998)
  [arXiv:hep-th/9802109].
  %%CITATION = PHLTA,B428,105;%%
}

\lref\barstt{
I.~Bars and G.~Quelin,
  %``Dualities among 1T-Field Theories with Spin, Emerging from a Unifying
  %2T-Field Theory,''
  arXiv:0802.1947 [hep-th].
  %%CITATION = ARXIV:0802.1947;%%
}

%\BeisertEZ
\lref\staudacher{
  B.~Eden and M.~Staudacher,
  %``Integrability and transcendentality,''
  J.\ Stat.\ Mech.\  {\bf 0611}, P014 (2006)
  [arXiv:hep-th/0603157].
  %%CITATION = JSTAT,0611,P014;%%
  N.~Beisert, B.~Eden and M.~Staudacher,
  %``Transcendentality and crossing,''
  J.\ Stat.\ Mech.\  {\bf 0701}, P021 (2007)
  [arXiv:hep-th/0610251].
  %%CITATION = JSTAT,0701,P021;%%
}
%\HofmannQF
\lref\hofmann{
  R.~Hofmann,
  %``Operator product expansion and local quark-hadron duality: Facts and
  %riddles,''
  Prog.\ Part.\ Nucl.\ Phys.\  {\bf 52}, 299 (2004)
  [arXiv:hep-ph/0312130].
  %%CITATION = PPNPD,52,299;%%
}
%\ManoharTZ
\lref\manohar{
  A.~V.~Manohar,
  %``An introduction to spin dependent deep inelastic scattering,''
  arXiv:hep-ph/9204208.
  %%CITATION = HEP-PH/9204208;%%
}
%\RandallVF
\lref\rstwo{
  L.~Randall and R.~Sundrum,
  %``An alternative to compactification,''
  Phys.\ Rev.\ Lett.\  {\bf 83}, 4690 (1999)
  [arXiv:hep-th/9906064].
  %%CITATION = PRLTA,83,4690;%%
}

%\GeorgiSI
\lref\georgi{
  H.~Georgi,
  %``Another Odd Thing About Unparticle Physics,''
  Phys.\ Lett.\  B {\bf 650}, 275 (2007)
  [arXiv:0704.2457 [hep-ph]].
  %%CITATION = PHLTA,B650,275;%%
  H.~Georgi,
  %``Unparticle Physics,''
  Phys.\ Rev.\ Lett.\  {\bf 98}, 221601 (2007)
  [arXiv:hep-ph/0703260].
  %%CITATION = PRLTA,98,221601;%%
}

%\BernEW
\lref\amplitudes{
  Z.~Bern, M.~Czakon, L.~J.~Dixon, D.~A.~Kosower and V.~A.~Smirnov,
  %``The Four-Loop Planar Amplitude and Cusp Anomalous Dimension in Maximally
  %Supersymmetric Yang-Mills Theory,''
  Phys.\ Rev.\  D {\bf 75}, 085010 (2007)
  [arXiv:hep-th/0610248].
  %%CITATION = PHRVA,D75,085010;%%
  R.~Britto, F.~Cachazo, B.~Feng and E.~Witten,
  %``Direct proof of tree-level recursion relation in Yang-Mills theory,''
  Phys.\ Rev.\ Lett.\  {\bf 94}, 181602 (2005)
  [arXiv:hep-th/0501052].
  %%CITATION = PRLTA,94,181602;%%
  F.~Cachazo, P.~Svrcek and E.~Witten,
  %``MHV vertices and tree amplitudes in gauge theory,''
  JHEP {\bf 0409}, 006 (2004)
  [arXiv:hep-th/0403047].
  %%CITATION = JHEPA,0409,006;%%
  }

 \lref\fajm{
 L.~F.~Alday and J.~M.~Maldacena,
  %``Gluon scattering amplitudes at strong coupling,''
  JHEP {\bf 0706}, 064 (2007)
  [arXiv:0705.0303 [hep-th]].
  %%CITATION = JHEPA,0706,064;%%
}

%\StermanWJ
\lref\weinsterman{
  G.~Sterman and S.~Weinberg,
  %``Jets From Quantum Chromodynamics,''
  Phys.\ Rev.\ Lett.\  {\bf 39}, 1436 (1977).
  %%CITATION = PRLTA,39,1436;%%
}
%\NojiriMH
\lref\odintsov{
  S.~Nojiri and S.~D.~Odintsov,
  %``On the conformal anomaly from higher derivative gravity in AdS/CFT
  %correspondence,''
  Int.\ J.\ Mod.\ Phys.\  A {\bf 15}, 413 (2000)
  [arXiv:hep-th/9903033].
  %%CITATION = IMPAE,A15,413;%%
}
\lref\shenkertwo{
  M.~Brigante, H.~Liu, R.~C.~Myers, S.~Shenker and S.~Yaida,
  %``The Viscosity Bound and Causality Violation,''
  arXiv:0802.3318 [hep-th].
  %%CITATION = ARXIV:0802.3318;%%
}

\lref\cornalbae{
  L.~Cornalba,
  %``Eikonal Methods in AdS/CFT: Regge Theory and Multi-Reggeon Exchange,''
  arXiv:0710.5480 [hep-th].
  %%CITATION = ARXIV:0710.5480;%%
}

%\BlauVZ
\lref\bng{
  M.~Blau, K.~S.~Narain and E.~Gava,
  %``On subleading contributions to the AdS/CFT trace anomaly,''
  JHEP {\bf 9909}, 018 (1999)
  [arXiv:hep-th/9904179].
  %%CITATION = JHEPA,9909,018;%%
}

%\BriganteNU
\lref\shenkerone{
  M.~Brigante, H.~Liu, R.~C.~Myers, S.~Shenker and S.~Yaida,
  %``Viscosity Bound Violation in Higher Derivative Gravity,''
  arXiv:0712.0805 [hep-th].
  %%CITATION = ARXIV:0712.0805;%%
}

\lref\npcorrections{
  J.~C.~Collins and D.~E.~Soper,
  %``The Two Particle Inclusive Cross-Section In E+ E- Annihilation At Petra,
  %Pep And Lep Energies,''
  Nucl.\ Phys.\  B {\bf 284}, 253 (1987).
  %%CITATION = NUPHA,B284,253;%%
  D.~de Florian and M.~Grazzini,
  %``The back-to-back region in e+ e- energy energy correlation,''
  Nucl.\ Phys.\  B {\bf 704}, 387 (2005)
  [arXiv:hep-ph/0407241].
  %%CITATION = NUPHA,B704,387;%%
  Y.~L.~Dokshitzer, G.~Marchesini and B.~R.~Webber,
  %``Non-perturbative effects in the energy-energy correlation,''
  JHEP {\bf 9907}, 012 (1999)
  [arXiv:hep-ph/9905339].
  %%CITATION = JHEPA,9907,012;%%
  A.~V.~Belitsky, G.~P.~Korchemsky and G.~Sterman,
  %``Energy flow in QCD and event shape functions,''
  Phys.\ Lett.\  B {\bf 515}, 297 (2001)
  [arXiv:hep-ph/0106308].
  %%CITATION = PHLTA,B515,297;%%
W.~Y.~.~Crutchfield, F.~R.~.~Ore and G.~Sterman,
  %``Quark - Gluon Correlations And Vacuum Fluctuations,''
  Phys.\ Lett.\  B {\bf 102}, 347 (1981).
  %%CITATION = PHLTA,B102,347;%%
   G.~P.~Korchemsky, G.~Oderda and G.~Sterman,
  %``Power corrections and nonlocal operators,''
  arXiv:hep-ph/9708346.
  %%CITATION = HEP-PH/9708346;%%
  G.~P.~Korchemsky and G.~Sterman,
  %``Power corrections to event shapes and factorization,''
  Nucl.\ Phys.\  B {\bf 555}, 335 (1999)
  [arXiv:hep-ph/9902341].
   C.~W.~Bauer, S.~P.~Fleming, C.~Lee and G.~Sterman,
  %``Factorization of e+e- Event Shape Distributions with Hadronic Final States
  %in Soft Collinear Effective Theory,''
  arXiv:0801.4569 [hep-ph].
    C.~Lee and G.~Sterman,
  %``Universality of nonperturbative effects in event shapes,''
{\it In the Proceedings of FRIF workshop on first principles non-perturbative QCD of hadron jets, LPTHE, Paris, France, 12-14 Jan 2006, pp A001}
  [arXiv:hep-ph/0603066].
   }

%\MuellerRP
\lref\generalizedpdf{
  D.~Mueller,
  %``Generalized parton distributions: Theoretical review,''
  Nucl.\ Phys.\  A {\bf 755}, 71 (2005)
  [arXiv:hep-ph/0501158].
  %%CITATION = NUPHA,A755,71;%%
  A.~V.~Belitsky and D.~Mueller,
  %``Theory and phenomenology of generalized parton distributions: A brief
  %overview,''
  arXiv:hep-ph/0105167.
  %%CITATION = HEP-PH/0105167;%%
}

%\SchreierUM
\lref\schreier{
  E.~J.~Schreier,
  %``Conformal symmetry and three-point functions,''
  Phys.\ Rev.\  D {\bf 3}, 980 (1971).
  %%CITATION = PHRVA,D3,980;%%
}

%\TkachovKK
\lref\jetdef{
  F.~V.~Tkachov,
  %``Measuring Multijet Structure of Hadronic Energy Flow Or What IS A Jet?,''
  Int.\ J.\ Mod.\ Phys.\  A {\bf 12}, 5411 (1997)
  [arXiv:hep-ph/9601308].
  %%CITATION = IMPAE,A12,5411;%%
}

%\BeisertJJ
\lref\beisertdil{
  N.~Beisert,
  %``The complete one-loop dilatation operator of N = 4 super Yang-Mills
  %theory,''
  Nucl.\ Phys.\  B {\bf 676}, 3 (2004)
  [arXiv:hep-th/0307015].
  %%CITATION = NUPHA,B676,3;%%
}

%\AbreuUS
\lref\energycorrexp{
  P.~Abreu {\it et al.}  [DELPHI Collaboration],
  %``Energy-energy correlations in hadronic final states from Z0 decays,''
  Phys.\ Lett.\  B {\bf 252}, 149 (1990).
  %%CITATION = PHLTA,B252,149;%%
 M.~Z.~Akrawy {\it et al.}  [OPAL Collaboration],
  %``A Measurement of energy correlations and a determination of alpha-s (M2
  %(Z0)) in e+ e- annihilations at s**(1/2) = 91-GeV,''
  Phys.\ Lett.\  B {\bf 252}, 159 (1990).
  %%CITATION = PHLTA,B252,159;%%
   K.~Abe {\it et al.}  [SLD Collaboration],
  %``Measurement of alpha-s (M(Z)**2) from hadronic event observables at the Z0
  %resonance,''
  Phys.\ Rev.\  D {\bf 51}, 962 (1995)
  [arXiv:hep-ex/9501003].
  %%CITATION = PHRVA,D51,962;%%
   K.~Abe {\it et al.}  [SLD Collaboration],
  %``Measurement of alpha-s from energy-energy correlations at the Z0
  %resonance,''
  Phys.\ Rev.\  D {\bf 50}, 5580 (1994)
  [arXiv:hep-ex/9405006].
  %%CITATION = PHRVA,D50,5580;%%
  }

  \lref\osborn{
   H.~Osborn and A.~C.~Petkou,
  ``Implications of conformal invariance in field theories for general
  dimensions,''
  Annals Phys.\  {\bf 231}, 311 (1994)
  [arXiv:hep-th/9307010].
  %%CITATION = APNYA,231,311;%%
}
%\PolchinskiTT
\lref\jpms{
  J.~Polchinski and M.~J.~Strassler,
  ``Hard scattering and gauge/string duality,''
  Phys.\ Rev.\ Lett.\  {\bf 88}, 031601 (2002)
  [arXiv:hep-th/0109174].
  %%CITATION = PRLTA,88,031601;%%
}

%\BrowerEA
\lref\psregge{
  R.~C.~Brower, J.~Polchinski, M.~J.~Strassler and C.~I.~Tan,
  %``The Pomeron and Gauge/String Duality,''
  JHEP {\bf 0712}, 005 (2007)
  [arXiv:hep-th/0603115].
  %%CITATION = JHEPA,0712,005;%%
}

%\ParisiEG
\lref\cparameter{
  G.~Parisi,
  %``Super Inclusive Cross-Sections,''
  Phys.\ Lett.\  B {\bf 74}, 65 (1978).
  %%CITATION = PHLTA,B74,65;%%
}

%\PolchinskiJU
\lref\jpls{
  J.~Polchinski and L.~Susskind,
  ``String theory and the size of hadrons,''
  arXiv:hep-th/0112204.
  %%CITATION = HEP-TH/0112204;%%
}
%\WittenQJ
\lref\wittenhol{
  E.~Witten,
  %``Anti-de Sitter space and holography,''
  Adv.\ Theor.\ Math.\ Phys.\  {\bf 2}, 253 (1998)
  [arXiv:hep-th/9802150].
  %%CITATION = 00203,2,253;%%
}
\lref\klhigherloop{
 A.~V.~Kotikov and L.~N.~Lipatov,
  %``DGLAP and BFKL equations in the N = 4 supersymmetric gauge theory,''
  Nucl.\ Phys.\  B {\bf 661}, 19 (2003)
  [Erratum-ibid.\  B {\bf 685}, 405 (2004)]
  [arXiv:hep-ph/0208220].
  %%CITATION = NUPHA,B661,19;%%
   A.~V.~Kotikov, L.~N.~Lipatov and V.~N.~Velizhanin,
  %``Anomalous dimensions of Wilson operators in N = 4 SYM theory,''
  Phys.\ Lett.\  B {\bf 557}, 114 (2003)
  [arXiv:hep-ph/0301021].
  %%CITATION = PHLTA,B557,114;%%
 A.~V.~Kotikov, L.~N.~Lipatov, A.~I.~Onishchenko and V.~N.~Velizhanin,
  %``Three-loop universal anomalous dimension of the Wilson operators in N =  4
  %SUSY Yang-Mills model,''
  Phys.\ Lett.\  B {\bf 595}, 521 (2004)
  [Erratum-ibid.\  B {\bf 632}, 754 (2006)]
  [arXiv:hep-th/0404092].
  %%CITATION = PHLTA,B595,521;%%
}

 \lref\kloneloop{
 A.~V.~Kotikov and L.~N.~Lipatov,
  %``DGLAP and BFKL evolution equations in the N = 4 supersymmetric gauge
  %theory,''
  arXiv:hep-ph/0112346.
  %%CITATION = HEP-PH/0112346;%%
 }

\lref\horowitzitzh{
  G.~T.~Horowitz and N.~Itzhaki,
  %``Black holes, shock waves, and causality in the AdS/CFT correspondence,''
  JHEP {\bf 9902}, 010 (1999)
  [arXiv:hep-th/9901012].
  %%CITATION = JHEPA,9902,010;%%
}
 %\HottaQY
\lref\hotta{
  M.~Hotta and M.~Tanaka,
  %``Shock wave geometry with nonvanishing cosmological constant,''
  Class.\ Quant.\ Grav.\  {\bf 10}, 307 (1993).
  %%CITATION = CQGRD,10,307;%%
}
%\BraunRP
\lref\conformalqcd{
  V.~M.~Braun, G.~P.~Korchemsky and D.~Mueller,
  %``The uses of conformal symmetry in QCD,''
  Prog.\ Part.\ Nucl.\ Phys.\  {\bf 51}, 311 (2003)
  [arXiv:hep-ph/0306057].
  %%CITATION = PPNPD,51,311;%%
}

%\StrasslerBV
\lref\msrecent{
  M.~J.~Strassler,
  %``Why Unparticle Models with Mass Gaps are Examples of Hidden Valleys,''
  arXiv:0801.0629 [hep-ph].
  %%CITATION = ARXIV:0801.0629;%%
}

%\KeldyshUD
\lref\Keldysh{
  L.~V.~Keldysh,
  %``Diagram technique for nonequilibrium processes,''
  Zh.\ Eksp.\ Teor.\ Fiz.\  {\bf 47}, 1515 (1964)
  [Sov.\ Phys.\ JETP {\bf 20}, 1018 (1965)].
  %%CITATION = SPHJA,20,1018;%%
}

%\DaviesYV
\lref\twoparticle{
  P.~C.~W.~Davies and S.~A.~Fulling,
  %``Radiation From Moving Mirrors And From Black Holes,''
  Proc.\ Roy.\ Soc.\ Lond.\  A {\bf 356}, 237 (1977)
  %%CITATION = PRSLA,A356,237;%%
}

%\BakshiDV
\lref\maham{
  P.~M.~Bakshi and K.~T.~Mahanthappa,
  %``Expectation value formalism in quantum field theory. 1,''
  J.\ Math.\ Phys.\  {\bf 4}, 1 (1963).
  %%CITATION = JMAPA,4,1;%%
P.~M.~Bakshi and K.~T.~Mahanthappa,
  %``Expectation value formalism in quantum field theory. 2,''
  J.\ Math.\ Phys.\  {\bf 4}, 12 (1963).
  %%CITATION = JMAPA,4,12;%%
  }

\lref\kllong{
  A.~V.~Kotikov and L.~N.~Lipatov,
  %``DGLAP and BFKL equations in the N = 4 supersymmetric gauge theory,''
  Nucl.\ Phys.\  B {\bf 661}, 19 (2003)
  [Erratum-ibid.\  B {\bf 685}, 405 (2004)]
  [arXiv:hep-ph/0208220].
  %%CITATION = NUPHA,B661,19;%%
}

\lref\braun{  I.~I.~Balitsky and V.~M.~Braun,
  %``Evolution Equations for QCD String Operators,''
  Nucl.\ Phys.\  B {\bf 311}, 541 (1989).
  %%CITATION = NUPHA,B311,541;%%
}

\lref\veneziano{
  K.~Konishi, A.~Ukawa and G.~Veneziano,
  %``Jet Calculus: A Simple Algorithm For Resolving QCD Jets,''
  Nucl.\ Phys.\  B {\bf 157}, 45 (1979).
  %%CITATION = NUPHA,B157,45;%%
}
%\SveshnikovVI
\lref\tkachov{
  N.~A.~Sveshnikov and F.~V.~Tkachov,
  %``Jets and quantum field theory,''
  Phys.\ Lett.\  B {\bf 382}, 403 (1996)
  [arXiv:hep-ph/9512370].
  %%CITATION = PHLTA,B382,403;%%
}
\lref\necflat{
  G.~Klinkhammer,
  %``Averaged energy conditions for free scalar fields in flat space-times,''
  Phys.\ Rev.\  D {\bf 43}, 2542 (1991).
  %%CITATION = PHRVA,D43,2542;%%
   R.~Wald and U.~Yurtsever,
  %``General proof of the averaged null energy condition for a massless scalar
  %field in two-dimensional curved space-time,''
  Phys.\ Rev.\  D {\bf 44}, 403 (1991).
  %%CITATION = PHRVA,D44,403;%%
}

%\YurtseverWC
\lref\yurt{
  U.~Yurtsever,
  %``The Averaged null energy condition and difference inequalities in quantum
  %field theory,''
  Phys.\ Rev.\  D {\bf 51}, 5797 (1995)
  [arXiv:gr-qc/9411056].
  %%CITATION = PHRVA,D51,5797;%%
}

%\HorowitzBV
\lref\horsteif{
  G.~T.~Horowitz and A.~R.~Steif,
  %``Space-Time Singularities in String Theory,''
  Phys.\ Rev.\ Lett.\  {\bf 64}, 260 (1990).
  %%CITATION = PRLTA,64,260;%%
}

%\WittenQJ
\lref\wittenhol{
  E.~Witten,
  %``Anti-de Sitter space and holography,''
  Adv.\ Theor.\ Math.\ Phys.\  {\bf 2}, 253 (1998)
  [arXiv:hep-th/9802150].
  %%CITATION = 00203,2,253;%%
}

%\FreedmanTZ
\lref\freedmanthree{
  D.~Z.~Freedman, S.~D.~Mathur, A.~Matusis and L.~Rastelli,
  %``Correlation functions in the CFT($d$)/AdS($d+1$) correspondence,''
  Nucl.\ Phys.\  B {\bf 546}, 96 (1999)
  [arXiv:hep-th/9804058].
  %%CITATION = NUPHA,B546,96;%%
}

  %\MetsaevYB
\lref\tseytlinhigher{
  R.~R.~Metsaev and A.~A.~Tseytlin,
  %``CURVATURE CUBED TERMS IN STRING THEORY EFFECTIVE ACTIONS,''
  Phys.\ Lett.\  B {\bf 185}, 52 (1987).
  %%CITATION = PHLTA,B185,52;%%
}

%\GubserBC
\lref\GubserBC{
  S.~S.~Gubser, I.~R.~Klebanov and A.~M.~Polyakov,
  %``Gauge theory correlators from non-critical string theory,''
  Phys.\ Lett.\  B {\bf 428}, 105 (1998)
  [arXiv:hep-th/9802109].
  %%CITATION = PHLTA,B428,105;%%
}

%\AharonyDJ
\lref\AharonyDJ{
  O.~Aharony and Y.~Tachikawa,
  %``A holographic computation of the central charges of d=4, N=2 SCFTs,''
  JHEP {\bf 0801}, 037 (2008)
  [arXiv:0711.4532 [hep-th]].
  %%CITATION = JHEPA,0801,037;%%
}

\lref\brodsky{
  S.~J.~Brodsky and G.~F.~de Teramond,
  %``AdS/CFT and Light-Front QCD,''
  arXiv:0802.0514 [hep-ph].
  %%CITATION = ARXIV:0802.0514;%%
 S.~J.~Brodsky and G.~F.~de Teramond,
  %``Light-Front Dynamics and AdS/QCD: The Pion Form Factor in the Space- and
  %Time-Like Regions,''
  arXiv:0707.3859 [hep-ph].
  %%CITATION = ARXIV:0707.3859;%%
}
\lref\brodskylepage{
  G.~P.~Lepage and S.~J.~Brodsky,
  %``Exclusive Processes In Quantum Chromodynamics: Evolution Equations For
  %Hadronic Wave Functions And The Form-Factors Of Mesons,''
  Phys.\ Lett.\  B {\bf 87}, 359 (1979).
  %%CITATION = PHLTA,B87,359;%%
}

%complex angular momentum PDF.
%Also discussion of transverse PDF's. "Generalized parton distribution functions"

%\MuellerED
\lref\complexj{
  D.~Mueller and A.~Schafer,
  %``Complex conformal spin partial wave expansion of generalized parton
  %distributions and distribution amplitudes,''
  Nucl.\ Phys.\  B {\bf 739}, 1 (2006)
  [arXiv:hep-ph/0509204].
  %%CITATION = NUPHA,B739,1;%%
}

%\PolchinskiJW
\lref\psdis{
  J.~Polchinski and M.~J.~Strassler,
  %``Deep inelastic scattering and gauge/string duality,''
  JHEP {\bf 0305}, 012 (2003)
  [arXiv:hep-th/0209211].
  %%CITATION = JHEPA,0305,012;%%
}

\lref\hiddenvalley{
  M.~J.~Strassler and K.~M.~Zurek,
  %``Echoes of a hidden valley at hadron colliders,''
  Phys.\ Lett.\  B {\bf 651}, 374 (2007)
  [arXiv:hep-ph/0604261].
  %%CITATION = PHLTA,B651,374;%%
}

%\GreenSP
\lref\GSW{
  M.~B.~Green, J.~H.~Schwarz and E.~Witten,
  %``SUPERSTRING THEORY. VOL. 1: INTRODUCTION,''
%\href{http://www.slac.stanford.edu/spires/find/hep/www?irn=1755021}{SPIRES entry}
{\it  Cambridge, Uk: Univ. Pr. ( 1987) 469 P. ( Cambridge Monographs On Mathematical Physics)}
}

%\GrinsteinQK
\lref\keni{
  B.~Grinstein, K.~Intriligator and I.~Z.~Rothstein,
  %``Comments on Unparticles,''
  arXiv:0801.1140 [hep-ph].
  %%CITATION = ARXIV:0801.1140;%%
}

\lref\ColemanYG{
  S.~R.~Coleman and B.~Grossman,
  %``'T Hooft's Consistency Condition As A Consequence Of Analyticity And
  %Unitarity,''
  Nucl.\ Phys.\  B {\bf 203}, 205 (1982).
  %%CITATION = NUPHA,B203,205;%%
}

\lref\MikhailovBP{
  A.~Mikhailov,
  %``Notes on higher spin symmetries,''
  arXiv:hep-th/0201019.
  %%CITATION = HEP-TH/0201019;%%
}

\lref\DobrevRU{
  V.~K.~Dobrev, V.~B.~Petkova, S.~G.~Petrova and I.~T.~Todorov,
  %``Dynamical Derivation Of Vacuum Operator Product Expansion In Euclidean
  %Conformal Quantum Field Theory,''
  Phys.\ Rev.\  D {\bf 13}, 887 (1976).
  %%CITATION = PHRVA,D13,887;%%
}

 %\FerraraYT
\lref\FerraraYT{
  S.~Ferrara, A.~F.~Grillo and R.~Gatto,
  %``Tensor representations of conformal algebra and conformally covariant
  %operator product expansion,''
  Annals Phys.\  {\bf 76}, 161 (1973).
  %%CITATION = APNYA,76,161;%%
}

% N=1 superconformal

%% Discusses two parameters in the correlators of N=1 supercurrents. Also a single two point function.
%\OsbornQU
\lref\onlyosborn{
  H.~Osborn,
  %``N = 1 superconformal symmetry in four-dimensional quantum field theory,''
  Annals Phys.\  {\bf 272}, 243 (1999)
  [arXiv:hep-th/9808041].
  %%CITATION = APNYA,272,243;%%
}

%% Apparently wrong...
%\ParkBQ
\lref\ParkBQ{
  J.~H.~Park,
  %``N = 1 superconformal symmetry in 4-dimensions,''
  Int.\ J.\ Mod.\ Phys.\  A {\bf 13}, 1743 (1998)
  [arXiv:hep-th/9703191].
  %%CITATION = IMPAE,A13,1743;%%
}

% BOSONic T and V
%\ErdmengerYC
\lref\ErdmengerYC{
  J.~Erdmenger and H.~Osborn,
  %``Conserved currents and the energy-momentum tensor in conformally  invariant
  %theories for general dimensions,''
  Nucl.\ Phys.\  B {\bf 483}, 431 (1997)
  [arXiv:hep-th/9605009].
  %%CITATION = NUPHA,B483,431;%%
}

%% Discussion of a and c and proof that J RR goes as c-a
%\AnselmiAM
\lref\freedmanc{
  D.~Anselmi, D.~Z.~Freedman, M.~T.~Grisaru and A.~A.~Johansen,
  %``Nonperturbative formulas for central functions of supersymmetric gauge
  %theories,''
  Nucl.\ Phys.\  B {\bf 526}, 543 (1998)
  [arXiv:hep-th/9708042].
  %%CITATION = NUPHA,B526,543;%%
   D.~Anselmi, J.~Erlich, D.~Z.~Freedman and A.~A.~Johansen,
  %``Positivity constraints on anomalies in supersymmetric gauge theories,''
  Phys.\ Rev.\  D {\bf 57}, 7570 (1998)
  [arXiv:hep-th/9711035].
  %%CITATION = PHRVA,D57,7570;%%
}

% Values of a and c
%\ChristensenGI
\lref\duff{
  S.~M.~Christensen and M.~J.~Duff,
  %``Axial And Conformal Anomalies For Arbitrary Spin In Gravity And
  %Supergravity,''
  Phys.\ Lett.\  B {\bf 76}, 571 (1978).
  %%CITATION = PHLTA,B76,571;%%
}

%\cite{Duff:1977ay}
\lref\duffobs{
  M.~J.~Duff,
  %``Observations On Conformal Anomalies,''
  Nucl.\ Phys.\  B {\bf 125}, 334 (1977).
  %%CITATION = NUPHA,B125,334;%%
}

%%%% NEw refs

%\ErdmengerYC
\lref\erdosbon{
  J.~Erdmenger and H.~Osborn,
  %``Conserved currents and the energy-momentum tensor in conformally  invariant
  %theories for general dimensions,''
  Nucl.\ Phys.\  B {\bf 483}, 431 (1997)
  [arXiv:hep-th/9605009].
  %%CITATION = NUPHA,B483,431;%%
}

%\ErdmengerXX
\lref\gravanomaly{
  J.~Erdmenger,
  %``Gravitational axial anomaly for four dimensional conformal field
  %theories,''
  Nucl.\ Phys.\  B {\bf 562}, 315 (1999)
  [arXiv:hep-th/9905176].
  %%CITATION = NUPHA,B562,315;%%
}

%\BelitskyYS
\lref\begrkr{
  A.~V.~Belitsky, A.~S.~Gorsky and G.~P.~Korchemsky,
  %``Gauge / string duality for QCD conformal operators,''
  Nucl.\ Phys.\  B {\bf 667}, 3 (2003)
  [arXiv:hep-th/0304028].
  %%CITATION = NUPHA,B667,3;%%
}

%%% REFS ENERGY CORREL

\lref\ellis{  C.~L.~Basham, L.~S.~Brown, S.~D.~Ellis and S.~T.~Love,
  ``Energy Correlations In Electron-Positron Annihilation In Quantum
  Chromodynamics: Asymptotically Free Perturbation Theory,''
  Phys.\ Rev.\  D {\bf 19}, 2018 (1979).
  %%CITATION = PHRVA,D19,2018;%%
 C.~L.~Basham, L.~S.~Brown, S.~D.~Ellis and S.~T.~Love,
  ``Energy Correlations In Electron - Positron Annihilation: Testing QCD,''
  Phys.\ Rev.\ Lett.\  {\bf 41}, 1585 (1978).
  %%CITATION = PRLTA,41,1585;%%
  C.~L.~Basham, L.~S.~Brown, S.~D.~Ellis and S.~T.~Love,
  %``Electron - Positron Annihilation Energy Pattern In Quantum Chromodynamics:
  %Asymptotically Free Perturbation Theory,''
  Phys.\ Rev.\  D {\bf 17}, 2298 (1978).
  %%CITATION = PHRVA,D17,2298;%%
   }

 \lref\jm{
    J.~M.~Maldacena,
  %``The large N limit of superconformal field theories and supergravity,''
  Adv.\ Theor.\ Math.\ Phys.\  {\bf 2}, 231 (1998)
  [Int.\ J.\ Theor.\ Phys.\  {\bf 38}, 1113 (1999)]
  [arXiv:hep-th/9711200].
  %%CITATION = IJTPB,38,1113;%%
}

%%%%%%%%%%%%%%%%%%%%%%%%%%%%%%%%%%%%%%%%%%%%%%%%%%%%%%5

%%%%%%%%%%%%%%%%%%%%%%%%%%%%%%%%%%%%%%%%%%%%%%%%%%%%%%%%%%%%%%%%%%%%%%%%%

%%%%%%%%%%%%%%%%%%%%%%%%%%%%%%%%%%%%%%%%%%%%%%%%%%%%%%%%%%%%%%%%%%%%

\Title{
% \vbox{\baselineskip-12pt \hbox{ } \hbox{ } }
} {\vbox{\centerline{ Conformal collider physics:    }
\centerline{ Energy and charge correlations  } }}
\bigskip

% \bigskip
\centerline{Diego M. Hofman$^{a}$ and Juan Maldacena$^b$}
\bigskip
\centerline{\it $^a$   Joseph Henry Laboratories, Princeton
University, Princeton, NJ 08544, USA}

\centerline{ \it  $^b$School of Natural Sciences, Institute for
Advanced Study} \centerline{\it Princeton, NJ 08540, USA}

\vskip .3in \noindent
%%%%%%%%%%%%%%%%%%%%%%%%%%%%%%%%%%%%%%%%%%%%%%%%%%%%%%%%%%%%%%%%%%%%%%%%%%%%%%%%%%%%%%%%%%%%

We study  observables in a conformal field theory which are very closely related
to the ones used to describe hadronic events at colliders. We focus on the correlation
functions of the energies deposited on   calorimeters placed at a large distance from the collision.
We consider
initial states produced by an operator insertion and we study some general properties of the
energy correlation functions for  conformal field theories.  We  argue that
the small angle singularities of energy correlation functions are
controlled by the twist of  non-local light-ray operators with a definite spin.
We relate the charge two point function to a particular moment of
the parton distribution functions appearing in deep inelastic scattering.
The one point energy correlation
functions are   characterized by a few numbers.
For ${\cal N}=1$ superconformal theories the one point function
for states created by the R-current or the stress tensor  are determined
by the two parameters $a$ and $c$ characterizing the conformal anomaly.  Demanding
that the measured energies are positive we get   bounds on $a/c$.
We also give a prescription for computing the energy and charge correlation functions in theories
that have a gravity dual. The prescription amounts to probing the falling string state as it
crosses the $AdS$ horizon with gravitational shock waves. In the leading, two derivative,
gravity approximation the
energy is uniformly distributed on the sphere at infinity, with no fluctuations. We compute
the stringy corrections and we show that they lead to small, non-gaussian, fluctuations in the
energy distribution. Corrections to
the one point functions or antenna patterns are related to
higher derivative corrections in the bulk.

%%%%%%%%%%%%%%%%%%%%%%%%%%%%%%%%%%%%%%%%%%%%%%%%%%%%%%%%%%%%%%%%%%%%%%%%%%%%%%%%%%%%%%%%%%%%%%%%%%%%%%%%%%%%%%%%%%%%%%%%%%%%%%%%%%%%%%%%%%%%%%%%%%%%%%%%%%%%%%%%%%%%%%%%%%%%%%%%%%%%%%%%

 \Date{ }

%%%%%%%%%%%%%%%%%%%%%%%%%%%%%%%%%%%%%%%%%%%%%%%%%%%%%%%%%%%%%%%%%%%%%%%%%

\newsec{Introduction}

In this paper we consider conformal field theories and we study physical processes that
are closely related to the ones studied at particle colliders. In some sense we will be studying
``conformal collider physics''.
We consider an external perturbation that is localized in space and time near $t \sim \vec x \sim 0$.
This external perturbation couples to some operator ${\cal O}$ of the conformal field theory and
produces a localized excitation in the conformal field theory.
This excitation then grows in size and propagates
outwards. We want to study the properties of the state that is produced.
For this purpose we consider idealized ``calorimeters''
that  measure
the total flux of energy per unit angle far away from the region where the localized perturbation
was concentrated.
As a particular example one could have in mind   a real world process
 $e^+ e^- \to \gamma^* \to $hadrons\foot{For early work on the applications of scale invariance to strong interactions and, in
particular, $e^+e^-$ collisions, see \polscale .}, where we produce hadrons via an intermediate off shell photon. We can treat the process to
lowest order in the electromagnetic coupling constant and to all orders in the strong coupling constant. The QCD computation reduces to studying
the state created on the QCD vacuum by the electromagnetic current $j^\mu_{em}$. From the point of view of QCD this current is simply a global
symmetry. In this case the theory is not conformal, but   at high enough energies we can  approximate the process as a conformal one to the
extent that we can ignore the running of the coupling and the details of the hadronization process. In this paper we will analyze  similar
processes but in conformal field theories.

%%% TO PUT FIGURES INSERT:
\ifig\calorimeters{ A localized excitation is produced in a conformal field theory and its
decay products are measured by calorimeters sitting far away.
  } {\epsfxsize2.5in\epsfbox{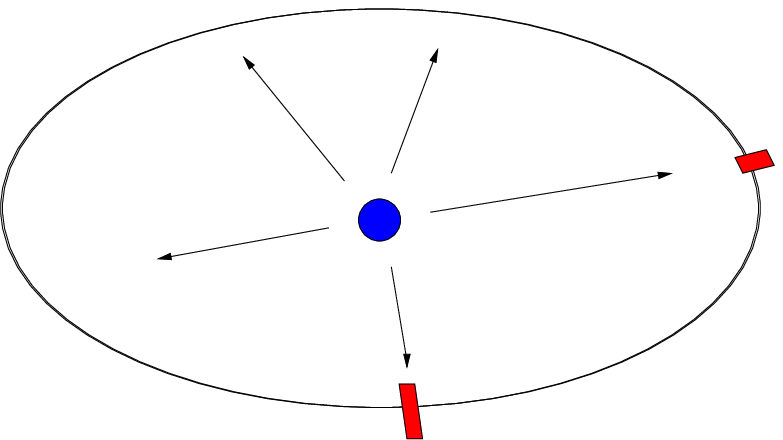}}

Our goal is to describe features of the produced state. For example, at weak coupling
we expect to see a certain number of fairly well defined jets. At strong coupling we expect
to see a more spherically symmetric distribution \refs{\psdis,\fallingmass,\msrecent}.
We need suitably inclusive variables
which are IR finite.
 In QCD this is commonly done using inclusive
 jet observables \weinsterman , see \stermanreview\ for a review.
 In this paper we
study a particularly simple set of  inclusive observables which are the
 energy correlation functions,
originally introduced in \ellis .
They are defined as follows. We place calorimeters at angles $\theta_1, \cdots , \theta_n$ and
we measure the total energy per unit angle deposited at each of these angles. We multiply
all these energies together and compute the average over all events.
These are also inclusive, IR finite observables which one could use to study properties of
the produced state.
Energy correlation functions for hadronic final states
have been measured experimentally and they are  one of the ways
of making precise determinations of $\alpha_s$
  (see \energycorrexp\ for example).

A nice feature of  energy correlation functions  is
that they are  defined in    terms of correlation functions of  local
 gauge invariant operators.
They  are given   in terms of the stress tensor operator \tkachov .
 More precisely, consider
the expression for the integrated energy flux per unit angle
at a large sphere of radius $r$
\eqn\energfl{
{\cal E}(\theta) = \lim _{r \to \infty} r^2  \int_{-\infty}^\infty dt \, n^i T^0_{~ i}(t, r \vec n^i)
}
where $n^i$ is a unit vector in $R^3$ and it specifies the point on the $S^2$ at infinity where
we have our ``calorimeter''.  If we integrate this quantity over all angles we get the
total energy flux which is equal to the energy deposited by the operator insertion.
Energy correlation functions are defined as the quantum expectation value of a
product of energy flux operators on the state produced by the localized operator insertion
\eqn\encorr{
\langle {\cal E}(\theta_1) \cdots {\cal E}(\theta_n) \rangle \equiv  { \langle 0 | {\cal O}^\dagger
 {\cal E}(\theta_1) \cdots {\cal E}(\theta_n) { \cal O} |0 \rangle \over
 \langle 0 | {\cal O}^\dagger
   { \cal O} |0 \rangle }
 }
where ${\cal O}$ is the operator creating the localized perturbation.
 Note that the operators are ordered as written, they are not time ordered. Notice, also, that the
 expectation values in the left hand side of \encorr\ are defined on the particular state created by
 the operator ${\cal O}$ and they are {\it not} vacuum expectation values.
The energy operators are very far away from each other and they  commute with each other. 
This will become more clear below when we think of the
operators as acting  on null outgoing infinity, sometimes called ${\cal J}^+$.
  Of course, we usually think of the energy deposited at various calorimeters as
  commuting
  observables, since we measure them simultaneously.
 Notice that when we compute an $n$ point function we place calorimeters at $n$ points but
 we also allow energy to go through the regions where we have not placed calorimeters.

 In this paper we will assume that
 we have a conformal field theory. There are several motivations for doing so.
 First, the conformal case is simpler because it has more symmetry and, at the same time, 
 it allows us to
 consider theories that are strongly coupled. There are some interesting statements
 that can be made using conformal symmetry.  Second, we could have a theory for new physics beyond
 the Standard Model
 which is conformal, as in the Randall-Sundrum II \rstwo\ or  the unparticle \georgi\ scenarios, or
 approximately conformal, as in the
 ``hidden valley''  scenario \hiddenvalley . One would like to describe the events in these
 theories.
 In order for energy correlations to be observable to us
 we need some way to transfer the energy from the new sector  back to the
 standard model, as in \hiddenvalley . Depending on the details, this conformal breaking and
 conversion process might or might not destroy the energy correlations one computes in the
 conformal theory. We will not discuss this problem here.
  A similar issue arises in QCD.
 For a sample of references on the influence of hadronization on energy
 correlations for QCD see \refs{\ellis,\npcorrections}.
  The final motivation is a more theoretical one, which is to understand better the AdS/CFT
  correspondence \refs{\jm,\gkp,\wittenhol}. Energy correlations are natural observables
  on the field theory side which one would like to understand using gravity and string theory
  in AdS.
  %Energy correlations are intimately related to jet observables and they are an interesting
  %and physical way to measure a state in a conformal field theory.
  We will see that on the
  gravity side, energy correlations translate into the probing of a string state, created by the
  localized perturbation, with a gravitational shock wave as it falls into the AdS horizon. Thus, the problem becomes a high
  energy scattering calculation in the bulk.

This paper is organized as follows. In section two we make some general remarks on energy correlation functions in conformal field theories. By
making conformal transformations we can picture the problem in various ways. We also make some remarks on the small angle behavior of the
correlators  when two of the energy operators come close together. We point out that this small angle behavior can be analyzed by means of an
operator product expansion which involves non-local light-ray  operators which are closely related to the ones that appear
 in the discussion of
deep inelastic scattering. We also relate a  moment of the deep inelastic cross section, or parton distribution function, to a particular
energy two point correlation function. Finally, we consider the general form of the energy one point function $\langle {\cal E}(\theta) \rangle
$ and relate it to vacuum expectation values of three point functions.

In section three  we study conformal field theories that have a gravity (or string theory) dual and we describe a prescription for computing the
energy correlation functions. The  procedure amounts to taking a ``snapshot'' of the wavefunction of the   state produced by the
  operator insertion. In the gravity approximation
we find that the energy correlation functions are perfectly spherically symmetric
 as was expected from the very rapid fragmentation that one expects at
strong coupling. This phenomenon was originally analyzed in deep inelastic processes in  \psdis (see also \msrecent ).

In section four we discuss the leading stringy corrections. They amount to small fluctuations in the energy distribution of
order $1/\sqrt{\lambda}$.
We also consider these corrections for charge correlations which have interesting features in the case that the charges are carried by flavor
symmetries. Finally, we study the regime where two of the angles come close together and find
that the result is determined by the energy of
peculiar non-local string states which are dual to the light-ray operators that appeared in the general field theory discussion. These operators
have a high conformal dimension at strong coupling going like $\Delta \sim \lambda^{1/4}$.

In section five we present a summary, conclusions and a discussion of open problems.

\newsec{ Energy correlations in  conformal field theories  }

In this section we study energy correlation functions in general conformal field theories. The discussion in this section is valid for any value of the
coupling.

\subsec{Energy correlations in various coordinates systems}

The goal of this subsection is to think about energy correlations in various coordinate systems in order to make manifest its various  properties and also in
order to simplify later computations.

It is interesting to take a step back and think about the energy density as follows. For any generator, $G$, of the conformal group there is an associated
conformal killing vector
  $\zeta_G^\mu$ ($x^\mu \to x^\mu + \zeta_G^\mu $). The associated conserved charge
  can be written
  as the integral of a conserved current,  constructed by contracting  $\zeta^\mu_G$ with
   the stress tensor, over
a spatial hypersurface \eqn\genexp{ Q_{G} =  \int_{\Sigma_3} *_4  j_G ~,~~~~~~~~~~~~ j_G^\mu \equiv T_{\mu \nu} \zeta^\nu_{G} } where the normalization of the
stress tensor is chosen so that $T_{\mu \nu} =  -{ 2 \over \sqrt{g} } { \delta S \over \delta g^{\mu \nu} } $.
% We will be interested in the limit
% in which the spatial hypersurface $\Sigma_3$ becomes null.
This expression of the charges is covariant under conformal transformations.  It is also invariant under Weyl transformations of the four
dimensional metric\foot{We are ignoring the conformal anomaly since it only contributes as a $c$-number, independent of the quantum state of the
field theory.}  $ g_{\mu \nu} \to \Omega^2 g_{\mu \nu}$, $T_{\mu \nu} \to \Omega^{-2} T_{\mu \nu}$.

It is convenient to understand clearly the symmetries of the problem. We are interested in measuring the flux of energy at large distances.
Thus, we focus our attention on the boundary of Minkowski space $R^{1,3}$. The conformal generators that leave the boundary fixed
 are the dilatation and   the Poincare generators,  including the translations $P^\mu$ and the $SO(1,3)$ lorentz
transformations.  In other words,
  we have the whole conformal group except the special conformal transformations.
In order to see that the large $r$ limit in \energfl\ is well defined, and also to gain some more
insight into the problem, it is convenient to perform a
conformal transformation from the original coordinates $x^\mu$ to new coordinates $y^\mu$. The new coordinates are such that the future boundary of the
original Minkowski space is mapped to the null surface $y^+=0$. The explicit change of coordinates is\foot{This type of coordinates has also been studied in \cornalbae .} \eqn\explictch{ y^+ = -{ 1 \over x^+} ~,~~~~~~y^- =  x^-
- { x_1^2 + x_2^2 \over x^+} ~,~~~~~ y^{1} = {  x^{1} \over x^+} ~,~~~~~~y^{2} = {  x^{2} \over x^+} } where $y^\pm = y^0 \pm y^3$, and similarly for $x^\pm$.
The inverse change of coordinates is given by the same expressions with $x\leftrightarrow y$. The advantage of the new coordinates is that now the energy
is expressed in terms of an integral over the surface at $y^+=0$ and we do not have to take any limit, such as the large $r$ limit in \energfl .  Actually, to
be more precise, the surface $y^+=0$ corresponds to the future lightlike boundary of Minkowski space. The  energy correlation function \energfl\ involves an
integral over the past and the future boundaries of Minkowski space. However, in the physical situation we are interested in, where we have the vacuum  in the
past, there is no contribution from the past light-like boundary and we can focus only on the future boundary.
Of course, one could also directly define the energy flux operator in terms of an integral over only
the future boundary.

In order to switch between different coordinate systems it is convenient to think about $R^{1,3}$ as
 follows. We introduce the six coordinates $Z^M$
subject to the identification $Z^M \sim \lambda Z^M$ and the constraint\foot{Note that $Z^{-1}$ is the ``minus one'' component of the vector $Z$ and it does
not denote the inverse of $Z$. Hopefully, this notation will not cause confusion.    } \eqn\constrz{ - (Z^{-1})^2 - (Z^0)^2 + (Z^1)^2 + (Z^2)^2+ (Z^3)^2+
(Z^4)^2   =0 } The usual coordinates on $R^{1,3}$ are projective coordinates
$x^\mu = { Z^\mu \over Z^{-1} + Z^4 } $, $\mu =0,1,2,3$.
 The metric induced on this surface,
\constrz , by the $R^{2,4}$ metric is fixed up to an overall $x$-dependent factor.
  We can choose a metric by choosing a ``gauge condition'' such as $Z^{-1} + Z^4 =1$.
   Different ``gauge conditions''  lead to metrics that
differ by a Weyl rescaling.   The coordinates $y^\mu$ in \explictch\ correspond to the choice \eqn\cooth{
 y^0 = -{ Z^{-1} \over Z^0 + Z^3 } ,~~~~~y^3 = -{ Z^4 \over Z^0 + Z^3 } ~,~~~~~~y^{1} = { Z^{1}
 \over Z^0 + Z^3 }~,~~~~~~y^{2} = { Z^{2}
 \over Z^0 + Z^3 }
 }
 In fact, using \cooth\ and \explictch\ we can easily go between the two sets of coordinates.
 We have $dx^\mu dx_\mu = { dy^\mu dy_\mu \over (y^+)^2 } $.
 We also clearly see that \explictch\  amounts to a ${\pi \over 2 }$
  rotation in the [-1,0] plane and
 in the [4,3] plane of $R^{2,4}$, which is an element of the conformal group.
The boundary of Minkowski space is the null surface
given by
 $Z^{-1} + Z^4 =0$.  We can think of the various generators of the conformal group as the antisymmetric
matrices $M^{[MN]}$ which generate the transformations $\delta Z^N = M^{[NM]} Z_M$\foot{This $SO(2,4)$ manifestly invariant formalism has also been studied recently in \barstt .}. Defining $Z^\pm = Z^{-1} \pm Z^4$, we can see that all the generators that
leave the surface $Z^+ = 0$ invariant are all the ones with no $+$ index plus the generator $M^{[+-]}$. In this language the four momentum generators in the
$x$ coordinates correspond to $M^{[-\mu]}$, $\mu =0,1,2,3,4$. These generators have a particularly simple form at $Z^+=0$ \eqn\simplegen{ P_\mu \sim Z_\mu {
\partial \over \partial Z^-} - Z_{-} {\partial \over \partial Z^\mu } ~~~\longrightarrow ~~~\left. P_\mu \right|_{Z^+=0} \sim  Z_\mu { \partial \over \partial
Z^-} } (note that $Z_- = -Z^+/2$). Since the Killing vectors are all proportional to each other, then all four generators involve a single component of the
stress energy tensor. Using \genexp , \cooth\ and \simplegen\ we can write \eqn\exprg{  \eqalign{
  P_x^0+P_x^3 = &  \int  dy_1 dy_2 \, { \cal E}(y_1,y_2)
  \cr
  P_x^0-P_x^3 = & \int  dy_1 dy_2 \,    ( y_1^2 + y_2^2) { \cal E}(y_1,y_2)
  \cr
  P_x^1 = & \int  dy_1 dy_2 \,   y^1   { \cal E}(y_1,y_2)
  \cr
  P_x^2 = & \int  dy_1 dy_2 \, y^2 { \cal E}(y_1,y_2)
\cr
  { \cal E}(y_1,y_2) \equiv &  2 \int_{-\infty}^\infty  dy^- T_{--}(y^-,y^+=0,y^1,y^2)
    } }
We see that they are all determined by $T_{--}$ thanks to the simple form of the generators at $Z^+=0$ \simplegen . The conclusion is that we are computing
correlation functions of $T_{--}$ and these determine all the components of the energy and the momentum. These expression have the advantage that no limit is
involved but they have the disadvantage that the $SO(3)$ rotation symmetry is not manifest. Since no limit is involved, it is clear that the expectation values
of \exprg\ will be finite. In fact, we are considering an external operator insertion which is localized in $x$ space. This implies, in particular, that it is
localized near $x^+ \sim 0$ so that it is far enough from $y^+=0$ which is the point where we insert the operators \exprg .

We should note that the dilatation symmetry of the original coordinates $x^\mu \to \lambda x^\mu$ becomes a boost in the $y^+,y^-$ plane in the $y$ coordinates
\explictch . Similarly the dilatation transformation in the $y$ variables becomes a boost  in the $x^\pm$ plane.

%The $x$-momentum generators $P^\mu$ correspond to the following
%In terms of the $y$ coordinates \explictch\ these correspond to the following generators
%\eqn\gener{ P_x^+ = P_x^0 + P_x^3 = - P_y^+ ~,~~~~~~~ P_x^- = P_x^0 -
%P_x^3 =  K_y^- ~,~~~~P_x^i = L_y^{[-,i]}, ~~i=1,2 } where $L_y$ is a Lorentz transformation  and $K_y$ a special conformal transformation in the $y$
%coordinates.  The subscripts $x,y$ indicate that the generators are viewed as acting on the $x$ or $y$ coordinates respectively. Once we have this
%identification of generators we can write the corresponding Killing vectors at  $y^+=0$ and obtain the expression of the $x$-translation generators in the $y$
%coordinates

%%% TO PUT FIGURES INSERT:
\ifig\coordinates{ (a) Penrose diagram of flat Minkowski space. The doted line is a surface at constant $r$ where we measure the energy flux. In the large $r$
limit this becomes the light-like boundary, ${\cal J}^+$,  of Minkowski space. We consider only the future part of the boundary. The semicircle represents a
localized operator insertion.  In (b) we extend the coordinates to the conformal completion of Minkowski space, which gives us $S^3 \times R$. The future
boundary of the original space is simply the light-cone of the point at spatial infinity, $i^0$.
  } {\epsfxsize1.5in\epsfbox{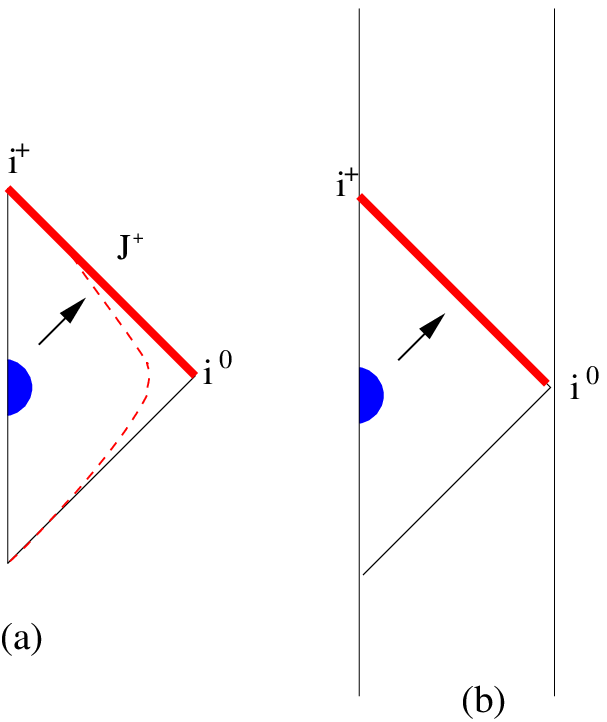}}

An alternative point of view is the following. We write the original coordinates as \eqn\origco{ ds^2 = -dt^2 + dr^2 + r^2 d\Omega_2^2 = r^2 \left[ { - dt^2 +
dr^2 \over r^2 } + d\Omega_2^2 \right]
% r^2 \left[ { - d\tau^2 + d \sigma^2 \over \sin^2 \sigma } + d\Omega_2^2 \right]
} The original metric and the bracketed metric in \origco\ differ by a Weyl transformation, but such a transformation leaves the physics of the CFT invariant.
So we can view our CFT as defined on an extremal black hole: $AdS^2 \times S^2$. Then, the boundary of Minkowski space corresponds to the  black hole horizon
situated at $ t, r= \infty$. We see that we can view our measurement as one done at the horizon of an extremal black hole. (Of course we can also consider
other coordinates related by Weyl transformations which would suggest other pictures.)
 By introducing new coordinates we can write the $AdS_2$ metric in \origco\ as
\eqn\metrnewc{ ds^2 = { -dt^2 + dr^2 \over r^2 } = { - d\tau^2 + d\sigma^2 \over \sin^2 \sigma } ~,~~~~~~~~~~t = { \sin \tau \over \cos \tau + \cos \sigma }
~,~~~~~ r = { \sin \sigma \over \cos \tau + \cos \sigma  } } The horizon is at $\tau^+ \equiv \tau + \sigma = \pi$.   We also define $\tau^- = \tau - \sigma$.
We can then write the generators \exprg\ as \eqn\genersym{ \eqalign{ P^0 = &    \int d\Omega_2 \,  {\cal E}(\vec n) \cr P^i = & \int d\Omega_2   \,
  n^i \, {\cal E}(\vec n)
 \cr
 {\cal E}(\vec n) \equiv  &  2 \int_{\tau^+ = \pi}  d\tau^-   \left(  \cos {\tau^- \over 2 }
\right)^2  \, T_{\tau^- \tau^-}
 }}
 where $n^i$ is a unit vector in $R^3$ and specifies a point on $S^2$.
  In these coordinates the $SO(3)$ rotation symmetry is
 manifest. The fact that the energy flux and the momentum flux is related to the same operator,
 $T_{--}$, is indeed what we would naively expect in a theory of massless particles.
 Namely,  if   at some
point of the sphere  we have energy ${ \cal E}(\theta)$ then we have momentum $P^i = n^i {\cal E} (\theta)$.  Here we have shown that this also holds for a
general interacting CFT. This is due to the simple form of the Killing vector \simplegen\ at $Z^+=0$.

Note also that the SO(1,3) Lorentz symmetry acts on the 2-sphere as the $SL(2,C)$ group of conformal transformations of $S^2$. Our problem however, does not
reduce to computing correlators in a 2d CFT, since the state we are considering breaks the $SL(2,C)$ invariance.
Under these transformations the
operator ${\cal E}$ transforms as  a dimension three operator. The easiest way to see this it to recall that these $SL(2,C)$ transformations are the ordinary
Lorentz transformations of the original coordinates. In particular we have seen that $x^\pm$
boosts become dilatation operators in the $y$ variables. In those
variables it is clear that $\int dy^- T_{--}$ has dimension three. In particular, one can find the relation between
 the operator ${\cal E}(y^1,y^2)$ which is defined on a plane  to
the one on the sphere,  ${\cal E}(\vec n)$, by following the coordinate transformation between the plane and the sphere at $Z^+=y^+=0$ \eqn\coordstrans{
\eqalign{ y_1 + i y_2 = &
 { \sin \theta e^{ i \varphi} \over ( 1 + \cos \theta) } = \tan { \theta \over  2} e^{ i \varphi}
\cr
 dy_1^2 + dy_2^2  = &  { d \theta^2 + \sin^2 \theta d \varphi^2 \over ( 1 + \cos \theta )^2 } \equiv
 \Omega^2 ds^2_{S^2}
 \cr
 { \cal E}(y_1,y_2) = & \Omega^{ -3} { \cal E}( \vec n) = {   ( 1 + \cos \theta)^3 } {\cal E}
 ( \vec n )
}}

Physically, we expect that our idealized calorimeters will measure positive energies. Therefore, the expectation values of ${\cal E}(\vec n)$ should be
non-negative. In quantum field theory the expectation value of the stress tensor can  be negative in some spacetime region.
 However,
in our case we are integrating the stress tensor along a light like direction. In a free field theory one can show that  the expectation value \eqn\positcond{
\int dy^-  \langle T_{--} \rangle \geq 0 }
 is positive on any state \necflat
 \foot{The curved space analog of this condition
  has also been explored  for free fields in curved space, since
  it plays a role in proving singularity theorems in general relativity. }.
We expect that the same should be true in an interacting field theory. In appendix A we recall the argument in free field theories and give a handwaving
argument suggesting that this should be true in general. We will later see that this condition implies interesting constraints on certain field theory
quantities, so it would be nice to be able to give a more solid argument for the positivity of \positcond\ than the one we give in the appendix.

Notice that the energy flux operators ${\cal E}(\theta)$ commute with each other since operators at different values of $\theta$ are separated by spacelike
distances. This is most clear when we express the operators in terms of the $y$ coordinates as in \exprg . Thus, we can certainly consider the probability that
we measure specific energy functions ${\cal E}(\theta) = f(\theta)$ and derive the probability functional that governs the process. Once can also impose some
cuts on the energy distribution and compute such probabilities. This is done when jet cross sections are computed, as in \weinsterman .  In fact, a specific
Feynman diagram with $n$ particles coming out at angles $\theta_1, \cdots , \theta_n$ gives a contribution to the case where the energy function $f(\theta)$ is
a delta function localized at these points. The energy correlation functions we have defined correspond to average energies where we also allow extra particles
that come out and do not go into the calorimeters we are choosing to focus on.

Besides
 putting a detector at infinity that measures energy we can also put a detector
 that measures charge. In that case we have the charge flux operator
 \eqn\charfg{
 { \cal Q}(\vec n) = \lim_{r\to \infty} r^2 \int_{-\infty}^\infty dt \, n^i j_i(t, r \vec n)
 }
 where $j$ is the current associated to a global $U(1)$ symmetry of the field theory.
In the coordinates \explictch\ this becomes  $ {\cal Q}( y^1,y^2 ) = \int dy^- j_-(y^-,y^+=0,y^1,y^2))$.
 Under $SL(2,C)$ transformations ${\cal Q}$
transforms as a field of conformal dimension two. We can similarly compute energy and charge correlation functions. One can also easily consider non-abelian
global symmetries, and measure the components of various charges, as long as we do not put two charge insertions at the same point.

Now let us make some remarks on the operator ordering. Since the energy flux operators commute with each other for different $\theta$, then, it
does not matter how they are ordered. However, it is important that they are inserted between the operator, ${\cal O}$,
 that creates the state and the one annihilating it, as in \encorr .
This is the standard ordering when we compute expectation values. If we use perturbation theory to compute them it is important that we do {\it not} use
Feynman propagators since those are for time ordered situations. However, to do perturbation theory it is very convenient to use Fenymann propagators. In such
a case we have to be careful to remember that we should use the in-in \refs{\Keldysh,\maham}
 formalism to evaluate the expectation value. This consists
in choosing a contour that starts with the initial state, goes forward in time to the times where the stress tensor operators are evaluated and then goes
backwards in time.

In a conformal field theory we could also consider the following. Minkwoski space can be mapped to a finite region of $R \times S^3$. In fact, $R\times S^3$
can be split into an infinite number of regions, each of which is mapped to Minkowski space. In that case we can consider one of the regions as the original
Minkowski space and the region immediately to the future as the region parametrizing the part of the
Schwinger-Keldysh contour that goes back in time, as long as we
transform the wavefunction of the in state in the bra appropriately. We found this picture useful for gaining intuition, but not particularly useful for doing
computations.

\subsec{Small angle singularities and the operator product expansion}

\ifig\colinear{ (a) Singularities in the energy correlation functions arise when we place two calorimeters very close to each other, at a small angle $\theta$.
  (b) At the level of Feynman diagrams such singularities
come from colinear radiation.
  } {\epsfxsize2.5in\epsfbox{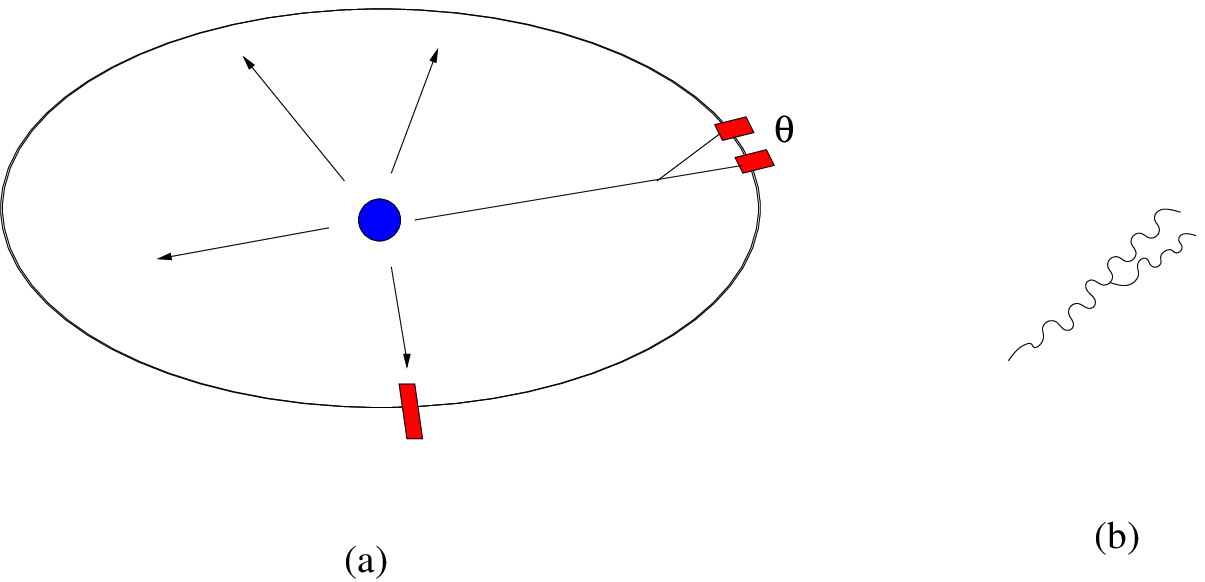}}

The energy correlation functions develop  singularities when two of the energy operators are evaluated at very similar angles $\theta_1 \sim \theta_2$, see
\colinear a. Such singularities are related to collinear radiation. To leading order in the gauge theory coupling $\lambda = 4 \pi \alpha_s N $  the leading
singularity goes like ${\cal E}(\theta_1 ) {\cal E}(\theta_2 ) \sim { C \lambda  \over \theta_{12}^2 } $  \refs{\ellis} and it comes from a Feynman diagram
like the one shown in \colinear b.

It is clear that such a limit should be characterized by some sort of operator product expansion.
 In this section we will make some remarks on the type of
operators  that appear in this  expansion.

It is simpler to think about the problem in the $y^\mu$ coordinates introduced in \explictch . We should, then, compute   the OPE of operators of the form
\eqn\operoc{
{ \cal E}( y^1,y^2) {\cal E}( 0,0) \sim \int dy^- T_{--}(y^-,y^+=0,\vec y) \int d{y'}^- T_{--}({y'}^-,y^+=0, \vec 0) }
 The two operators are sitting at two different points
in the transverse directions. We have set one at zero for convenience and the other at $\vec y = (y_1,y_2)$. Note that the distance between the two stress
tensor insertions is $|{\vec y}|$ irrespective of the values of $y^-,{y'}^-$. This distance is  spacelike, so one expects to be able to perform an operator
product expansion when $\vec y \to \vec 0$. Nevertheless, since the two stress tensors are sitting at two very different points in the $y^-$ directions,
  the operators appearing in the OPE are
not local operators.   To leading order the operator is specified by two points that are light-like separated
\lcstring .
% This type of OPE is
% sometimes called a ``light-cone'' OPE \lcopeearly \lcstring .
Such operators are useful for thinking about many high energy processes in QCD \refs{\lcopeearly,\muellerlightray,\hofmann}. They are sometimes called ``string
operators'' or ``light ray'' operators. Various ``parton distribution'' functions are defined in terms of matrix elements of such operators, see
\muellerlightray\ for example. It is important to note that these operators are non-local along one light-like direction but they are perfectly local in all
remaining three directions.

In order to characterize these non-local
 operators
it is useful to label them according to their transformation properties under the conformal group \braun\
 (see \refs{\conformalqcd}
 for a
review). Let us define the twist generator to be $T = \Delta - j$, where $j$ is
the spin (really a  boost generator) in the $y^+,y^-$ plane\foot{Note that we define $j$ to be the
spin in the $y^+,y^-$ plane only, not the total spin. The spin in the transverse directions is another
generator which does not appear in the definition of the twist.}.
 More explicitly,
the twist transformation is
 $(y^+,y^-, \vec y) \to (\lambda^2 y^+,y^- ,\lambda \vec y)$.
 The spin is the transformation $ (y^+,y^-,\vec y) \to ( \eta y^+, \eta^{-1}
 y^- , \vec y) $.
As it is well known, at zero coupling,
 one can consider twist two operators which correspond to primary operators
of higher spin. For example, if we have a scalar field, $\phi$, in the adjoint representation, then we can schematically define the operators
\eqn\confbil{
{\cal U}_j = Tr[ \phi  {\overleftarrow{\partial}_-\!\!\!\!\!\!\!\!\!\overrightarrow{\phantom{\partial}}}^j \phi] } This is schematic because there is a precise
combination of derivatives that makes it a conformal
 primary\foot{The precise form   is
 $ {\cal U}_j = \sum_{ k =0}^j  { (-1)^k \over [ k ! ( j-k)! ]^2 }
  Tr[( \partial^k_- \phi ) \partial_-^{j-k} \phi]
 $, where $\phi$ is a scalar field
\refs{\MikhailovBP,\DobrevRU,\FerraraYT}. }. Such conformal primaries exist only if $j$ is even. One is sometimes interested in extending the definition of
such operators to  generic, real or complex,  values of $j$.
 This problem was considered in detail in
 \braun . There, it was found that one could start with the  operators
\eqn\biloc{ {\cal U}(y^-,{y'}^-) = Tr[ \phi(y^-) W( y^-, {y'}^-) \phi({y'}^-) ] = Tr[ \phi(y^-) P e^{ \int_{y^-}^{{y'}^-} A } \phi({y'}^-) ] }
 where $W$ is an adjoint Wilson line along a null
direction. All operators are inserted at the same values of $y^+, y^1$ and $y^2$ (but of course, at different values of $y^-$). We can also replace $\phi$ by a
fermion or a gluon operator $F_{-i}$.
 Under twist transformations $y^-$ remains invariant but the transverse
 coordinates are rescaled.  In the quantum theory this scaling transformation mixes the operator \biloc\ with operators with
 other values of $y^-, {y'}^-$.
   By thinking about the
 action of the collinear  conformal group (the $SL(2,R)$ set of transformations of $x^-$)
  it is possible to diagonalize the action of the twist generator.
 To leading order the operators are diagonalized by considering suitable combinations of these
 light-ray
 operators \refs{\braun,\conformalqcd}.
 These operators are labeled by their center of mass
 momentum $k_-$ along the $y^-$ direction and
 their spin.
For our purposes we will be interested only in operators which are integrated
 over the center of mass position along the $y^-$ direction so that they carry zero momentum
 along $y^-$.
 In that case the operators of arbitrary spin
 constructed from scalar fields can be written as
 \eqn\operaj{
  { \cal U}_{j-1}  = \int_{-\infty}^{\infty}
   dy^- \int_0^\infty { du \over u^{ j+1} }  Tr[ \phi(y^- +u) W( y^-+u, {y}^--u) \phi({y}^--u) ]
} The subindex of ${\cal U}$ denotes the total spin and $j$ denotes the spin before we do the $y^-$ integration. This is an expression that  makes sense for
arbitrary complex
 values of $j$. When $j$ approaches an even integer  we find a pole in $j$ coming from a
 logarithmic
 divergence in the integral at small $u$ of the form $\int { du \over u} $. The coefficient of this
 divergent term contains  the ordinary local operator \confbil ,
  see \refs{\braun,\conformalqcd} for more details.
  There are similar expressions for operators constructed from two fermions
 or two Yang-Mills field strengths.
 One can compute the value
of the twist for these operators and one finds \braun\
 $\tau(j) = 2 +  \gamma(j) $, where $\gamma(j)$ is the anomalous dimension.
One can also consider higher twist operators which
 contain more field insertions or extra derivatives with respect to the transverse direction or
 $y^+$.
 In that case, in order to diagonalize the matrix of anomalous
 dimensions, it is not enough to give the total spin of the operator. Nevertheless, this can be
 done, see \conformalqcd .

The OPE has the schematic form
\eqn\energ{ { \cal E } ( \vec y ) {\cal E}(\vec 0)
\sim \int dy^- T_{--}(y^-,\vec y) \int d{y'}^- T_{--}({y'}^-, 0) \sim
\sum_n |\vec y|^{\tau_n-4} \left. {\cal U}_{j-1,n} \right|_{j=3} }
 where the sum is over all operators which are local in $y^+, \vec y$, but not necessarily
local in $y^-$, which have total spin $j-1=2$, (or $j =3$) and twist $\tau_n$.
The spin is determined since
 the total  spin of the left hand side is one for each of the two energy insertions.
Equation \energ\ is schematic because we have not explicitly indicated the fact that the
operators in the right hand side could carry spin in the transverse directions. A more precise
expression has the form
\eqn\energ{ { \cal E } ( \vec y ) {\cal E}(\vec 0)
\sim   \sim
\sum_{k,n} y^{(i_1} \cdots y^{i_k)} |y|^{\tau_{n,k} - k -4} \left. {\cal U}_{(i_1 \cdots i_k);j-1;n}
\right|_{j=3}
}
where we have now considered operators that carry spin in the transverse directions, the indices
$i_1, \cdots i_k$ are symmetric and traceless.

Among the operators which have twist two  at zeroth  order there are only a few that have $j=3$. For example, in QCD there are only two, a
bilinear in fermions and a bilinear in the gluon field strength. Thus,  for the given spin we are considering ($j=3$)  we will have to
diagonalize a finite matrix of anomalous dimensions.

In summary: {\it The small angle behavior of the energy
 correlation functions is determined by the spin $j=3$ non-local
 operators that appear in the OPE }
\eqn\correlf{
\langle {\cal E}(\theta_1)
{\cal E}(\theta_2)
 \cdots \rangle \sim \sum_n  |\theta_{12}|^{\tau_n -4 } \langle {\cal U}_{3-1,n}(\theta_2)
  \cdots \rangle
 } where the
dots denote other energy insertions and $|\theta|$ is the angle between the two energy insertions that are getting close to each other. The sum over $n$ runs
over all the higher twist operators that can appear. We will see that
in ${\cal N}=4$ super Yang Mills these operators develop large
anomalous dimensions  at strong coupling.

Note that the spin symmetry in the $y^+,y^-$ plane,
 that we used to select the operators that contribute,
 is the dilatation symmetry of the original Minkowski
space. This symmetry ensures that the energy correlation functions scale properly as we rescale the total energy (or rescale the variables
$x^\mu$). In other words, there can be no anomalous dimensions under total energy rescalings  since that would conflict with energy
conservation. This is the physical reason why we are forced to select particular operators in this OPE.

In the case of QCD the  small angle behavior of energy correlation functions  was computed  a long time ago in \refs{\parisipetr,\veneziano}
 using a slightly different
language. They also needed to include the effects of the running coupling.

Let us now turn to the case of ${\cal N}=4$ super Yang Mills at weak coupling.
 The weak coupling computation of
the leading twist anomalous dimensions was done in
\refs{\kloneloop,\klhigherloop} (see also \refs{\staudacher,\begrkr}).
We should consider operators which are invariant
under all the symmetries that leave  the particular component of the stress tensor in \energ\ invariant.
 These include the $SO(6)$ R-symmetry
% the $SO(2)$
%symmetry of the transverse space
and a parity symmetry.
We can classify the operators according to their transformation properties under the $SO(2)$ group
that transforms the transverse coordinates.  All operators  are made out of a pair of
  scalars, fermions or gauge field strengths. The local
  operators with zero transverse
  spin in $SO(2)$  and spin $j$ ($j$ even) in the $+-$ directions are
  \refs{\kloneloop} \eqn\operatf{ Tr[ \phi
 {\overleftarrow{\partial}_-\!\!\!\!\!\!\!\!\!\overrightarrow{\phantom{\partial}}}^j \phi]  ~,~~~~~~Tr[ F_{- i }
{\overleftarrow{\partial}_-\!\!\!\!\!\!\!\!\!\overrightarrow{\phantom{\partial}}}^{j-2} F_{-i }] ~,~~~~~~ Tr[ \psi \Gamma_-
{\overleftarrow{\partial}_-\!\!\!\!\!\!\!\!\!\overrightarrow{\phantom{\partial}}}^{j-1} \psi] }
Supersymmetry relates these three towers of operators. Since
supersymmetry carries spin,
 the various members of the supermultiplet have different spin. However, the anomalous
 dimension for all the members of the supermultiplet is the same and it is given by
 a function which has the weak coupling expansion \refs{\kloneloop,\dolanosborn}
 \eqn\andimme{
\gamma(j) = { \lambda \over 2 \pi^2 } [ \psi(j-1)-\psi(1) ]  + \cdots } where $\psi  = \Gamma'(z)/\Gamma(z)$. This was computed also to two and three loops in
\klhigherloop .
  The fact that $\gamma(j=2) =0$
corresponds to the fact that the stress tensor is not renormalized.

Since we are interested in operators with a definite spin, we conclude that
 the three operators
that diagonalize the anomalous dimension matrix are in three different multiplets. For spin three operators we have the anomalous dimensions \refs{\kloneloop}
\eqn\threeitg{\eqalign{ \tau_1 -2= & \gamma(j=3) ~,~~~~~~~~\tau_2 -2= \gamma(j=5) ~,~~~~~~~~~
\tau_3 -2= \gamma(j=7) \cr \tau_1 -2= & { \lambda \over 2 \pi^2 }
~,~~~~~~~~~~~~~\tau_2 -2= { 11 \lambda \over 12 \pi^2 }
  ~,~~~~~~~~~~~~~~\tau_3 -2= { 137 \lambda \over 120 \pi^2 }
  }}
where we just gave the first order expression. We see from \threeitg\ and \andimme\
 that all three anomalous dimensions  in \threeitg\ are positive and $\tau_1-2$
is the smallest one which will give us the leading order singularity. However, for weak coupling all
three contributions
are similar.

In addition to the operators we discussed, we can also have operators which have non-zero transverse
 spin. At twist two, the only one consistent with the symmetries
is the spin two operator
\eqn\transversp{
{ \cal U}_{(il); j } = Tr[ F_{- (i}
 {\overleftarrow{\partial}_-\!\!\!\!\!\!\!\!\!\overrightarrow{\phantom{\partial}}}^{j-2}
F_{- l)} ]
}
where the indices $i,l=1,2$ are symmetrized and traceless. In the ${\cal N}=4$ theory these
operators are in the same supermultiplet as the ones considered above \beisertdil .
For this reason their
anomalous dimension is also given in terms of the same formula
\eqn\anomt{
\tilde \tau_j -2 = \gamma(j + 2)  ~,~~~~~~~~~~~\tilde \tau_3   = 2 + { 11 \lambda \over 12 \pi^2 }
}

Thus we expect to have a small angle singularity of the form
\eqn\opercof{
 \langle {\cal E}(\vec y ) {\cal E}( 0)  \cdots \rangle
 \sim \sum_{a=1}^3 |y|^{-2 + (\tau_a -2)  }  c_a \langle {\cal U}_a  \cdots \rangle
 +  y^{(i}y^{l) }  |y|^{-4 + (\tilde \tau_3 -2)}
   \tilde c    \langle {\cal U}_{(il)}
 \cdots \rangle
 }
 where the operators ${\cal U}_a$ are the linear combinations that diagonalize the anomalous
 dimension matrix for the operators with zero transverse spin. $c_a$ and $\tilde c$ are
 coefficients that can be obtained by performing
  the operator product expansion explicitly. These constants
 are independent of the state for which we compute the energy correlation. Of course, the terms $\langle {\cal U}_a  \cdots \rangle$ and  $\langle {\cal U}_{(il)}
 \cdots \rangle$ do depend on the
 state on which we compute the energy correlation function.
The coefficients $c_a$, $\tilde c$ start at order $\lambda$ at weak
coupling since it is easy to check explicitly that at tree level there is no contribution to the
operator product expansion of two energy flux operators.

In QCD one can do a similar analysis, including the effects of the beta function, see   \veneziano .
% \foot{
%In eqn. (5.27) of \veneziano , with $n=2$,
% one can see the one loop (resumed) QCD result for the small angle behavior of the energy-energy
% correlator. \veneziano\ also discusses other correlation functions. }.
  In that case, the operator made out from scalars in \operatf\ does not contribute.

Having done one OPE, we
  could also do a further OPE of the resulting operator with a third energy flux
operator. That would give  an operator of total spin $j-1 =3$, or $j=4$,  and so on. More generally, we can consider the case where $n$ energy operators come
close together. If we keep the ratios of angles between these $n$ points fixed, then the small angle behavior is given by the anomalous dimension of the
operator of spin $j = n +1$. The structure of a jet at weak coupling is largely controlled by these
operator product expansions.

Note that, at weak coupling,
 after we consider the effects of the anomalous dimensions, the energy correlators
have small angle singularities that are integrable. In fact,
 if we do the integral over a small angle $\theta_0$ of the energy two point function we find
schematically \eqn\integran{
 \int d^2 \theta { \cal E}(\theta) {\cal E}(0) \sim
 \int_0^{\theta_0} d^2 \theta { \lambda \over \theta^ { 2 - \gamma_* \lambda } }
 \sim (\theta_0)^{ \gamma_* \lambda}
 }
 where the anomalous twist of the spin three operator is $\tau -2 = \gamma_* \lambda$ and
 $\gamma_*$ is a numerical constant. This expression is schematic because at weak coupling we have
 to include all the terms in \opercof .
 If $\theta_0$ is fixed and $\lambda \to 0$ then we see that the integral gives a finite
 answer.
 This is to be expected since the total integral of one energy insertion over the
 whole sphere should give the total energy, independently of $\lambda$ \foot{Here we are 
 also assuming that the energies are locally  positive. For charge correlators
  we cannot make the same argument because the charge can be positive or negative. }.
 The fact that we get
 a finite contribution from this region is consistent with the idea that the energy is going
 out in localized jets.
 We can also estimate the angular size of jets, by  finding a  $\theta_0$ in such a way that we get a fixed fraction, $f$,
 of the total energy in the jet.
 This gives an estimate $\theta_0 \sim e^{ - c/ \lambda }$, where $c$ depends on $f$. This
 was originally discussed in \weinsterman, see also  \veneziano\ for a more detailed discussion.

 We could also do
an OPE of two charge operators, each of which has spin zero, after we integrate the spin one current over $y^-$ \charfg . In this case we get operators with
total spin $j-1=0$, or $j=1$. Some of these  have negative anomalous dimensions. In fact, we expect that charge correlators would be more singular at small
angles due to the fact that a gluon can create a pair of oppositely charged particles fairly easily and there is no reason that we couldn't get a divergence
when we integrate the charge correlator at small $\theta$.

Finally, we should mention that in QCD
 the energy-energy correlation two point
 function was computed for all angles in\foot{The results presented in the following references show some disagreements. For a detailed comparison between these results see
 \nasonqcd . We thank S. Catani for pointing this out to us.}
  \refs{\ellis,\ellissecond,\kunsztqcd,\gloverqcd,\cataniqcd}. It was
also compared to experiment in
 \energycorrexp , where it was used as a way to measure $\alpha_s$.

\subsec{Energy flux one point functions}

In this section we will make some simple and general remarks about the energy flux one point function \eqn\energf{ \langle {\cal E}(\vec n) \rangle = { \langle
0 | {\cal O}_q^\dagger
 {\cal E}(\vec n )  { \cal O}_q |0 \rangle \over
 \langle 0 | {\cal O}^\dagger_q
   { \cal O}_q |0 \rangle }
 }
 These one point functions are determined up to a few coefficients by Lorentz symmetry, even
 in non-conformal theories. Here we will consider these in the CFT context in order
 to make contact with other results in conformal field theories.

The energy flux one point function \energf\ amounts to computing a three point function in the CFT.    Three point functions in a generic CFT are determined up
to a few numbers by conformal symmetry \refs{\schreier,\osborn,\erdmengerb}.

Let us start with   the case that we create   the external state with a scalar operator with energy $ q $ and zero momentum. Strictly speaking such an operator
is not a localized insertion. Thus, more precisely, we will be considering operators of the form \eqn\opeffs{ { \cal O }_q  \equiv  \int d^4 x { \cal O}(x) e^{
- i q x_0} \exp\{ - { x_0^2 + x_1^2 + x_2^2 + x_3^2 \over \sigma^2 } \}
 ~,~~~~~~~~~~~q \sigma \gg 1
 }
where the last inequality ensures that the operator is localized,
 has finite norm and  has four momentum approximately ${\tilde q}^\mu =   (q, \vec 0) +
o(1/\sigma ) $. In particular we have $q^0 \sim q$.
 Once we know this precise
form of the operator we see that we can also write it in  other coordinate systems by
   performing the suitable conformal transformation and taking into account the conformal transformation
   properties of ${\cal O}(x)$.

In what follows we will consider field theory states produced by scalar operators, ${\cal O } \sim {\cal S}$, conserved currents ${\cal O} \sim \epsilon_i j_i
$, and the stress tensor, ${\cal O} \sim \epsilon_{ij} T_{ij}$.
 In all cases we consider
states with essentially zero spatial momentum as in \opeffs .
 The case where $q^\mu$ is a  generic four
vector  can be obtained by performing a simple boost of the configurations we discuss.

In the case that we insert a scalar operator it is clear by $O(3)$ symmetry that the energy one point function is constant on the two sphere. In addition the
integral over the angles should give the total energy. Thus, for a scalar operator we have \eqn\integra{ \langle {\cal E}(\vec n) \rangle  = { q \over 4 \pi }
}

Even though we know the answer already, it is possible to do the calculation explicitly by writing down the unique general expression for the three point
function of two scalars and the stress tensor \osborn . Its normalization is fixed by a Ward identity in terms of the two point function of the two scalars. This Ward
identity is another version of the energy conservation argument that we used above. Writing down the three point function and doing the   integrals in the
limit \energfl\ we indeed obtain \integra . One has to be careful about the operator ordering. In appendix C we do this explicitly.

We now turn to the case where the external perturbation couples to a conserved current in the CFT. In that case the operator is given by ${\cal O}_{\epsilon,q}
\sim \epsilon^\mu j_\mu(q) $ where $\epsilon^\mu$ is a constant
 polarization vector.
Due to the current conservation condition we can identify
 $\epsilon^\mu \sim \epsilon^\mu  + \lambda q^\mu $. So we can
choose $\epsilon$ to point in the spacelike directions. In this case $O(3)$ symmetry and the  energy conservation condition constrain the form of the one point
function to \eqn\nepar{\eqalign{ \langle {\cal E }(\vec n) \rangle = & { \langle 0| ( \epsilon^* \cdot j^\dagger ) \,  {\cal E}(\vec n) \, ( j \cdot \epsilon )
|0  \rangle \over
 \langle 0|  ( \epsilon^* \cdot j^\dagger ) \,   (j \cdot \epsilon ) |0  \rangle}  =
 {  q \over  4 \pi }
  \left[  1  + a_2
  \left( { | \vec \epsilon \cdot \vec n |^2 \over | \vec \epsilon |^2 }  - { 1 \over 3 }  \right) \right]
  \cr
 = &  {  q \over  4 \pi }
  \left[  1  + a_2
  ( \cos^2 \theta  - { 1 \over 3 }  ) \right]
  }}
where $\theta$ is the angle between between the point on the $S^2$, labeled by $n^i$,
 and the direction of the
polarization vector $\epsilon^i$.

The fact that we have one free parameter is in agreement with the general analysis of the three point function of two conserved currents and the stress tensor.
In fact in \osborn\ it was shown that
 the three point function is determined by conformal symmetry up to two
parameters and one of them is fixed by the Ward identity of the stress tensor.

Note that $a_2$ in \nepar\ obeys a constraint that comes from demanding  that the expectation value of the energy ${\cal E}(\theta)$ is positive, see
\positcond .
  This condition
leads to the constraint \eqn\positcon{
 3 \geq a_2 \geq - { 3 \over 2 }
 }

This one point function was computed for the electromagnetic current in QCD in \ellis . To first order in $\alpha_s$ the result is \eqn\valatqcd{ a_2 = -{ 3
\over 2} + { 9 \alpha_s \over 2 \pi } + \cdots
 }
 To the order written
 in \valatqcd\ we can approximate the QCD computation by a conformal field theory with
 the value of the coupling set by the energy of the process $\alpha_s =
 \alpha_s(|q^0|)$.
 %  This has been computed to higher orders in \ellissecond .

In the application to   $e^+e^-$ collisions that produce a gauge boson which in turns couples to a
 current  the polarization vector of the current  depends
  on the polarization states of the $e^+$ and $e^-$
as well as the type of gauge boson we are considering ($\gamma$ or $Z$, $Z'$, etc). In the case that we  consider unpolarized electrons  we can express the
answer in terms of the angle with respect to the beam axis, $\theta_b$. ($ \cos \theta_b = \vec n . \hat z $ where $\hat z$ is the beam axis). The polarization
vectors for the current are orthogonal to the beam direction and we should average over them.  After doing this average,
 we find that \nepar\ becomes
\eqn\anglrebem{ \langle {\cal E }(\vec n) \rangle = { \sum_s \langle 0| ( \epsilon_s^* \cdot j^\dagger ) \,  {\cal E}(\vec n) \, ( j \cdot\epsilon_s )
 |0  \rangle \over
 \sum_s  \langle 0|  ( \epsilon_s^* \cdot j^\dagger ) \,   (j \cdot \epsilon_s ) |0  \rangle}
 ={  q \over  4 \pi }
  \left[  1  + a_2
  ( { 1 \over 2} \sin^2 \theta_b   - { 1 \over 3 }  ) \right]
  }
  where we sum over polarization vectors transverse to the beam.
For a current that couples to free fermions we find the familiar $(1 + \cos^2 \theta_b)$ distribution, as we can check from the leading order QCD result
\valatqcd .

 For a current that couples to  free complex bosons of charges $ q^b_i$ and
 Weyl fermions of charges $q^{wf}_i$    we
 get
 \eqn\valat{
 a_2^{free} = { 3  } {  \sum_i (q_i^b)^2 - (q_i^{wf})^2 \over
 \sum_i (q_i^b)^2 +  2 (q_i^{wf})^2  }
 }
 where we sum  over  both  left and right Weyl fermions.
 Note that the case where we only have bosons
  saturates the upper bound in \positcon\ and free fermions saturate the lower bound in
  \positcon .  In fact, going back to \anglrebem\ we see that we get the well known
   distributions
  proportional to $\sin^2 \theta_b$ or $(1 + \cos^2\theta_b)$ for free bosons and fermions
  respectively.

 We can consider a similar problem now in an ${\cal N}=1$ superconformal theory.
  If
 the current is a global symmetry that commutes with supersymmetry (a non-R symmetry) then one
 can see that $a_2=0$. In a free supersymmetric theory we see from \valat\ that
  the bosons and Weyl fermions cancel each other.
  For an interacting theory
  this follows  from the fact
 that such a current is in the same multiplet as a scalar operator,
  and for a scalar operator we
 do not have any arbitrary parameters \onlyosborn .
  Thus   the
  value of $a_2$  is fixed by superconformal
 symmetry. However, since we got $a_2=0$ in a free theory, we have $a_2=0$ for any global symmetry
 of a SCFT.

 On the other hand we can get a non-zero value of $a_2$ if we consider the $R$ current\foot{We
 thank Scott Thomas for pointing this out. }.
 The $R$ current
 is in a different supermultiplet. In fact, it is in a supermultiplet
 with the stress tensor. All three point functions among elements of this supermultiplet are
  determined by two numbers, $c$ and $a$ \onlyosborn .
   These numbers also characterize the
 anomalies of the $R$ current, which are encoded in parts of the $jjj$ and $jTT$ three point functions
 \refs{\freedmanc,\onlyosborn}. They also contribute to the conformal anomaly on a general background,
 \eqn\tracean{
T^\mu_\mu = { c  \over  16 \pi^2 } W_{\mu\nu\delta\sigma} W^{\mu\nu\delta\sigma} - {  a  \over 16 \pi^2 } E~, ~~~~~~~~~  E = R_{\mu \nu \delta \rho}R^{\mu \nu
\delta \rho} - 4 R_{\mu \nu} R^{\mu \nu} + R^2 }  where $W$ is the Weyl tensor and $E$ the Euler density.
%and $E$ is the Euler density\foot{$ W^2 = R_{\mu\nu\delta\sigma}R^{\mu\nu\delta\sigma} -2
%R_{\mu \nu}R^{\mu\nu} + 1/3 R^2 $, $E = R_{\mu\nu\delta\sigma}R^{\mu\nu\delta\sigma} - 4
%R_{\mu \nu}R^{\mu\nu} +  R^2 $.}
 The $c$ coefficient is the only constant that
  appears in  the  two point functions of the currents and the stress tensor \onlyosborn .
  Thus $c$  appears in the part of the three point function
  that is fixed by
 the Ward identity. The coefficient $a_2$  is given by
  a linear combination of $a$ and $c$. The particular linear combination is
  independent of the theory. It is fixed by supersymmetry.
   We can compute the precise combination by considering the particular case of free
  field theories.   As we explain in more detail below
 we find that
  \eqn\refnone{
  \langle {\cal E }(\theta)\rangle  = 1 + {3  } { c-a \over c}
   ( \cos^2 \theta - { 1 \over 3 } )
  }
  This formula was obtained as follows.
 We used that for a free supersymmetric theory with $n_V$ vector multiplets and $n_S$ chiral
 multiplets we have
 $48 a = 9 n_V + n_S$, and $24 c = 3 n_V + n_S$  \refs{\freedmanc,\duffobs,\duff}. The vector multiplet
 has one  Weyl fermion of charge $1$ and the scalar multiplet in a free theory
  has a Weyl fermion of charge $-1/3$
 and a boson of charge $2/3$.  Then using \valat\ we obtain \refnone . Note that even though
 we used free field theories to fix the numerical coefficients, the final result \refnone\ is
 true for a general interacting ${\cal N}=1$ SCFT.

  In ${\cal N}=4 $ super Yang Mills $a=c$ and the result for the one point function
   is spherically symmetric. Of course
  this is not a surprise since U(1) subgroups of the SO(6) symmetry group can also be viewed
  as global symmetries from the point of view of ${\cal N}=4$ written as an ${\cal N}=1$ theory. Thus, in
  ${\cal N}=4$ super Yang mills the result is independent of the coupling.

  The positivity constraint \positcon , together with   \refnone\ gives
  $  { 3  c \over 2 }  \geq a \geq  0
  $.

  We can also consider the energy one point function in the case that the state is created by
  the stress tensor.
  As above, we take the momentum of the inserted operator in the time like direction.
  Then the operator that we are considering is characterized by a symmetric polarization tensor
  $\epsilon^{ij}$ which we take to have indices in the purely spacelike directions by using
  the conservation equations. Since the stress energy tensor is traceless, we also
  take $\epsilon^{ii} =0$. By O(3) invariance
  we see that the most general form of the three point function is
  \eqn\mostgengav{
  \langle {\cal E}(\theta) \rangle = { \langle 0 | \epsilon^*_{ij} T_{ij}
  {\cal E}(\theta) \epsilon_{lk} T_{lk } |0 \rangle \over
   \langle 0 | \epsilon^*_{ij} T_{ij}
   \epsilon_{lk} T_{lk } |0 \rangle } = { q^0 \over 4 \pi }
    \left[
   1 + t_2
   \left( { \epsilon_{ij}^* \epsilon_{il} n_i n_j \over   \epsilon_{ij}^* \epsilon_{ij}} - { 1 \over 3 }
   \right)
 + t_4
 \left( {|\epsilon_{ij} n_i n_j |^2 \over   \epsilon_{ij}^* \epsilon_{ij}} - { 2 \over 15 } \right)
  \right]
  }
  We see that we have two undetermined coefficients. This agrees with the general analysis of
  stress tensor three point functions in \osborn , where they found that conformal symmetry determines
  the three point function of the stress tensor up to three coefficients, one of which is fixed
  by a Ward identity. We have chosen the constants in the last two terms in \mostgengav\ in such a
  way that the corresponding terms integrate to zero on the sphere.

  By demanding that $\langle {\cal E} \rangle$ is positive we get the constraints
\eqn\constts{ \eqalign{ & ( 1 - { t_2 \over 3} - { 2  t_4  \over 15 }) \geq 0 \cr & 2 ( 1 - { t_2 \over 3} - { 2  t_4  \over 15} ) + t_2 \geq 0
\cr &  {3 \over 2 } ( 1 - { t_2 \over 3} - { 2  t_4  \over 15 } ) + t_2 + t_4 \geq 0 }} We obtain these constraints by using $O(3)$ invariance
in \mostgengav\ to set $\vec n = \hat z$. We then view the resulting equation as a bilinear form on the space of $\epsilon$'s. This space can be
divided into three orthogonal parts according to their $O(2)$ invariance properties (the spin of $\epsilon $ along the $\hat z$ axis). We have
an $SO(2)$ scalar, a vector and a symmetric traceless tensor. On each of these subspaces we get each of the constraints \constts . Each of this
limits in saturated in a free theory with no vectors, no fermions or no bosons respectively. The fact that the first equation is saturated in a
theory without vectors is clear. In that case, if we consider a stress tensor insertion with spin +2 in the $\hat z$ direction we cannot have
emission of bosons or fermions in the $\hat z$ direction due to the orbital angular momentum wavefunctions. It is also possible to write a
general bound on the two coefficients that appear in the conformal anomaly \tracean , see appendix C.

 In an ${\cal N}=1$ supersymmetric
  theory we find that
  \eqn\ttwotfour{
  t_2 = 6 ( c-a)/c ~,~~~~~~~~~t_4 =0
  }
  By requiring that \mostgengav\ is positive
  for all choices of traceless $\epsilon_{ij}$ we find
  \eqn\implic{
  { 3 \over 2 } c \geq a \geq { c \over 2 }
  }
  Of course $c>0$ due to the positivity of the two point functions.
   The bounds are saturated by free theories with only vector supermultiplets (upper bound) or only chiral
   supermultiplets (lower bound). It is interesting that the lower bound that we obtain in this way is
   precisely the same as the bound obtained in \shenkertwo (see also \shenkerone ) based on
   causality for a gravity theory that contains only the Einstein term and a $R^2$ term\foot{In order to see this one has to set $\lambda_{GB} \to 9/100$ (the bound in \shenkerone )
   into the expressions for $a$ and $c$ \odintsov\ (see eqn. (5.1) of \shenkerone ).}. In the
   theory considered in \shenkertwo\ one would also have $t_4=0$, though it is not clear whether
   it corresponds to any dual quantum field theory. Here we have only used general field theory
   considerations.

   For a non-supersymmetric theory it is also possible to
   derive a bound from   \constts .
%\foot{ We thank Y. Tachikawa for prompting us to compute this bound.}.
As explained in appendix C we
 find
\eqn\boundar{
 { 31 \over 18}   \geq {  a  \over c }  \geq { 1 \over  3 }
 }
 where the lower bound is saturated by a free theory with only scalar bosons and the upper bound
 by a free theory with only vectors. Note that the bound in supersymmetric theories \implic\ is
 more stringent than in non-supersymmetric theories \boundar . 
 Let us also add that the results from appendix C also allow us to
  calculate this bound for ${\cal N}=2$ supersymmetric theories. In this case we can obtain the 
  bound by taking the operator ${\cal O}$ to be
   one of the $SU(2)$ R-symmetry generators and demanding that the energy one point function is positive. 
     The result is in this case
 \eqn\boundarntwo{
 { 5 \over 4}   \geq {  a  \over c }  \geq { 1 \over  2 }
 }

This is a smaller window than for the ${\cal N}=1$ case, as expected. 
The upper bound corresponds to a free theory with vector supermultiplets 
only while the lower bound corresponds to a free theory with hypermultiplets only.
 This agrees with results in \takshap .

  We can make similar remarks for operators that involve charge correlations.
 For example, we could consider a theory with an $SU(2)$ global symmetry and then select one
 $U(1) \subset SU(2)$ to form the charge flow operator ${\cal Q}$ that we measure at infinity.
 We could consider a charged state state created by the current $\epsilon \cdot J^+$, where the
 plus indicates
 that it carries charge plus one. As in the energy correlations the charge correlations have
 a form
 \eqn\chargecor{
 \langle {\cal Q}(\vec n ) \rangle = { \langle 0 | \vec \epsilon^* \cdot {\vec J}^- \, { \cal Q}(\vec n) \,
 {\vec \epsilon } \cdot {\vec J}^+ | 0 \rangle \over \langle 0 | \vec \epsilon^* \cdot {\vec J}^-  \,
 {\vec \epsilon } \cdot {\vec J}^+ | 0 \rangle  } = { 1 \over 4 \pi } \left[ 1 + \tilde a_2
 \left( { |\vec \epsilon \cdot \vec n |^2 \over |\vec \epsilon|^2 } - { 1 \over  3 } \right) \right]
 }
 Again, the coefficient $\tilde a_2$ is related to the fact that there are two possible
 (parity preserving) structures
  for the three point function of three currents \refs{\schreier,\osborn}. One of them is fixed by the Ward identities in terms
  of the two point functions, a fact we used in \chargecor . We note that in a supersymmetric
  theory where these currents are global symmetries  $\tilde a_2=0$. One can show this as follows.
  First note from \onlyosborn\ that there is only one parity preserving structure for current three
  point functions in supersymmetric theories. This means that the value of $\tilde a_2$ is
  fixed by supersymmetry. One can show that is vanishes by computing it in a particular theory, such
  as a free theory or  ${\cal N}=4$ super Yang Mills at strong coupling.

  There are some cases where there are parity odd structures that can contribute. Such parity
  odd parts of three point functions are related to anomalies. For example, in the case of
  three currents these are related to the usual anomaly \schreier .
  Consider the case that we have an external state produced by a current and
  we measure a charge one point function distribution.
 For example,  we can consider a $U(1)$ current that has a cubic anomaly. A
 concrete example is the $R$ current in superconformal theories.
  We consider a state obtained by acting with this current on the vacuum and we measure the
  charge flux far away. We find
  \eqn\chargfl{
 \langle   { \cal Q}(\vec n) \rangle  =
 { \langle 0 | \vec \epsilon^* \cdot {\vec j}^\dagger \, { \cal Q}(\vec n) \,
 {\vec \epsilon } \cdot {\vec j} | 0 \rangle \over \langle 0 | \vec \epsilon^* \cdot {\vec j}^\dagger  \,
 {\vec \epsilon } \cdot {\vec j} | 0 \rangle  } = i \, \alpha \,   \epsilon_{jlk} \epsilon_j^* \epsilon_l n_k
 \sim \alpha \cos \chi
 }
 Note that this is non-zero only if $\epsilon$ is complex. This happens, for example, when we
 consider a circularly polarized state. Then $\chi$ is the angle between the direction of the spin
 of the current and the calorimeter. This leads to a charge flow asymmetry.
 Such asymmetries are extensively studied in $e^+ e^-$ collisions that produce a $Z$ boson that then decays.
 Here we are pointing out that the charge flow asymmetries are related to the anomaly.
 Of course the full electroweak symmetry is not anomalous. But if one focuses only in decays of
 the $Z$ into leptons, then the fact that the purely leptonic theory is anomalous leads to the
 charge flow asymmetry. The fact that tree level processes plus unitarity fix the anomaly was
 pointed out in \ColemanYG .

  The three point function of  two stress tensors and a current has a term that reflects the mixed
  gravitational anomaly \gravanomaly .
  Consider the case where the stress tensor creates the state and we measure the charge
  distribution of the current that has a mixed anomaly.
  The charge distribution has the structure
  \eqn\anomal{
  \langle {\cal Q}(\theta) \rangle = { \langle 0 | \epsilon^*_{ij} T_{ij}
  {\cal Q}(\theta) \epsilon_{lk} T_{lk } |0 \rangle \over
   \langle 0 | \epsilon^*_{ij} T_{ij}
   \epsilon_{lk} T_{lk } |0 \rangle } = i \beta
    { \epsilon_{ljk} \epsilon^*_{rl} \epsilon_{sj}  n^r n^s n^k
   \over |\epsilon_{ij}|^2 }
   }
   where $\beta$ is related to the anomaly coefficient.
   For a supersymmetric theory, and for
  ${\cal Q}$ given by the R-current,  we have that $\beta \sim
  (a-c)$ \refs{\freedmanc,\gravanomaly}. Notice that there is, in principle,
   another tensor structure consistent with $O(3)$ symmetry that could
  have contributed to \anomal,
  namely $ \epsilon_{ljk} \epsilon^*_{rl} \epsilon_{rj} n^k$. This term is,
  however, absent from the three point
  function once conformal symmetry and the Ward identities are imposed \gravanomaly .
 Thus in a theory with a gravitational mixed anomaly there is charge asymmetry for the states
 produced by the graviton.

\subsec{ Relation to deep inelastic scattering }

In this section we explore the relation between the energy correlation functions and the deep inelastic scattering cross sections.

The deep inelastic cross section for the scattering of an electron from a proton can be factorized into the electromagnetic process and the strong interactions
process. At lowest order in the electromagnetic coupling, but exactly in $\alpha_s$, the strong interactions part of the cross section can be written in terms
of the expectation value of two currents in the state of the target (which is traditionally a proton, but can be generalized to any other particle)
\eqn\vevcurs{ \eqalign{ \tilde W^{\mu \nu} =  & \int d^4 y e^{ i q y }
 \langle p | J^\mu(0) J^\nu( y) | p \rangle =
\cr = & \tilde F_1(x, q^2/p^2) \left( g_{\mu \nu} - { q_\nu q_\nu \over q^2 } \right) + { 2 x \over q^2 } \tilde F_2( x , q^2/p^2 ) \left( p_\mu + { q_\mu
\over 2 x } \right) \left( p_\nu + { q_\nu \over 2 x } \right) \cr
 & ~~~~~{\rm where}~~~~~ x \equiv  - { q^2 \over 2 p.q } ~,~~~~~~~~q^2 > 0
}} We are imagining that we have a plane wave state in the $y$ coordinates with timelike
momentum $p^2<0$.
The tensor \vevcurs\ is nonvanishing only if we create a timelike state
$  s = - ( q + p) \geq 0$ with the current. For these values of $p$ and $q$ our definition of $\tilde W^{\mu \nu}$, \vevcurs,
 coincides with the ordinary one \manohar , which involves a commutator of the currents.

We would like to relate these formulas to the ones appearing in the energy correlators. Let us consider the charge operator $Q$ evaluated in the $y$
coordinates \eqn\defq{ {\cal Q}(\vec y) = \int dy^- j_-(y^+=0, y^-, \vec y) }
 where $\vec y$ denotes two transverse dimensions.
These two transverse dimensions are related to the angles on the two sphere by \coordstrans . In the $y$ coordinates, we can Fourier transform this operator.
We, then, have something similar to the current appearing above, except that the current in \vevcurs\ is in momentum space also in the $y^+$ direction. Note
also that $q_-=0$, due to the $y^-$ integral in the definition of the charge flux operator ${\cal Q}$.
Since $q_-$ is zero, we find that $q^2 = ( \vec q)^2 >0$
and independent of $q_+$.
 However $x$ depends on $q_+$ since \eqn\valex{ { 1 \over x } = { - 2 p.q \over q^2 } = {  -(p+q)^2 + p^2 + q^2 \over q^2 } = { 4 p_-
q_+ \over q^2 } + \cdots } where the dots indicate terms that are independent of $q_+$. In order to produce the $\delta(y^+)$ that is present in the charge
flux operator we need to integrate over $q_+$. This integral translates into an integral over $x$. The range of integration can be determined by the condition
that $ -( p+q)^2 \geq 0$. Thus we end up integrating between $x=0$ and $x_{max}$ with $1/x_{max} = 1 + p^2/q^2$. In the limit $p^2/q^2 \to 0$ we get the usual
boundary $ x =1$.

We then have the following relation between the two quantities \eqn\claimres{ \eqalign{ \int d^2 \vec y e^{ i \vec q . \vec y } ~_p\langle  { \cal Q}(\vec 0) {
\cal Q}(\vec y) \rangle_p  = & \int_{-\infty}^{\infty }  { dq_+ \over 2 \pi }  W_{--} ( q_+, q_- = 0, \vec q ; p) = \cr =  & { ( - p_-)  \over 4 \pi  }
\int_0^{x_{max}}  { dx \over x }  F_2( x, q^2/p^2 ) }} This is a particular moment of the parton distribution functions. More precisely it is the moment
$M_1^{(2)}$. As it is well known, the even moments $M_{2 k}$ can be expressed in terms of the expectation values of local operators with spins $j = 2 k$
\manohar , via a dispersion relation argument. In fact, the moments $M_j^{(2)}$ can also be expressed in terms of the expectation values of the non-local
light-ray operators with spin $j$ for any $j$,
 see \complexj\ for a general discussion.

In \claimres\ the charge correlation is evaluated on a state with definite momentum in the $y$ coordinates. This implies, in particular, that the charge two
point function is also translation invariant in the transverse space,
 $ ~_p\langle {\cal Q}(\vec y) {\cal Q}(y') \rangle_p =
  ~_p\langle {\cal Q}(\vec y- \vec y') {\cal Q}(0) \rangle_p $.

In this article we have been mainly considering states which are in momentum eigenstates in the $x$ coordinates, related to the $y$ coordinates via \explictch
. This does not lead to momentum eigenstates in the $y$ coordinates. However, they do have definite momentum in the $p_-$ direction. To the extent that we can
neglect other components of the momentum in the $y$ coordinates we see that the charge correlator has a simple relation to the deep inelastic scattering
amplitude and the ordinary parton distribution functions. In the general case we will need to evaluate expectation values of the form $ \langle p' | J J |p
\rangle$. These require generalized parton distribution functions
 \generalizedpdf . Thus, if we have a state with definite momentum in the original
$x$-coordinates, we will have a supersposition of momenta in the $y$ coordinates and the charge two point function will be related to
 integrals over
generalized parton distribution functions. We will not write a detailed expression here.

Notice that the integral over $x$ is divergent at small $x$. We think that this is due to the fact that the integral over $\vec y$ is also divergent for the
charge correlator  since the small angle singularity is not integrable. This divergence, though, is
 local in $\vec q $ and can probably be extracted without
changing the overall picture. We have not checked this in detail. This problem is not present if we consider the energy correlation functions
and the relation to the deep inelastic amplitudes probed by gravitons. In that case all quantities are manifestly finite.

The fact that in our problem we do not have ordinary plane wave wavefunctions in $y$ has
an interesting consequence. It was shown in \psdis\ that, in the gravity regime,
 the leading power of $q$, which governs the short distance behavior
  in $\vec y$, is controlled by a double trace operator. We will show
below that
 this contribution is
  highly suppressed for operators that have definite momentum in $x$-space.
   We expect that this double trace
contribution will also be suppressed at weak coupling when we consider plane wave states in $x$-space.

Of course, everything we said here can be repeated for the energy correlation function, except that we should consider a deep inelastic process where we
scatter gravitons from the field theory excitations.

\subsec{Energy correlations and the $C$ parameter  }

Let us make here a side comment on the relation between the energy correlators
and other usually considered event shape variables. Event shape variables are certain
functions of the four momenta of the observed particles which are infrared safe.
One concrete example is the $C$ parameter,
  defined as \cparameter\
 \eqn\cparmde{
 C =  { 3 \over 2 E^2  }
  \int d^2\Omega_1   d^2 \Omega_2   {\cal E}(\vec n_1) {\cal E}(\vec n_2) \sin^2{\theta_{12}}
}
 where $E$ is the total energy (and we assume that the total momentum vanishes).
 We see that the expectation value of $C$ is given by an integral over the energy two point
 correlation function.

On the other hand, it is common to compute the cross section as a function of $C$ (see for example \catani ).
 This is just the
probability of measuring various values of $C$,   $ {d\sigma \over d C}$. This calculation involves more input than the two point correlation
function, since we would need to know all the moments of $C$, $\langle C^n \rangle$, to reconstruct ${ d\sigma \over d C}$.

The point of this short remark is to stress that, even though the $C$ parameter is given by a product of energies, the computation of the
cross-section as a function of $C$ involves knowledge of the $n$ point energy-correlation functions. Of course, in practice, ${ d \sigma \over d
C}$ is computed directly rather than going through the energy correlation functions.

  \newsec{ Energy correlation functions in theories with gravity duals}

  In this section we consider  energy correlation functions in conformal
  field theories that have gravity duals. We first start with some general remarks on the energy correlators and the basic ingredients necessary
  to calculate them. Then, we will present explicit calculations for the energy one point functions, which are given in terms of three point functions in
  the gravitational theories. Finally, we add a general prescription for computing arbitrary $n$ point
  functions.

\subsec{General remarks and basic ingredients of the calculation}

   The general prescription for computing correlation functions of local operators
   in the
   CFT using the gravity dual was derived in \refs{\wittenhol,\gkp}. Computations of the
   expectation values of the stress tensor for falling objects include
   \refs{\fallingmass,\fallingblackhole}.
    Since
    energy flux correlation
   functions are given in terms of stress tensor correlators,
    we simply need to  perform the integral
   over time and take the limit in \energfl .
    In order to simplify the computations it is useful to consider
   some coordinate changes.

Let us start by
 writing $AdS_5$ using the coordinates
\eqn\adsfn{ - (W^{-1})^2 - (W^0)^2  + (W^1 )^2 + (W^2 )^2 + (W^3 )^2  + (W^4 )^2 =-1 } The boundary of $AdS_5$ corresponds to the region where $W^M \to
\infty$. In that regime we can forget about the $-1$ in \adsfn\ and we recover the coordinates $Z^M$ that we described around \constrz . It will be convenient
to introduce three possible sets of coordinates which are natural from different points of view. The first two are
\eqn\coordsy{\eqalign{ { \rm Original }:  &
~~ { 1 \over z } = W^{-1} + W^4  ~,~~~W^\mu ={ x^\mu \over z} ~,~~~\mu =0,1,2,3 \cr { \rm Easy} : & ~~ { 1 \over y_5} \equiv W^0 + W^3 ~,~~~~~~W^{-1} =- y^0
~,~~~~W^4 =- y^3
 ~,~~~~W_{1,2} = { y_{1,2} \over y_5 }
}} Of course the metrics are simply \eqn\coordsymet{\eqalign{ { \rm Original}: & ~~ds^2 = { dx^2 + dz^2 \over z^2 } \cr { \rm Easy} : & ds^2 = { - dy^+ dy^- +
dy_1^2 + dy_2^2 + dy_5^2 \over y_5^2 } }} It is also convenient to introduce a third set of coordinates, which is defined as follows. We first choose three
coordinates describing the $H_3$ subspace $ - (W^0)^2 + (W^1 )^2 + (W^2 )^2 + (W^3 )^2 = - r^2$ for a fixed $r^2$. The two other coordinates are chosen as
\eqn\wpamf{ W^\pm = W^{-1} \pm W^4 }
 Then $r^2 = 1 - W^+ W^-$ and
the metric is \eqn\metrhyp{ {\rm  Hyperbolic }:  ~~~ ds^2 = - dW^+ dW^- - { 1 \over 4 } { ( W^- dW^+ + W^+ dW^-)^2 \over 1 - W^+ W^-}
 +
(1 - W^+ W^-) ds^2_{H_3} } The advantage of this coordinate system is that it makes the
 $SO(1,3)$ symmetry of the problem manifest. This $SO(1,3)$ symmetry are the isometries of $H_3$.
  In addition, the dilatation symmetry in the
 original coordinates becomes a boost in the $W^\pm$ plane, which is also a clear symmetry
 of the metric in this parametrization.

The surface that is at the boundary of four dimensional Minkowski space can be extended to the
 interior in a unique way so that it is
invariant under the symmetries that preserve the boundary of Minkowski space. In fact, this surface is simply given by $W^+ =0$.

 The insertion of the stress tensor operator corresponds to a non-normalizable perturbation
   of the metric in the bulk. It will be convenient to derive first the expressions
   for the momentum in the $y$ coordinates introduced in \explictch .
Since all the generators in \exprg\ are given in terms of the integral of $T_{--}(y') $ over a line along ${y'}^-$, let us compute this first.
We insert $T_{--}(y')$ on the boundary at ${y'}^+=0$ and ${y'}^1={y'}^2=0$ but at an arbitrary value of ${y'}^-$. We denote the boundary points
with primes and the bulk points without primes. This induces the following fluctuation in the metric of $AdS_5$, $g_{MN} \to g_{MN} + h_{MN} $,
\eqn\metrf{
 h_{MN} dx^M dx^M \sim     (dy^+)^2 { y_5^2 \over  [ - y^+ (y^- - {y'}^-)  + y_1^2 + y_2^2 + y_5^2 +
 i \epsilon ]^4 }
  }
We can now perform the integral over ${y'}^-$. We
 use the formula \eqn\formint{ \int_{-\infty}^\infty d {y'}^- { 1 \over [  y^+ {y'}^- + A + i
\epsilon]^4 } \sim \delta(y^+) { 1 \over A^3 } }
 Now, by a simple translation, we can set the energy operator at any
other value of ${y'}^{1,2}$. We obtain
  \eqn\metrflyc{ \eqalign{
  h_{MN} dX^M d X^N \sim&    \delta( y^+) ( dy^+)^2
   { y_5^2 \over ( y_5^2 + ( y_1 - y_1')^2 + ( y_2 - y_2')^2 )^3 } dy'_1 dy'_2 =
   \cr \sim&
   \delta(W^+) ( dW^+)^2
   { y_5^3 \over ( y_5^2 + ( y_1 - y_1')^2 + ( y_2 - y_2')^2 )^3 }dy'_1 dy'_2
 \cr
  \sim  & \delta(W^+) ( dW^+)^2
  { (Z^0 + Z^3)  \over ( W \cdot Z)^3 } dZ_1 dZ_2
   } }
where in the last line we have represented the boundary coordinates using \cooth , with $Z^+=0$, in order to get the answer in a form that will make it easy to
make coordinate changes.

  For example, if we
wish to express the result in the hyperbolic coordinates \metrhyp\ all we need to do is to express $W$ and $Z$ in terms of such coordinates. Let us define
$\vec n = (n_1,n_2,n_3)$ to be a unit vector on a two sphere. On the surface $W^+=0$ we can parametrize $W^0 =   \cosh \zeta$, $W_i = \sinh \zeta n_i$.  It is
then natural to take boundary coordinates $Z^0 =1, ~Z^i = {n'}^i $.
  We take the limit $\zeta \to \infty$ and we then have that
  \eqn\foust{ {W}^0 + W^3 \sim e^{\zeta} ( 1 + n_3) \sim e^\zeta \left(Z^0 + Z^3\right)
~,~~~~~~~~~y'_{1,2} = { Z_{1,2} \over {Z}^0 + Z^3 } = { n'_{1,2} \over (1 + n'_3 ) }
 }
The last equation is simply the change of coordinates \coordstrans .
  We then find the following expression for the
generators
 \eqn\metrfl{\eqalign{
 E  ~~ \longrightarrow ~~  &
 h^E_{MN} dX^N dX^M \sim  \delta(W^+) ( dW^+)^2       { 1 \over ( W^0 - W_i n'_i )^3 } d\Omega'_2
  \cr
  { P}^i ~~ \longrightarrow ~~ &
  h^{P^i}_{MN} dX^N dX^M \sim
  \delta(W^+) ( dW^+)^2    n'_i  { 1 \over ( W^0 - W_i n'_i )^3 } d\Omega'_2
 \cr
 { \cal E}(\vec n')  ~~ \longrightarrow ~~ &
  h^{{ \cal E}(\vec n')}_{MN} dX^N dX^M \sim  \delta(W^+) ( dW^+)^2       { 1 \over ( W^0 - W_i n'_i )^3 }
 }}
In summary, we have computed the metric fluctuation  that corresponds to the integrated insertion of the stress tensor that measures the energy
deposited in the idealized calorimeters that we are placing at infinity at some position $\vec n'$ on the two-sphere.
 In the original AdS coordinates the corresponding insertion
would be localized on the horizon of $AdS$ in Poincare coordinates ($z = \infty$).
 We found it convenient to express the
results in a couple of different coordinate systems that are regular at $z=\infty$ in order avoid having to take a limit. In these other coordinates we see
that we are performing a measurement on the $W^+=0$ surface.
This
 amounts to sampling the wavefunction of the particles in the bulk at $W^+=0$. At $W^+=0$ we have
 an $H^3$ subspace plus the null direction parametrized by $W^-$. The boundary of $H^3$ corresponds
 to the two sphere at infinity where we place the calorimeters.

The form of the equations \metrfl\ does not make explicit the Lorentz covariance of the expressions. In order to see this explicitly we can rewrite them as
 \eqn\metrfllo{
 P^\nu  ~~ \to ~~ h^{P^\nu}_{MN} dX^N dX^M \sim
  \delta(W^+) ( dW^+)^2
    { Z^\nu \over ( -W.Z)^3 } {dS^0_{\phantom{0}\lambda} Z^\lambda \over Z^0
    }}
  \noindent where we sit at $Z^+=0$ and we are integrating over a spacelike surface inside
  $ \sum_{\mu=0}^3 Z_\mu Z^\mu =0 $.   The integration surface differential is defined to be
  such that $dS^{\nu}_{\phantom{0}\lambda} Z^\lambda$ is parallel to $Z^\nu$.
  Therefore the transformation of $dS^0_{\lambda} Z^\lambda$ cancels the
  transformation of $Z^0$  \foot{This is completely analog to the fact in classical
  electrodynamics that the power radiated by an accelerating charge is a
  Lorentz scalar.} .

   In the same way that we have discussed the graviton associated to energy flux measurements we can
   also consider the U(1) gauge field configurations associated to
   charge flow measurements on the boundary theory.
   The operator that corresponds to putting a counter at infinity at some specific location and
   measuring the charge corresponds to the following bulk gauge field configuration
   \eqn\charg{ \eqalign{
   { \cal Q}(\vec n') \to &  A_M dx^M  \sim  dW^+ \delta(W^+)  { 1 \over ( W^0 - W^i n'_i )^2 }
   }}
   %k(\vec W,\vec n')
   %\cr
   % k(\vec W,\vec n' ) &  = { 1 \over ( W^0 - W^i n'_i )^2 }
   % }}
   % Notice that this looks like a plane wave for the gauge field\foot{Note that
   %  $k$ obeys the Laplace
   % equation in $H_3$ with eigenvalue 2, $\nabla_w k = 2 k $. However, in this case it is not true
   % that any such solution is a full solution to the non-linear problem.  }.

%In order to see that they transforms correctly under $SO(1,3)$ we can write these expressions as $ P^\mu \to \delta(W^+) ( dW^+)^2  h^\mu  $,
%where we write $h^\mu$ in an explicitly invariant fashion as \eqn\writehm{
 %h^\mu = \lim_{\epsilon \to 0 }  \epsilon
  %\int d^4 W' \theta( W'^0) 2 \delta( W'^2) { W'}^\mu { 1 \over ( -  W . W' )^{ 3 - \epsilon } }
 %e^{   W. W' }
%} Where $W'$ are four unconstrained variables and $W$ parametrizes an $H_3$ subspace of $AdS_5$,
 %$ W_\mu W^\mu =-1$, $\mu =0,1,2,3$. Here $\theta$ denotes the usual step function. We can
 %first do the integral over ${W'^0}$ and then the $\delta$ function fixes it to
  %$|\vec W'|$. The integral over this absolute value then has a pole whose residues
  %give the expressions \metrfl . From this expression we see that $SO(1,3)$ Lorentz transformations
  %act as the ordinary isometries of the $H_3$ subspace of $AdS_5$ that we are focusing on.

Having discussed the properties of the probe gravitons or gauge fields that represent our measurement, let us now turn to the field in the bulk that describes
the state that we insert with the operator ${\cal O}$.
 We can think of a
scalar source to make things simpler, although our results will be quite general. We are interested in obtaining the field configuration, $\phi$,  in the bulk
of $AdS_5$ created by the insertion of the operator ${\cal O}$ of dimension $\Delta$. If we insert the operator $\int d^4x'  \phi_0(x') { \cal O}(x')$ on the
boundary theory, then the bulk field configuration is given by \refs{\wittenhol,\gkp}
\eqn\orignexpr{\eqalign{
 \phi(x, z) = & \int d^4 x'  \phi_0(x')  { z^\Delta \over [ ( x- x')^2 + z^2 ]^{\Delta } } = \cr = &
\lim_{ z' \to 0 } \int d^4 x'  \phi_0(x')   {1 \over (z')^{\Delta} ( W. W' )^{\Delta } } \cr
\phi_q(W^+=0, W^-,W^\mu ) = &  \int d^4 x { e^{ i q.x' }  \over
\left[ - { W^- \over 2 } - W^0 {x'}^0 + W^i {x'}^i- i \epsilon
 \right]^\Delta }
}}
 where we have first rewritten the result in a way that
allow us to easily change coordinates. In the last line we wrote the expression for the bulk field at $W^+=0$ in the case that $\phi_0(x') = e^{ i q.x'} $.
Notice that we only need the wavefunction at $W^+=0$ since that is where the graviton perturbation is localized. It is not hard to do the integral in
\orignexpr\ explicitly. There is, however, a very simple way to see what the answer should be. We are creating a state that is a momentum eigenstate. For the
moment let us set $q^\mu = (q^0,\vec0)$. The momentum generator corresponds to a bulk isometry generated by a Killing vector that becomes simpler at $W^+=0$,
 \eqn\isomet{
 \left. P_x^\mu \right|_{W^+=0}  =
 % i \partial_x^\mu$ inside the bulk and require that $\phi_b(x,
%)$ is also an eigenstate. At $W^+=0$ the generator is, in $W$ coordinates \adsfn, $
 - 2 i W^\mu \partial_{W^-}
} Of course, this is similar to the corresponding boundary statement \simplegen .
 A wave function that
diagonalizes all four of these operators has to be a plane wave in $W^-$ and should be
 localized in the $W^\mu$ coordinates. In other words we have
 \eqn\phipw{
\phi_q(W^+=0,W^-,W^\mu) \sim  \left(q^0\right)^{\Delta-4} \, e^{i q^0  W^-/2} \delta^3 (\vec W) } \noindent where $\vec W$ refers to a parametrization of the
hyperboloid given by
 $W^i$ with $i=1,2,3$. In these coordinates, $W^0$ is just a
function of $\vec W$ since $W_\mu W^\mu = -1$ at $W^+=0$.
 $\left(q^0\right)^{\Delta-4}$ is a normalization constant that can be obtained from \orignexpr\
 by considering the dilatation operator acting on both sides. In other words,
$\phi_0(x')$ is dimensionless so that $\phi$  scales as
 $\left(q^0\right)^{\Delta-4}$. The overall constant in \phipw\ cancels out when we compute
 energy correlations. Note that in the $y$-AdS coordinates \coordsy\ the wavefunction with definite
 $x$-momentum, \phipw ,
  is localized at $y_1=y_2=0$, $y_5=1$
 when $y^+=0$.

%We are interested in the wavefunction at  surface $W^+=0$. We find \eqn\finans{ \phi_b(W) =   \int d^4 x'  \phi_0(x')
 % {1 \over  ( -{ W^- \over 2 }  - W^{0} t' + W_i . x'_i   + i \epsilon )^{\Delta } }
%} CHECK SIGN OF $i \epsilon$.
 % We have a similar
%expression for $\phi^*_b$. But notice that in $\phi_b^*$ we will have the opposite
 % $i\epsilon$ prescription. These expressions of the wavefunction allows us to compute the
  %gravity results. Of course we can consider any wavefunction we like. In particular, we
  %were interested in a wavefunction $\phi_0$ with fairly well defined momentum.
  %Let us assume it is a plane wave of the form $\phi_0 = e^{ - i q^0 t'} $. In that
  %case we find that the wavefunction is
  %\eqn\phipw{ \eqalign{
  %\phi_b = & \int dt' d^3 x' { e^{ - i q^0 t' } \over
  % ( -{ W^- \over 2 }  - W^{0} t + W_i . x_i   + i \epsilon )^{\Delta } }
 % \cr
 % = & \int d^3x' e^{ i q^0 ( W^-/2 - W_i x'_i)/W^0 } (q^0)^{\Delta -1} =
 % e^{ i q^0  W^-/2} (q^0)^{\Delta -4} \delta^3 (  { \vec W \over W^0} )
 % }}

  Therefore, the general result is that an incoming plane wave
  (with no spatial momentum) gives us a very peculiar wavefunction that is $\delta$ function
  localized in the $H_3$ subspace at the origin $W^i=0$.
   In addition we find that the momentum in the
  $W^-$ direction is proportional to  the original energy. The wavefunction for an external
  operator with a generic value of the momentum $q^\mu$ can be obtained by performing
  a boost of this solution. The end result is again a wavefunction that is localized at a
  point in $H_3$. It is localized at ${ \vec W \over W^0} = { \vec q \over q^0} $.
  The momentum in the $W^-$ direction is now
  $-{\sqrt{-q_\mu q^\mu} \over 2}$.
   The wavefunctions corresponding to plane waves have a divergent norm since
  a plane wave wavefunction has a divergent norm. One can consider the regularized external
  wavefunction in  \opeffs. In that case we find a finite norm. We discuss this case in more
  detail in appendix B. As expected, one finds that the delta function is smeared over a
  region $ | \vec W | \sim { 1 \over \sigma q } $. We will continue to discuss wavefunctions
  for plane waves, but having in mind that we will eventually smear the $\delta$ function in
  \phipw , as in \opeffs .

Once we have the bulk wavefunction we can compute the energy flux one point function. By considering the effects of the metric perturbation \metrfl , and
considering an operator that creates the bulk wavefunction $\phi$, we obtain the expression \eqn\stresstc{\eqalign{ \langle { \cal E}( \vec n' ) \rangle = &
N^{-2} \int dW^- d\Sigma_3 { 1 \over 4 \pi ( W^0 - \vec W . \vec n'
  )^3 } \left. [ ( 2i \partial_{W^-} \phi^*) ( - 2 i
\partial_{W^-} \phi ) ] \right|_{W^+=0} \cr
 N^2 = &  \int dW^- \int d\Sigma_3   \left.
  [   \phi^* (-i \partial_{W^-} \phi ) + c.c.]\right|_{W^+=0}
 }}
 where $\Sigma_3$ denotes the integral over the three dimensional hyperbolic
 space parametrized by $W^\mu$, with $W_\mu W^\mu = -1$.
 The last factor, $N^2$,  is simply the total production cross
 section and it is related to the two point function of the
 operator insertion.
In other words, the two point function $ \langle 0 | {\cal O}_q^\dagger { \cal O}_q | 0 \rangle = || {\cal O}_q | 0 \rangle ||^2$ is the norm of the state.
This norm is given in the bulk by the expression for $N^2$.
  When we insert the wavefunction \phipw\ we
 see that a single point in the integral over  hyperbolic space
   contributes. We finally get the expected result $\langle
   {\cal E} \rangle = { q \over 4 \pi }$, \integra .

  \subsec{Energy flux one point functions in theories with gravity duals}

Using the results above we are ready to calculate the energy one point functions for
 different type of sources. For a scalar source the symmetries  imply
the result \integra. Because there are no free parameters we know this is the correct result and we did not need to go throught the previous discussion. The
situation is more interesting for current sources. In a theory that has a gravity dual the three point function of two currents and a stress tensor
  can be computed from the bulk interaction between two bulk photons and a bulk graviton that
  follows from the bulk  Maxwell action
  \eqn\maxact{
  S = - { 1\over 4 g^2 } \int  d^5 x \sqrt{g}  F^2
  }
  where $g$ is the bulk gauge coupling. This term in the action also determines the
  two point function. Thus, we can see that the three point function will be determined and
  we will get a particular value for $a_2$ in \nepar. We can find this value by noticing that for
  ${\cal N}=4$ Super Yang Mills we had $a_2=0$. Thus, any theory that has a gravity dual
  gives us $a_2=0$, as long as the two derivative approximation \maxact\ is
  valid. In general there will be higher derivative corrections to this action.
  Up to field redefinitions there is a unique higher order operator that can contribute
  to the three point function
  \eqn\unicont{
 S = - { 1\over 4 g^2 } \int  d^5 x \sqrt{g}  F^2 +
  { \alpha_1 \over g^2 M_*^2 }  \int  d^5 x \sqrt{g}
  W^{\mu\nu \delta \rho} F_{\mu \nu} F_{\delta \rho}
  }
  where $W^{\mu\nu\delta\rho}$ is the Weyl tensor. Here
  $M_*$ is some mass scale in the gravity theory determining the strength of the correction relative
  to the strength of the Maxwell term in the action. In order to see that this is the only operator that contributes we
  consider the possible three point vertices between two photons and a graviton
  in flat space.
  It turns out that there are only two possible structures.
  This onshell vertex is
  so constrained because there is no kinematic invariant that we can make purely
  with the external momenta, which all square to zero. In fact the two possible interaction vertices
  consistent with gauge invariance are
  \eqn\interver{ \eqalign{
  v_1 = & \epsilon_{\mu \nu} \left[\epsilon_1^\mu   k_2^\nu  (\epsilon_2 . k_1) +
  ( 1 \leftrightarrow 2)    -
  k_1^\mu k_2^\nu (\epsilon_1 . \epsilon_2) \right]
  \cr
   v_2 = & \epsilon_{\mu \nu} \,  k_1^\mu k_2^\nu \, (\epsilon_1 . k_2) \, (\epsilon_2 . k_1)
  }}
  where $\epsilon_{1,2}^\mu$ are the polarization vectors of the gauge bosons and $\epsilon_{\mu \nu}$
  the polarization vector of the graviton. They are all transverse
   $\epsilon_1 \cdot k_1 =0$ and $\epsilon_{\mu \mu } =0$.
  The first arises from the quadratic action \maxact\ and the second from the higher order correction
  in  \unicont\ .

  We expect that the higher derivative corrections give us a deviation from a perfectly spherical
  energy distribution for the state created by the currents.
  Notice that the higher order correction will not contribute to the angle independent term in the
  energy one point function. The reason is that this term is related by the Ward identities to the
  two point function and the two point function is not corrected by the presence of the higher
  order operator in \unicont\ because the Weyl tensor vanishes is $AdS$.
  A  detailed computation in
  appendix D shows that the higher order term indeed contributes to the
  anisotropic contribution to the energy correlation function
  \eqn\valatwo{
  a_2 = - { 48 \alpha_1 \over R_{AdS}^2 M_*^2 }
  }

  Notice, in particular, that in non-supersymmetric weakly coupled $QCD$ we expect that
  the higher derivative corrections are comparable to the radius of $AdS$ since $a_2$ is of
  order one for weak coupling \valatqcd .

  This anisotropy is intimately related to the anisotropy in
   the gravitational field that is produced by a
  fast moving photon. Let us consider a photon with high momentum $|p_-| \gg 1$.
 We focus on the problem in flat space for the moment.
  Such a fast moving particle
  produces a metric of the form
  \eqn\metrif{
   ds^2 = - dx^+ dx^- + d\vec x^2 + \delta(x^-) (dx^-)^2 h(\vec x)  ~,~~~~~~~~\nabla^2 h =0
   }
   where $\nabla^2$ is the flat laplacian in the transverse directions.
   This metric is an exact solution of Einstein's equations (with zero cosmological constant)
   and arbitrary higher derivative corrections \horsteif .
    The particular form of the solution for $h$
   depends on the coupling of the photons to the graviton.
   For the lowest order action \maxact\ we
   find that
   $h_0 \sim { p_- \over |x| }$ which is independent of the spin of the photon.
  Here we are focusing on the five dimensional case, so that
   we have three transverse directions $\vec x$.
   On the other hand, the
   second interaction \unicont\ gives a function $h$ of the form
   \eqn\funch{
    h_1 \sim  { p_- \over M_*^2} \epsilon^*_i \epsilon_j \partial_i \partial_j { 1 \over |x| } =
    { 3 p_- \over M_*^2 }
    { |n_i \epsilon_i|^2
    - { 1 \over 3 } |\epsilon|^2 \over |x|^3 }
   }
   where $n_i = x_i/|x|$. Note that this contribution to the gravitational field is sensitive
   to the spin of the photon and it has a quadrupole form. In the case that we have a large number
   of photons, this quadrupole tensor would be proportional to the polarization density matrix
   of the photons.

   Even though we've discussed the case of a photon, all that we have said so far can be extended to the
   case that we have a non-abelian gauge theory in the bulk, which corresponds to a non-abelian
   global symmetry in the boundary theory.

   We have a similar story in the case that the inserted external operator is the stress tensor
   itself. Then there are three possible vertices and three parameters specifying the stress
   tensor three point function. One of these parameters is fixed by the Ward identities and it
   multiplies the three point function that we expect from the gravity action. The other two
   parameters multiply higher order gravity corrections.
   In fact, the three possible gravity vertices in five dimensions are
   \eqn\threegrav{\eqalign{
   v_1 = &  k_2^\mu \epsilon^1_{~\mu \nu} \epsilon^{2 \nu}_{~~ \delta} \epsilon^{ 3 \delta}_{~ ~\rho}
   k_2^\rho + { 1 \over 4 } \epsilon^1_{~\mu \nu} \epsilon^{2 \mu \nu} \epsilon^3_{~\delta\rho} k_1^\delta
   k_2^\rho + { \rm cyclic }
    \cr
   v_2 = &  (k _3^\mu \epsilon^1_{~\mu \nu} \epsilon^{2 \nu}_{~~\delta} k_3^\delta) \, (
   \epsilon^3_{~\rho \sigma}
   k_1^\rho k_2^\sigma ) + {\rm cyclic}
   \cr
   v_3 = & (\epsilon^1_{~ \mu \nu} k_2^\mu k_2^\nu) ( \epsilon^2_{~\delta \sigma} k_3^\delta k_3^\sigma )
    (\epsilon^3_{~\rho \gamma} k_1^\rho k_1^\gamma )
    }}
  Such vertices arise from terms in the action of the form
   \eqn\threeposver{ S = { M_{pl}^3 \over 2 } \left[ \int d^5 x \sqrt{g} R + { \gamma_1 \over M_{pl}^2} W_{\mu \nu \delta \sigma }
   W^{\mu \nu \delta \sigma}  + { \gamma_2 \over M_{pl}^4 } W_{\mu \nu \delta \sigma }
   W^{\delta \sigma \rho \gamma} W_{\rho \gamma}^{\phantom{\rho \gamma} \mu\nu}
   \right] }
   This is one way to parametrize the higher derivative corrections. In principle we can have
   another curvature cubed term but it does not contribute to the three point function \tseytlinhigher .

   In fact, in \tseytlinhigher\ such corrections were computed for various string theories. They
   found that $\gamma_1$ and $\gamma_2$ are non-zero in the bosonic string, only
   $\gamma_1$ is nonzero in the
   heterotic string and both $\gamma_1=\gamma_2=0$
   in the type II superstrings. Incidentally, ${1 \over N}$ corrections to this action, yielding an $R^4$ term, were computed for type IIB superstrings in
    $AdS_5 \times S^5$ in \banksgreen\ and their effect on the 3 point function of stress energy tensors in $\cal{N} =$ 4 SYM was discussed.

   %We will later rederive this fact in a simple way.

   One can compute the contributions of the higher order terms in the action to
   $t_2$ and $t_4$ as defined in \mostgengav . After performing some
    calculation described in appendix D, we find
   \eqn\valuesfos{ \eqalign{
   t_2 = & { 48 \gamma_1 \over R_{AdS}^2 M_{pl}^2 } + o({\gamma_2  \over R_{AdS}^4 M_{pl}^4})
   \cr
   t_4 = & { 4320 \gamma_2 \over R_{AdS}^4 M_{pl}^4 }
   }}
 to leading order in the $\gamma_i$. In addition we have assumed
 that the contribution of the $W^3$ operator to $t_2$ will be
 smaller than the one from the $W^2$ operator. This is expected in
 the large radius limit because $W^3$ has more derivatives.
 For an ${\cal N}=1$ supersymmetric theory $t_4=0$. Using \valuesfos\ and \ttwotfour\ we get the
 expression for the $R^2$ coefficient that was derived in \refs{\bng,\odintsov}.

    Notice  that the presence of the first correction to the action \threeposver\
   can also change the angular independent part of the energy flux one point function.
   This
   change should be compensated precisely by a change in the stress tensor
   two point function in order to obey the   Ward identity.
   % Also, the third operator might contribute to $t_2$.

   We could also consider charge one point functions. In that case there are two (parity
   preserving) structures for the three point function \refs{\schreier,\osborn,\freedmanthree}.
     The
   coefficient of one of them is determined by the Ward identities
   and arises only when we have a non-abelian gauge symmetry in the
   bulk. It comes from the usual bulk term of the form $\int Tr[ F^2 ]$.
   The second structure arises from a bulk term of the form
   $\int Tr[ F_{\mu \nu} F^{\nu \delta} F_{\delta}^{~~\mu}]$, or
   more generally,
    from a bulk coupling of the form $\int f_{abc} F^a_{\mu \nu} {F^b}^{\nu \delta}
   {F^c}_{\delta}^{~~\mu} $ with a totaly antisymmetric $f_{abc}$. Notice that these terms do not necessarily come
   from a non-abelian gauge symmetry. They could come from a coupling between three
   different $U(1)$ gauge field strengths in the bulk.

   The parity odd terms come from Chern Simons couplings in five dimensions
    \wittenhol \freedmanthree . For example, for a gauge field we can have
  $\int A \wedge F \wedge F $ or its non-abelian generalization.

\subsec{Comments on the $n$ point functions}

After this discussion on one point functions, let us move on to $n$ point functions.
  All we need to do is to consider metric fluctuations which contain several insertions
  of the energy flux operator.
   In general we would have to worry about the bulk tree level interactions
  among the bulk gravitons corresponding to the insertions of the operators. In our case,
  there is an important simplification. This is due to
  the fact that
   the following plane wave solutions\foot{For applications of this type of solutions in confining backgrounds see \nastaconf .} are exact solutions of Einstein's bulk
   equations \refs{\hotta,\horowitzitzh}
  \eqn\bulso{
  ds^2 = ds^2_{AdS_5} + ( d W^+)^2 \delta(W^+) h(w)
  }
  where $h(w)$ is a function defined on the transverse space, which in this case is a
  hyperbolic space $H_3$ of radius one, given by
  $ - (W^0)^2 + (W^1)^2 + (W^2)^2 + (W^3)^2 =-1$.
   The function $h(w)$ obeys the Laplace equation on this hyperbolic space
   \eqn\lapfc{
   {\vec \nabla}^2_w h = 3 h
   }
   Of course, one can check that\foot{Here the normalization factor is fixed such that we obtain the total energy upon integration.}
   \eqn\valhi{
   h_{\vec n'}  = { 2 i  \over 4 \pi ( W^0 - W^i {n'_i } )^3 }
   }
    is a
   solution and so will be
   an arbitrary superposition
   \eqn\superp{
   h =  \sum_{j=1}^n h_{\vec n'_j}
   }
   which represents the insertion of $n$ calorimeters at angular positions given by
   $\vec n'_j $, $j=1, \cdots, n$. This is summing all the gravity tree diagrams.
   We should now consider the propagation of the wavefunction on the background of this
   plane wave geometry. We want to consider the effects of each $h_{\vec n'_j}$ to first
   order in its strength but the combined effect of all of them.
   Let us recall how we would analyze this problem in flat space first.
   We consider a flat space plane wave of the form
   \eqn\fspwd{
   ds^2 = - dx^+ dx^- + (dx^+)^2 f(x^+) h(\vec x ) + d \vec x^2
   }
   and we will eventually take the limit where $f(x^+) \to  \delta(x^+)$.
   The scalar field   obeys the equation
   \eqn\eqnsca{
   - 4 \partial_- \partial_+ \phi - 4 f(x^+) h(\vec x)   \partial_-^2 \phi + {\vec \nabla}^2
   \phi - m^2 \phi  = 0
   }
   We now
   assume that $f(x^+)$ is nonzero only within some small neighborhood of the
   origin $- \epsilon < x^+
   < \epsilon$. This implies that $f(x^+)$ varies rapidly. Thus we assume that this rate
   of variation is much faster than the rate of variation of the wavefunction along
   the rest of the coordinates.
   In that region we can then approximately solve the wave equation \eqnsca\ as
   \eqn\approxso{
    \phi(x^+ = \epsilon) = e^{ - \int_{-\epsilon}^\epsilon f(x^+) h \partial_- } \phi(x^+ = -\epsilon)
    \to \phi(x^+ = \epsilon  ) = e^{- h \partial_- } \phi(x^+ = -\epsilon)
    }
   Generalizing this method to our case of interest we find that
   \eqn\geneourc{
    \phi(W^+ = \epsilon , W^-, W^\mu) =  e^{ -h \partial_{W^-} }  \phi(W^+ = - \epsilon , W^-, W^\mu)
    }
   where $W^\mu$ denotes a point in $H_3$. This is nothing else than a translation of magnitude $h$ in the $W^-$ direction.
  The same type of behavior was observed for scattering of particles off shock waves in  \shockthv\ and was used to study four point functions in the context of the AdS/CFT correspondence in \ccps .

   The computation we want to do involves the overlap of the final state with the
   initial state in the background deformed by the insertion of the plane wave. In addition,
   we need to divide by the norm \stresstc .
   If we write
   %\foot{Once again, the factor $2i$ is chosen such that the expansion yields the properly normalized $n$ point functions.}
   $h= \sum_j  h_{\vec n'_j}$ and we expand in
  each of the $h_{ \vec n' j}$ to first order  we get the $n$ point function
  \eqn\finalcorr{ \eqalign{
  \langle
  \prod_j
  {\cal E}(\vec n'_j)  \rangle =
  & N^{-2} \int_{W^+=0} dW^- d \Sigma_3 \,  \, \left[ ( i \partial_{W^-} \phi^*)
     \prod_{j=1}^n h_{\vec n'_j}(W) [
    ( -  \partial_{W^-} )^n  \phi ] + c.c. \right]
    }}
    where the integral over $d\Sigma_3$ is over the hyperboloid $W^\mu W_\mu =-1$, $\mu =0,1,2,3$.
  Let us specialize this expression  to the case that we
   have a plane wave external state, which leads to    \phipw .
   In that case we find that all the $h_{\vec n'_j}$
   are evaluated at $\vec W=0$ so that they become independent of the angle. Thus
   we get that not only the one point function is uniform but also all the $n$ point functions are uniform as well.
   This implies that there are no fluctuations in
    the energy and an observer would see a uniform
   energy deposition in all the detectors. In other words, we get
 \eqn\nptfnc{ \eqalign{
  \langle
  {\cal E}(\vec n'_1) \cdots {\cal E}(\vec n'_n) \rangle = \left( { q \over 4 \pi } \right)^n
  }}
  This is what we would expect by thinking that fragmentation
   is very rapid at strong coupling as suggested in \refs{\psdis,\msrecent}.
For a state with a generic, but definite, momentum we find \eqn\finarmpl{
 { \cal E}_{q^\mu}(\vec n') = { 1 \over 4 \pi }
 { ( q^2)^2 \over ( q^0 - \vec q . \vec n' )^3 }
 }
 which is simply the boosted version of the uniform distribution ${ \cal E} = { q^0 \over 4 \pi }$
 that we get for the case where $\vec q =0$.

 This is the result for plane wave states. If one considers a generic state, then there
 can be fluctuations, but such fluctuations are parametrized by the fact that we have a
 wavefunction for momentum.
 In other words, one can write the wavefunction $\phi_0(x)$ appearing
 in \orignexpr\ in
 momentum space as $\tilde \phi_0(p) \equiv \int d^4 x e^{ - i p.x} \phi_0(x)$. We consider
 only wavefunctions which are nonvanishing in the forward light-cone $ p^2 < 0, ~ p^0 > 0 $.
 We could consider other wavefunctions but the corresponding operators vanish when they act
 on the vacuum and will not contribute. Thus, in the formulas below we imagine that $p^\mu$
 is restricted to be in the forward light-cone.
 % Then one gets a superposition of the results in
 %\finarmpl\ with a probability given by
 %\eqn\probabil{
 %\rho(p) = N^{-2}  (p^2)^{\Delta -2 } |\tilde \phi_0(p) |^2  ~,~~~~~~~~~N^2 = \int
 % d^4 p (p^2)^{\Delta -2 } |\tilde \phi_0(p) |^2
%}
%The factor of $(p^2)^{ \Delta -2}$ appears when we consider the norm of a state that
%has momentum $p$,  see appendix A. This factor is determined by the dilatation operator.
  Then we can write the bulk wavefunction as
 \eqn\bulkwfnew{
 \phi(W^+=0, W^-, W^\mu) = \int_0^\infty {  d\lambda  \over \lambda}
  \lambda^{\Delta } e^{ i \lambda W^-/2 } \tilde \phi_0( \lambda W^\mu )
 }
 We see that for a plane wave with purely timelike momentum we should set $\tilde \phi_0 =
 \delta( p^0 - q^0) \delta^3(\vec p)$ and we recover \phipw .
 Inserting \bulkwfnew\  into \finalcorr\ we obtain
\eqn\finalcorrint{ \eqalign{
  \langle
  \prod_{i=1}^n
  {\cal E}(\vec n'_j)  \rangle = & N^{-2} \int d^4 p \, \rho(p) \, \prod_{i=1}^n { p^4 \over
  4 \pi ( p^0 - \vec p \vec n'_i )^3  }
  \cr
  \rho(p) = &  N^{-2}  (p^2)^{\Delta -2 } |\tilde \phi_0(p) |^2
   ~,~~~~~~~~~N^2 = \int
  d^4 p (p^2)^{\Delta -2 } |\tilde \phi_0(p) |^2
    }}
 The factor of $(p^2)^{ \Delta -2}$ appears when we consider the norm of a state that
 has momentum $p$,  see appendix A. This factor is determined by the dilatation operator.
  In appendix B we compute $\rho$ for the localized wavefunction $\phi_0(x)$ given in \opeffs .

 The final picture is that for a generic operator insertion we have a superposition of the
 results for each momentum, given by \finarmpl\ with a probability weight given by $\rho(p)$
 which is giving us  the  probability of exciting the mode with momentum $p^\mu$
 in the conformal field theory (or in the bulk gravity theory).

 For a generic  $\tilde \phi_0(p) $ \finalcorrint\ gives  non-trivial functions of the angles.
 The final picture for what we would see in each event is actually very simple.
 After we measure the energy on four of the calorimeters in each event, we can determine the
 value of $p$ that is contributing and, therefore, the energies in all other calorimeters is determined. See appendix B for a longer discussion of this
 point.
 In other words, from event to event, we have some random variations which are completely captured
 by the distribution of momenta $\rho(p)$. In the bulk picture, we have a pointlike particle in the
 bulk with some wavefunction. Measuring all the energies is tantamount to measuring the
 position of the particle on $H_3$ and its momentum in the direction $W^-$ when it crosses
 $W^+=0$. We can view this as the horizon of $AdS$. We can say that we are simply measuring the
 momentum of the particle as it crosses the $AdS_5$ horizon. In the approximation that
 we have a pointlike particle we have a small number of random (quantum) variables characterizing
 the event. We only have the position or momentum of the particle when it crosses the horizon.
 When we consider a string we have an infinite number of degrees of freedom and we can have
 much more variation in the energy deposition patterns.

   \newsec{Stringy corrections}

In this section we study stringy corrections to the gravity results. First we consider a flat space problem that is closely related to the problem we need to
solve in $AdS$. We then use these results to compute the leading order stringy corrections to the gravity results. Finally we study the small angle behavior of
the two point function and we find the stringy version of the operator product expansion we discussed above.

\ifig\curvedtoflat{ (a) The AdS computation of the energy correlators involves gravitons that propagate from the boundary to the interior on an
$H_3$ subspace of the full $AdS_5$ space. The gravitons originate on the boundary of $H_3$,  at the point where the calorimeter is inserted,
and propagate on $H^3$  to the interior.
 (b) Since the falling  string state is
localized on   $H_3$ we can approximate the computation by a flat space one.
}{\epsfxsize2.5in\epsfbox{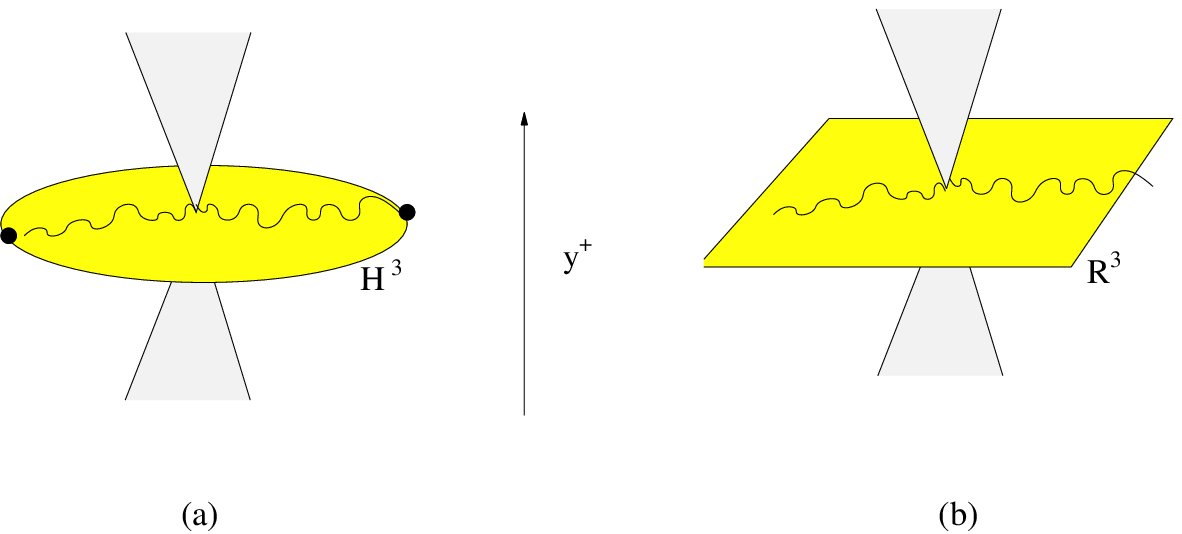}}

\subsec{ Strings probed by plane waves }

   Let us first make the approximation that the $AdS$ space is weakly curved and let us
   approximate the problem as that of strings in flat space, see \curvedtoflat .
  In fact, we have seen that the state created by the operator
  insertion is localized on the transverse surface, the $H_3$
  subspace. In addition, the energy flux operator corresponds to
   a graviton localized  at  $W^+=0$. Thus, we can just look at
   the problem in a neighborhood of this point and approximate it
   as a flat space problem where we have a particle, or a string,
   with nonzero  $p_-$ which  it is being probed by gravitons with
   $p_-=0$ that are localized along $y^+$. Note that the probe gravitons are extended in
   the $y^-$ direction.

More explicitly, we can consider a flat space problem where we
have a string with a non-zero value of $p_-$ which crosses a
gravitational plane wave of the form
   \eqn\gapwv{
   ds^2 = -d y^+ dy^- + (dy^+)^2 \delta(y^+) h + d \vec y^2
   }
where $h$ is a function of the transverse coordinates obeying $\vec \nabla^2 h =0$.
   Due to the symmetries of the problem it is convenient to choose light cone gauge where
   $y^+ = - 2 \alpha'  p_- \tau$. Recall that $p_-<0$ is the momentum conjugate to $y^-$.
   Following the usual steps that lead to light cone quantization we find that we
   get the following light cone gauge Lagrangian for the transverse dimensions
   \eqn\licgala{
   S = { 1 \over 4 \pi \alpha' } \int_{-\infty}^\infty  d\tau \int_0^{2 \pi }
     d \sigma d\tau  [( \partial_\tau \vec y )^2 - ( \partial_\sigma \vec y )^2]
    - { 1 \over 2 \pi }  p_- \int_0^{2 \pi }  d \sigma h( \vec y(\tau=0, \sigma) )
    }
    Notice that the $h$ dependent term is localized at $\tau =0$,
    at a single value of the worldsheet time. Thus, the string
    propagates freely in flat space away from $y^+=0$, or $\tau
    =0$. We will also assume that near $y^+\sim 0$ the string is
    localized near $\vec y =0$ in the transverse directions.
   We can then compute correlation functions from the expression
   \eqn\exprs{
   \langle \Psi | e^{  - i p_-    \int_0^{2 \pi } { d \sigma \over 2 \pi }
    h( \vec y(\sigma ))|_{\tau =0} }
   |\Psi \rangle
   }
   where $|\Psi \rangle $ the full wavefunction of the string state in the light cone gauge theory at
   $\tau =0$.
    We consider a function $h$ which is a sum of a finite number of
    plane waves,  $h = \sum_j   h_j  e^{ i \vec k_j \vec y} $,
    and  we expand to linear order in each
    perturbation $h_j$.
    The mass shell condition is  $\vec k^2_j =0$.  This implies that $\vec k$ is complex.
    (When we go back to the $AdS$ problem it will be natural to take the component of $k$ along the
    radial direction to be purely imaginary and the others to be real.)
    Let us assume that the string state corresponds to the ground state for the
    bosonic oscillators excitations,
    at least for the bosonic transverse directions where the
     momenta $\vec k$ are nonzero.
    We then find that we have to compute correlation functions of the form
    \eqn\correfn{ \eqalign{
    & (- i p_-)^n   \langle \psi_{cm} | \prod_j e^{ i \vec k_j \vec y }| \psi_{cm}  \rangle
    \langle 0| \prod_j \int { d \sigma_j \over 2 \pi } e^{ i \vec k_{j} \vec y_{osc}(\sigma ) }
    |0 \rangle
    \sim
    \cr
    & \sim (- i p_-)^n \langle \psi_{cm} | \prod_j e^{ i \vec k_j \vec y }| \psi_{cm}  \rangle
    \prod_j \int { d \sigma_j \over 2 \pi }  \prod_{ j < i } | 2 \sin { \sigma_i - \sigma_j
    \over 2 } |^{ \alpha' \vec k_i . \vec k_j }
    }}
    where we have separated out the contribution from the center of mass and the oscillators.
    Since the center of mass wavefunction is well localized
    we expect no contribution from it. Namely, we imagine a
    wavefunction which is localized near $\vec y =0$ an thus we
    simply need to evaluate the plane waves in \correfn\ at zero
    which just gives one. Namely,
    $ \langle \psi_{cm} | \prod_j e^{ i \vec k_j \vec y }| \psi_{cm} \rangle \sim 1 $.
    Note that if we neglect the oscillator contributions we recover the gravity result
    following from \geneourc .
    Therefore, the nontrivial contribution comes from the oscillators.
     Notice that these integrals are convergent
    if the $k$'s are all small enough.
    In the case of the two point function we have
    \eqn\twopsf{\eqalign{
    \int_0^{ 2 \pi } { d \sigma \over 2 \pi}  | 2 \sin { \sigma \over 2 } |^{ \alpha' k_1 . k_2 } = &
    { 2^{ \alpha' k_1 . k_2 } \over \sqrt{\pi} }
     { \Gamma( { 1 \over 2 } + { \alpha' k_1 . k_2 \over 2} ) \over
    \Gamma( 1+ { \alpha' k_1 . k_2 \over 2 } ) }
    = \cr
     = &1 + { \pi^2 \over 24 } ( \alpha' k_1 . k_2 )^2
    + \cdots
    }}
    In the second line, the 1 corresponds to the gravity result and the second term is the
    first correction. Naively, one might have expected the first correction to be of order
    $\alpha'$. However, the first term vanishes and the   order   $\alpha'^2$ term  is
    the first non vanishing one.

   It is convenient to rewrite this result in position space.
   We find that the gravity result plus the leading order correction can be written as
    \eqn\twopsn{ \eqalign{
   % \langle { \cal E }(\vec n'_1 ) { \cal E }(\vec n'_2 )
   % \rangle =
   N^{-2} \int_{y^+=0} dy^- d^3y
   &
   \left[ i \phi^* \partial_{y^-}^3 \phi + c.c.\right]
   % \times
   % \cr
   % &
   \left[ h_{1}(y)
    h_{2}(y) +  { \pi ^2 \over 24 }  \alpha'^2 ( \partial_i \partial_j  h_{1}(y)
    \partial_i \partial_j h_{2}(y) ) \right] ~,
   \cr
   { \rm where} & ~~~~~N^2 =   \int_{y^+=0} dy^- d^3 y
    \left[  i \phi^* \partial_{y^-} \phi + c.c.\right]
   }}
 where $\phi$ is the wavefunction of the center of mass of the
 closed string state and $h_1$ and $h_2$ are two graviton plane
 wave states. We have also normalized the result.

In fact, from our discussion we can easily see the origin of the $\alpha'$ corrections
  to the three
graviton vertex in various string theories. These were computed in \tseytlinhigher .
 We first consider the case where there is just one probe graviton in \correfn . $\alpha'$ corrections  can only
  arise if the initial state contains
bosonic oscillators in the transverse directions. For a graviton in the superstring we have no bosonic oscillators in the initial state, only fermion zero
modes. Thus the vertex is the same as the gravity one. In the case of the heterotic string the graviton contains fermion zero modes for the right movers and  a
bosonic oscillator for the left movers. Such an oscillator can give rise to momentum dependent terms of the from given by the second vertex in \threegrav , but
not like the third in \threegrav .
 Finally, in the case of the
bosonic string a graviton with indices in the transverse directions involves bosonic
oscillators for both left and right movers and  gives rise to a vertex like
the third in \threegrav\ (plus the first two, of course).

It is interesting that the string result is finite. One might have worried that
since we are using $\delta(y^+)$ wavefunctions we would obtain divergencies. As we will
see in more detail below, the Regge behavior of the scattering amplitudes in string theory
ensures that the results are finite.

%function
%It is interesting that the string corrections to the gravity result  are finite.
%In fact, one might have
%worried  that, since we are considering states with $\delta $ function wavefunctions in the
%$W^+$ direction,
%we would end up with infinities. In fact, we could consider higher dimension
%operators in the gravity theory which would lead to infinities. For
%example, consider the operator $R^{\mu \sigma \nu \delta}
%R^{\mu' ~ \nu'}_{~ \sigma ~ \delta}
%\partial_\mu \partial_\nu \phi \partial_{\mu'} \partial_{\nu'} \phi $. This leads
%to terms like
% $ R_{+i+j} R_{+i+j} \partial_-^2 \phi
%     \partial_-^2 \phi $ which contain a $\delta(y^+)^2 $ and are infinite.
%We conclude that such terms do not arise in string theory or that there are cancelations
%among various terms of this type.
%We will later return to this point and show that Regge behavior is important for
%understanding the finiteness of the answer. A related point is the following. We might
%have also thought that since the correction in the second line of
% \twopsf\ is polynomial in momenta, we would
%be able to write a higher order operator that would reproduce this answer. The correction,
%however, is non-local in $y^-$  ~ \foot{The covariant vertex would have an
%extra power of $p_-$. Thus we would get $ p_-^3 (k_1.k_2)^2$. But the polarizations of
%the gravitons have the form $\epsilon_{++}$, thus we would need four powers of $p_-$, rather than
%three, to construct a local operator.}. SHOULD WE LEAVE THIS COMMENT, OR EXPLAIN MORE...

\subsec{ Leading order $\alpha'$ corrections to the two point function  }

We now generalize this result to curved space. We can simply
replace the ordinary derivatives in \twopsn\ by covariant
derivatives. However, in the AdS context $h$ obeys an equation of
the form   $\nabla^2 h = 3 h$, or more precisely $\nabla^2 h = { 3
\over R_{AdS}^2 } h$, so that terms that would have been zero in
flat space are non-zero in $AdS$, so we seem to be faced with an
ambiguity. However, these ambiguities only affect terms
that do not have angular dependence, at least for the first
correction. Thus, such terms only correct the constant part of the
energy correlations. We can fix such corrections by demanding that
we obey the energy conservation conditions. It is convenient to
think about the problem in the hyperbolic coordinates \metrhyp . The graviton
wavefunction associated to the insertion of a calorimeter at $\vec n'$ on the
two sphere is given by
 \stresstc
\eqn\formofh{
h \sim  { 1 \over ( W^0 - \vec W. \vec n' )^3 }
\sim { 1 \over ( 1 + { |\vec W|^2 \over 2 } -
\vec W . \vec n' )^3 }
}
where we have have expanded the result around $\vec W\sim 0$.
We have already seen that the center of mass wavefunction is localized near the origin of
hyperbolic space if we have a state created by an operator with zero spatial momentum on $R^{1,3}$,
see \phipw .   Thus, we can
evaluate the derivatives in \twopsn\  and then set  $\vec W =0$.
  This gives
 \eqn\twoptcos{
 \langle { \cal E }(\vec n'_1 ) { \cal E}(\vec n'_2 )
    \rangle = ({ q^0 \over 4 \pi } )^2  \left[ 1 + { \pi^2 \over 24 }
    { {\alpha'}^2  \over R_{AdS}^4}
    \left( \partial_i \partial_j h \partial_i \partial_j h |_{\vec W =0}  + {\rm const} \right)\right]
}
where the constant is an angle independent term that we cannot compute purely in
flat space. It
can be fixed so that we obey the energy conservation condition.
In the end we find
    \eqn\endres{
    \langle { \cal E }(\vec n'_1 ) { \cal E}(\vec n'_2 )
    \rangle = ({ q^0 \over 4 \pi } )^2  \left[ 1 + {  6 \pi^2 \over \lambda  }
     ( \cos^2
    \theta_{12} - { 1 \over 3 } )  + \cdots \right]
     }
     for ${\cal N}=4$ super Yang Mills.
     where $\cos \theta_{12} = \vec n_1' . \vec n_2' $.
     We see that, as expected here, the distribution
     rises in the forward and backward regions. We have fixed the constant
      term in the correction
     by demanding that the integral over one of the angles gives the total energy.
     The dots in \endres\ denote higher order terms  in the $1/\sqrt{\lambda}$
     expansion.

In this derivation we have assumed that the state we are considering has no oscillators
excited along the three transverse $AdS$ directions. In the case of the superstring
we can still have a massless mode with indices in the transverse  $AdS$ directions
since those can be accounted for by fermion zero modes on the string worldsheet in
light cone gauge.  The result \endres\ is very general and holds for any theory with
     a ten dimensional weakly coupled dual with an $AdS_5$ factor if we replace
     $1/\lambda \to { {\alpha'}^2/R_{AdS}^4 } $, under the assumption that we are
     creating a ten dimensional massless closed string with the external operator.

\subsec{Corrections to the $n$ point function}

We now consider the $n$ point function $ \langle {\cal E}(\vec n_1) \cdots {\cal E}(\vec n_n)
\rangle $.
We have seen that the gravity result
is just a constant. Let us compute the stringy   corrections.  The leading deviation
can be computed by expanding the full expression \correfn\ up to quadratic order
in products of $k_i . k_j$. The resulting correction is basically the same as the one
contributing to the two point function \endres . In order to see something new we can
 go to cubic order in
the products $k_i . k_j$.
In the end this gives us a correction to the
$n$ point function which looks like
\eqn\correcn{\eqalign{
\langle {\cal E}(\vec n_1) \cdots {\cal E}(\vec n_n) \rangle = &
 \left( {q \over 4 \pi } \right)^n \left[ 1
 + \sum_{i < j }  {  6 \pi^2 \over \lambda  }
     [  (\vec n_i . \vec n_j )^2  - { 1 \over 3 } ]  + \right.
 \cr
 & \left. +      { \beta  \over \lambda^{3/2} } [
     \sum_{i < j <  k} (\vec n_i . \vec n_j)
     ( \vec n_j . \vec n_k)(\vec n_i . \vec n_k ) + \cdots ]
  + o (\lambda^{-2})  \right]
}}
where  $\beta$ is a numerical
 coefficient\foot{$\beta = - 1728 \int_0^{2 \pi } { d\sigma_1 d\sigma^2
\over ( 2 \pi)^2 } \log[ 2 \sin { \sigma_1 \over 2 } ]  \log[ 2 \sin { \sigma_2 \over 2 } ] \log[ 2 |\sin { (\sigma_1 - \sigma_2) \over 2 }| ]  \sim 518 \pm 5
$.} and the dots denote terms that are necessary to ensure that the integral over each of the angles  gives zero as well as a term that corrects the
coefficient of the $(\vec n_i \vec n_j)^2 $ term by an order $\lambda^{-3/2}$ amount.

Thus, we find that for a strongly coupled field theory the energy distribution is
uniform with small fluctuations which have an amplitude of order $1/\sqrt{\lambda}$.
In other words, $\delta {\cal E}/{\cal E} \sim {1 \over \sqrt{\lambda} }$. The two
point function of these fluctuations is given by the first non-constant term in \correcn .
One might have thought  that these flucutuations would be gaussian. However, we find that the
three point function of the fluctuations is of order $\lambda^{-3/2}$. Thus, when we
normalize the two point functions to one, the three point functions are of order one.
Thus we conclude that the fluctuations are not even approximately gaussian.
More explicitly, we can define a fluctuation operator
\eqn\flucop{
\delta = { { \cal E} - \langle { \cal E}  \rangle \over \langle { \cal E}  \rangle } ~,~~~~~~
{ \rm and } ~~~~ \hat \delta = \sqrt{\lambda } \delta
}
where the operator $\hat \delta$ is defined so that its two point function is independent
of $\lambda$.
We then have
\eqn\twoandthree{ \eqalign{
\langle \hat \delta(\vec n_1)   \rangle = & 0
\cr
\langle \hat \delta(\vec n_1) \hat \delta( \vec n_2 ) \rangle = &  {  6 \pi^2   }
     [  (\vec n_1 . \vec n_2 )^2  - { 1 \over 3 } ] [ 1 + o ( \lambda^{-1/2} )  ]
\cr
\langle \hat \delta(\vec n_1) \hat \delta( \vec n_2 )  \hat \delta( \vec n_3 ) \rangle =
&  { \beta }[
       (\vec n_1 . \vec n_2 )(\vec n_1 . \vec n_3 )(\vec n_2 . \vec n_3 )  + \cdots     ]
}}
We see that the three point function is not parameterically suppressed relative to the two point
function. Of course, they are both suppressed relative to the gravity result.

\subsec{Stringy corrections to  charge  two point functions }

\ifig\feynmandiv{ (a) Feynman diagrams that lead to energy correlation functions. The gravitons do not interact before they touch the falling state,  $\phi$.
We have also indicated the $t$ and $s$ channels as we define them in the text. (b) Diagram that leads to a divergence in the charge correlation function. The
intermediate state is a graviton and the $AAh$ vertex comes from the Maxwell term in the action.  This divergence is cured by going to string theory and
exploiting the Regge behavior of the amplitudes. In (c) we draw a diagram that can arise due to a higher derivative contact interaction in the gravity theory
which could lead to a divergence in the gravity approximation. }{\epsfxsize1.8in\epsfbox{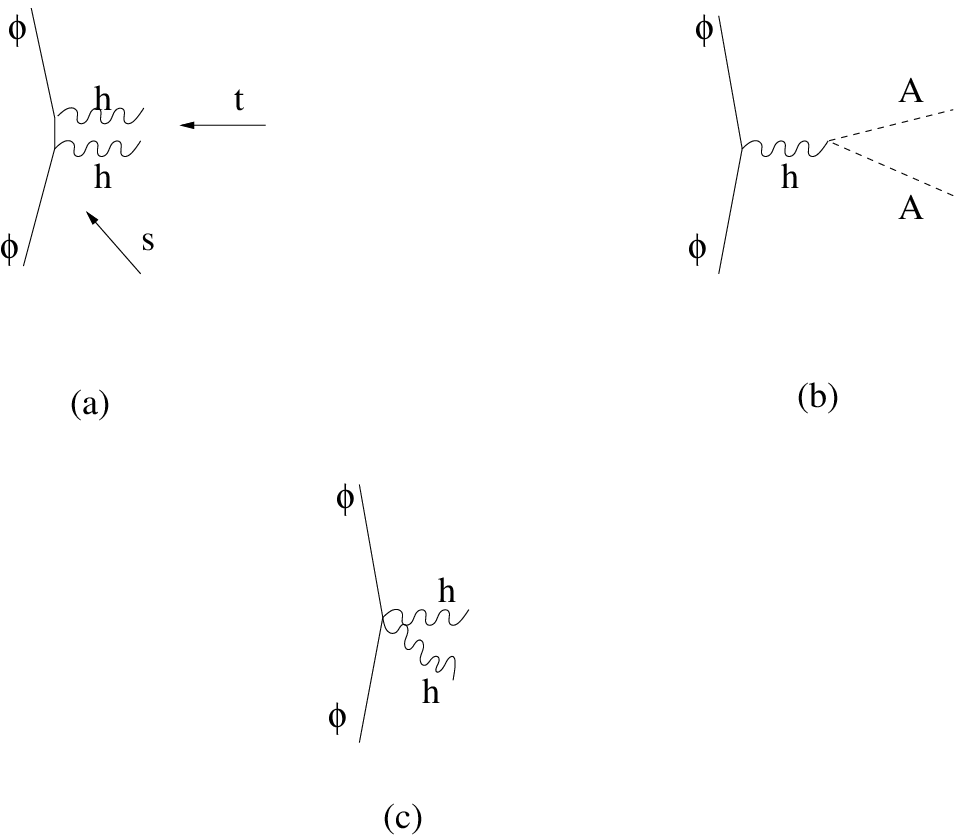}}

     We now consider  the two point function of a charge that
     is dual to a closed string mode. For example, we can pick one of the
       $SO(6) $ currents in ${\cal N}=4$ super Yang Mills. More generally we
       consider  a current
      associated to a symmetry that is carried by fields in the adjoint representation
      in the dual field theory.
      Imagine that the current comes from Kaluza Klein reduction.
     Then the corresponding vertex operator, in light cone gauge, has the form
      $A_+ \to  \partial_\tau \varphi e^{ i k. x } $  where $\varphi$ is one of the internal
      dimensions. We assume that the state we are measuring, $|\Psi\rangle$,
       does not have any oscillator excited
      in the $\varphi$ direction and that it is not charged.
             Thus
      the one point function of the charge, is zero.
    The two point function is actually infinite in the gravity approximation. This is due
    to the Feynman diagram in
       \feynmandiv (b).  This is intimately related
    to the fact that \charg\ is not an exact solution of the gravity equations, but it sources
    a gravitational plane wave proportional to $F_{+ i}^2$, which leads to the square of a
     $\delta(y^+)$ function.
     We did not run into this problem in the gravity theory
     because there are no diagrams of this form due to the fact that the gravitational shock wave
     is an exact solution of the theory. Even in the gravity theory we could have run into this
     problem if we had had a higher derivative contact interaction that brings together two gravitons, as in
     \feynmandiv (c).

     Let us now compute the two point function in string theory.
       The corresponding flat space expression is similar to
       \twopsf , but with an extra factor coming from the contractions of the $\partial_\tau \varphi$
       field coming from the two vertex operators.
        We get a result proportional to
            \eqn\twopsfve{
    \int_0^{ 2 \pi } { d \sigma \over (2 \pi ) }
    | 2 \sin { \sigma \over 2 } |^{ \alpha' k_1 . k_2 -2 }  =
    { 2^{ \alpha' k_1 . k_2 -2 } \over \sqrt{\pi} }
    { \Gamma( -{ 1 \over 2 } + { \alpha' k_1 . k_2 \over 2 }   ) \over
    \Gamma(   { \alpha' k_1 . k_2 \over 2  } ) }
    \sim - { \alpha' k_1 . k_2 \over  4 }
    + \cdots
    }
    We have defined the integral by analytic continuation in $k_1.k_2$
    % \foot{
    %The fact that for small $k_1.k_2$ the integral diverges might be related to
    %  to the fact that we expect that the charge correlators should diverge at
    %very small angles due to collinear divergences. }.
    We see that we get a perfectly finite answer in string theory. The string theory
    answer even goes to zero as we take the small momentum limit.
     Translating this flat space result \twopsfve\ to $AdS$ as we did above we get
    \eqn\chargc{
    \langle {\cal Q}(\vec n_1 ) {\cal Q}(\vec n_2 ) \rangle =  { \gamma \over \sqrt{ \lambda} }
    \vec n_1 . \vec n_2 ={ \gamma \over \sqrt{ \lambda} }
    \cos \theta_{12}
    }
where $\gamma$ is a positive numerical coefficient. This result has the angular dependence
that one would intuitively expect, with the two oppositely charged particles going in opposite
directions.

Let us emphasize once more an important point. Due to the fact that we are considering shock waves which are highly localized, with   $\delta(W^+)$
wavefunctions,  it is important to perform the computation in string theory rather than first taking the low energy limit of string theory and then doing the
computation. We will revisit this point later.

%%% TO PUT FIGURES INSERT:
\ifig\openstrings{ (a) Worldsheet vertex operator insertions for charge correlators associated
to closed string gauge fields. These charges are carried by the bulk of the worldsheet. In (b) and (c)
we consider charges carried by open strings. We consider an open string stretching between two different
branes called $A$ and $B$. In (b) we consider the charge two point function for the $U(1)_A$ living
on the   brane $A$. The two vertex operators are inserted at the same point. In (c) we consider the charge
two point function for charge living on the brane $A$, $U(1)_A$, and the charge living on  brane $B$,
$U(1)_B$. The vertex operators are inserted on different boundaries and the result is non-singular.
}{\epsfxsize2.5in\epsfbox{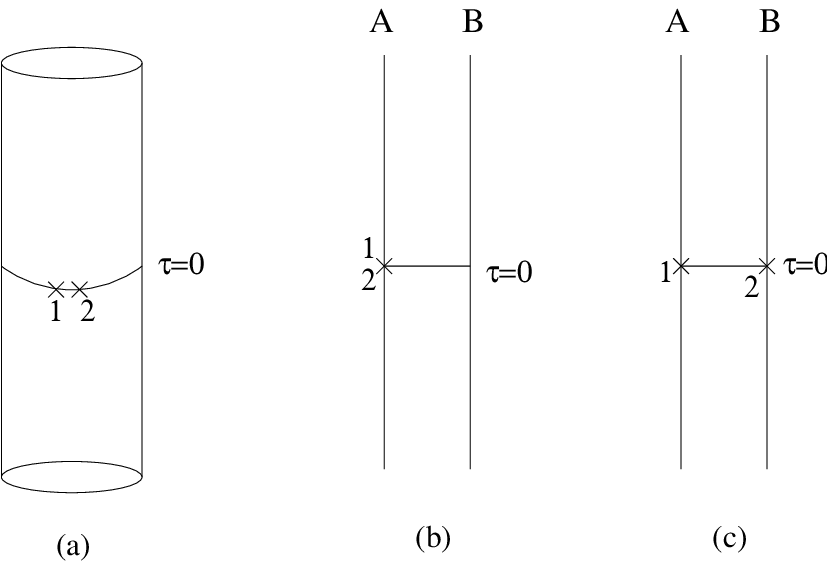}}

    Another interesting situation arises when we consider currents that act only on
    fields in the fundamental representation, such as flavor symmetries. In this case
    the currents live on D-branes in the bulk.
   For simplicity let us  assume that we have two D-branes with two
    different $U(1)$ gauge fields in the bulk. Let us call them $U(1)_A$ and $U(1)_B$.
    We could imagine a   QCD-like theory
     where $U(1)_A$ and $U(1)_B$ are different flavor number
    symmetries.
    At leading order in $N$ we   detect these charges in the detector only if we create a
     mesonic operator that contains the corresponding  quarks. Consider a situation
    where we have a  lorentz scalar
    meson where the quark is charged under $U(1)_A$ and the anti-quark under
    $U(1)_B$. In such a situation we expect to find only one charge of each  type in the
    detector.   In this case the charge one
    point functions are $\langle { \cal Q }_A (\vec n) \rangle =
    \langle { \cal Q }_B (\vec n) \rangle = { 1 \over 4 \pi} $.
    What is the charge two point function for $U(1)_A$? From the boundary field theory point of view
    we expect    it to be zero for generic  angles since the charged quark can be detected
    only at  one particular angle, since we are working to leading order in $N$ where we do not
    create quark anti-quark pairs.
    On the other hand, to leading order in $N$,
    from the gravity plus maxwell theory in the bulk   we get
    % the same
   %  infinity we discussed before, from the diagram in  \feynmandiv (b).
   get a  completely spherically symmetric distribution of charge. In this case, the divergent term
    coming from the Feynman diagram in
    \feynmandiv (b), is subleading in $N$ and we do not consider it (string theory ought to
   make this $1/N$ correction finite too).
   In other words,
    the two point function for the charges is
    $\langle { \cal Q }_A (\vec n) { \cal Q }_A (\vec n') \rangle_{gravity} = { 1 \over (4 \pi)^2 }$.
   This contradicts the field theory expectations. The resolution is that
    the stringy corrections are so large that they completely change  the gravity  result.
    Let us first see how this works in the flat space case. Here, we can quantize the open
    string in light cone gauge and we will get an action very similar to the one we had for the
    closed string except that the photon vertex operator, which is inserted at
     $\tau =0$, is also inserted at $\sigma =0$ at one of the boundaries of the open string.
     We find that
     \eqn\twopch{
       \langle \psi_{cm} |    e^{ i \vec k_1 \vec y }   e^{ i \vec k_2 \vec y } | \psi_{cm}
       \rangle
    \langle 0|  e^{ i \vec k_{1} \vec y_{osc}(0,0 )  } e^{ i \vec k_{2} \vec y_{osc}(0,0 )  }
    |0 \rangle
     }
     If we ignore the oscillators we go back to the gravity result.
     However, the contribution from the oscillators involves a singularity, since
     both vertex operators are evaluated at the same point.
     Formally, this gives
      a contribution of the form $ 0^{ 2 \alpha' k_1.k_2} $. If
     $k_1 . k_2 > 0$, then we see that this vanishes. Thus, if we define the answer by analytic
     continuation   we get zero for all values of $k_1.k_2$, including
      the physical values of $k_1. k_2$ for our problem (which are negative).
       The reason that stringy corrections have such a large
     effect is that we start with singular wavefunctions for the photon, which contain
     a $\delta(y^+)$. If we had started with a smooth wavefunction in the $x^+$ dimension we would
     have integrated the vertex operators along the $\tau$ direction on the boundary of
     the open string worldsheet and we would have obtained a non-vanishing function of $k_1.k_2$.

     On the other hand, if we compute the two point function for the two different $U(1)$ charges,
     the charge carried by the quark and the charge carried by the antiquark,
      then we get the vertex operators at opposite points of the string and we obtain a finite
     answer
     \eqn\twopchdif{
     \langle \psi_{cm} |  e^{ i \vec k_1 \vec y }   e^{ i \vec k_2 \vec y } | \psi_{cm}
       \rangle
    \langle 0|  e^{ i \vec k_{1} \vec y_{osc}(0,0 )  } e^{ i \vec k_{2} \vec y_{osc}(0,\sigma = \pi )  }
    |0 \rangle \sim  2^{ 2 \alpha' k_1 . k_2 }
     }
  In this case the leading order $\alpha'$ correction to the two point function reads
     \eqn\finaltwodiff{
      \langle {\cal Q}_A(\vec n_1 ) {\cal Q}_B(\vec n_2 ) \rangle =  { 1 \over (4 \pi)^2 }
      \left[ 1 - {  8 \log 2 \over \sqrt{ \lambda } } \cos \theta_{12} \right]
    }
   We see that there is a tendency for the two charges to go in opposite directions, as one
   naively expects.  Of course,
   at weak coupling the quark and the antiquark fly in opposite directions (if we have the simplest
   operator which contains only a quark anti-quark pair). If we were to consider a higher
   point function we would get zero again.

   The general lesson is that when we compute charge correlators  it is very important to understand
   the effects of stringy corrections.

   Once we consider finite $N$ corrections   we do not expect the two point
   functions for the same $U(1)$ to be exactly zero.

\subsec{ Small angle behavior of the two point function and the operator product expansion}

In this section we   study the small angle behavior of the two point functions using string theory.
The leading order correction to the energy flux two point function \endres\ is
analytic at small angles, i.e. when $\theta_{12} \to 0$. As we will explain below this
is no longer the case once all the $\alpha'$ corrections are included. We will show
that at small angles there is a non-analytic term of the form $ |\theta_{12}|^{p}$ with a
power, $p$, that we will compute. This power  is intimately related to the singularities
in the first line of \twopsf\ as a function of $ k_1.k_2 $.

Let us first   understand    how the
  singularities in the flat space answer
   \twopsf\ arise. These singularities are
    at $ \alpha' k_1.k_2 = -1 - 2 n$. We can rewrite this
   condition as
   \eqn\polepos{ t \equiv  -(k_1 + k_2)^2
   =  { 2 + 4 n \over \alpha' } ~,~~~~~ n=0,1,2, \cdots
   }
   Similar looking singularities are a well known feature of string scattering amplitudes
   and they arise when an invariant, such as $t$, is equal to the mass of a string state.
   In that case we can view them as arising from the production of an on-shell closed string
   state.
    In our case, however,  there are no states in the closed string spectrum with
    masses given by \polepos . Thus we seem to have a puzzle. We will argue that we indeed
    have certain string states, but of a non-local kind.

%%% TO PUT FIGURES INSERT:
\ifig\worldsheetope{ (a) The poles of the ordinary closed string amplitude arise from the region where the two vertex operators are close to each other but are
integrated over both $\tau$ and $\sigma$.   (b) Wordsheet OPE for the problem we are considering where we have wavefunctions localized in $x^+$. In light cone
gauge this results in operators localized at $\tau =0$. Thus we get singularities from the region of the integral where $\sigma_{12} \to 0$.
}{\epsfxsize2.5in\epsfbox{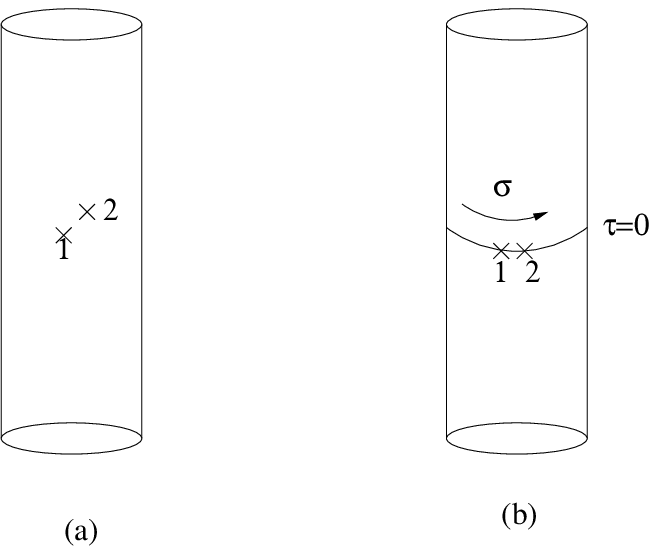}}

    Let us first understand the worldsheet  origin of these singularities.
    For concreteness, let us focus on the first singularity  at
   $\alpha' k_1 .k_2 = -1$. We see that at this point the integral in \twopsf\
   diverges at $\sigma =0$ like $\int   {d \sigma \over \sigma } $.
   At $\sigma \sim 0$ the two closed string vertex operators of the
   external gravitons come close together, see  \worldsheetope b.
   This looks similar to the ordinary OPE region
   of a closed string worldsheet which produces the  usual closed string state poles.
   The crucial
   difference  is that in our case the integral runs only over
   the sigma direction (see \worldsheetope b),
   while in the ordinary case it runs  over $\tau $ and $\sigma$ (see
   \worldsheetope a).  For this
   reason the position of the poles has been shifted compared to the ordinary closed string
   poles. Schematically we have
   \eqn\restc{ \eqalign{
   { \rm Usual~Case:} &~~~~\int dz^2 | z|^{ \alpha' k_1. k_2 } \sim { 1 \over \alpha' k_1 . k_2 +2 }
   \cr
   { \rm Our~Case:} & ~~~~~\int d\sigma
   | \sigma|^{ \alpha' k_1. k_2 } \sim { 1 \over \alpha' k_1 . k_2 +1 }
   }}

   We  have now understood how the singularities arise from the  worldsheet computation.
   Before moving on, let us clarify further a confusing aspect of these singularities.
   All we are doing is to scatter four string states: the two gravitons, the state we are
   measuring and its complex conjugate.
     So, why are the singularities different than singularities of
   the ordinary closed string scattering amplitude?
   What happens is that we are choosing very peculiar wavefunctions for the two
   gravitons. These
   wavefunctions contain $\delta(y^+)$ factors  which implicitly
   carry an infinite amount of momentum. More precisely, in order to go from the usual momentum space
   result to our expression for $\delta(y^+)$ wavefunctions  we
   should integrate the  momentum space result
   over $k_{1 \, +}$ and $k_{2 \, +}$.
    The four point amplitude
     is characterized by  $t = -(k_1 + k_2)^2$
  and $s = - ( p + k_1)^2 $, where $p$ is the momentum of the incoming
   closed string state with non-zero $p_-$.
  Since $k_{i \, -} =0$, we have that $t$ is independent of $k_{i \, +}$ and it continues to
  be given by
  the transverse components of $k_i$,  $t = - ( \vec k_1 + \vec k_2)^2 $.
  On the other hand $s$ contains a contribution of the form
  $s = 4 p_- k_{1 \, +} + \cdots $.
  For the polarizations of the gravitons that we are choosing here the
  amplitude has the form  \GSW \foot{We take the momenta to be nonvanishing only in the first
  five dimensions. The falling string state is taken to be a graviton which has indices in the
  remaining five dimensions. Of course the two probe gravitons have indices in the $++$ directions.}
  \eqn\amplit{
   { \cal A}_{4} = p_-^4 \left( { 1 \over s } + { 1 \over u} \right)
   { \Gamma( 1- { \alpha' s \over 4 } ) \Gamma( 1- { \alpha' t \over 4 } )
   \Gamma(1 - { \alpha' u \over 4 } ) \over
    \Gamma( 1+ { \alpha' s \over 4 } ) \Gamma(1+ { \alpha' t \over 4 } )
   \Gamma( 1+ { \alpha' u \over 4 } )} ~,~~~~~~~u = -s -t
   }
   In the gravity limit we recover the results that we expect from the diagram in \feynmandiv (a)
   and a crossed version.
   We can take $p_-$ to be fixed. Then the integral over $k_{1 \, +}$
  translates into an integral over $s$ \foot{The integral over $k_{2 \, +}$ simply gets
  rid of the momentum conservation delta function in the $k_+$ direction. }.
  At large $s$, the four point amplitude is controlled by   Regge behavior.
  The amplitude goes as
  \eqn\calamp{
  { \cal A}_4  \sim s^{  - 2 +  { \alpha' t \over 2 }  }
  }
  So the integral over $s$ converges at large values of $s$
  for small $t$ \foot{There are poles along the real $s$ axis. As usual,
  we give these poles a small positive or
  negative imaginary part so that we are analyzing the amplitude in the physical sheet. Thus, the
  poles along the real $s$ axis do not lead to divergences in the amplitude.}.
 As we increase $t$, the integral over $s$ first diverges when
  the amplitude goes like $1/s$, $ {\cal A}_4 \sim 1/s $.
  This condition is precisely the $n=0$ case in \polepos . We get
  the higher order ones by a similar reasoning by expanding the amplitude to higher orders in the
  $1/s$ expansion. One minor subtlety is that only even powers of $1/s$ give rise
  to $1/s$ terms in the full amplitude, after we adjust $t$ appropriately\foot{In other words, $A_4 \sim s^{- 2 + \alpha' t /2} \sum_{n=0}^\infty { c_n(\alpha't)/s^n }$, but
  $c_{ 2 k+1} ( \alpha' t = 4 k +4) =0$.}.  Odd powers of $1/s$ would
  lead to extra singularities beyond those given by \polepos , see more on this below.
   Thus, we see that the poles \polepos\ are associated to the high
  energy behavior of the string amplitude. This is to be expected since by probing the string
  at $y^+=0$ we are taking a snapshot of  the string state and this requires high energy
  scattering. The fact that the amplitudes we are computing are finite is related to the
  fact that the high energy scattering displays Regge behavior.
  In conclusion, the result we obtained in lightcone gauge
  is perfectly consistent with the usual structure
   of the Shapiro-Virasoro amplitude.

   A related remark that we can make at this point is the following. Let us go back to the
   case where we consider a neutral falling state probed by
    two closed string gauge bosons. Going
   to ten dimensions we can view the gauge bosons as Kaluza Klein gravitons. In that case
   the flat space amplitude is very similar to \amplit\ except that we now have
   \eqn\otheramp{
   { \cal A}_{g} = p_-^2
   { \Gamma( 1- { \alpha' s \over 4 } ) \Gamma( 1- { \alpha' t \over 4 } )
   \Gamma(1 - { \alpha' u \over 4 } ) \over
    \Gamma( 1+ { \alpha' s \over 4 } ) \Gamma(1+ { \alpha' t \over 4 } )
   \Gamma( 1+ { \alpha' u \over 4 } )}
   }
   If we take the small momentum limit of this amplitude we get a constant. If we integrate
   this constant with respect to $s$, in order to go to $\delta$ function wavefunctions, then
   we get an infinity. This is the infinity that we mentioned above as coming from the field
   theory diagram in \feynmandiv (b). Of course the full string amplitude is not constant. Thus, once
   we go to string theory we should integrate the full string amplitude \amplit , which
   goes as $ {\cal A}_g \sim s^{ -1 + { \alpha' t \over 2 } }$ and converges if $t$ is negative.
   We can define it by analytic continuation for other values of $t$. Thus, we see that for this
   particular case, taking the low energy limit of the amplitude first and then doing the $s$ integral
   gives a very different answer than doing first the $s$ integral and then the low energy limit,
   which gives \twopsfve .
   In the case of energy correlations, if we first take the low energy limit of \calamp\ and then
   we do the integral we get the same answer as doing first the integral and then the low energy
   limit. In this case this happens because the contribution is coming mainly
   from the $s\sim 0$ and $u\sim 0$
   region.

   It is useful to perform explicitly the worldsheet
    operator
   product expansion of the two graviton vertex
   operators, see figure \worldsheetope (b). We obtain\foot{In the following expression it is convenient to redefine the range of
   $\sigma$ to $[-\pi, \pi]$, such that insertions close to the operator at zero are given by small $|\sigma|$.}
   \eqn\opepr{
   p_- e^{ i k_1 . y(\tau =0,\sigma) } p_- e^{ i k_2 .y (0,0)} \sim  p_-^2
   |\sigma|^{ \alpha' k_1 .k_2 } [  e^{ i (k_1 + k_2 ) y(0,0) } + \ldots  ]
   }
   The pole  arises when the power of $\sigma$  is precisely $1/\sigma$.
   This gives rise to the $n=0$ case in
   \polepos . The operator that appears at this point has the form
   \eqn\poleoper{
    p_-^2 e^{ i k .y } ~, ~~~~~~~~~ m^2=- k^2 = { 2 \over \alpha ' }
    }
    This is the operator that appears on the worldsheet
    in light cone gauge. One is tempted to write
    an operator in conformal  gauge that would reduce to \poleoper\  in light-cone gauge.
    Due to its peculiar $p_-$ dependence,
    we are forced to write an expression of the form\foot{Notice that the extra power of $p_-$ appears to compensate the one appearing from the delta function
    $\delta(y^+) \sim {\delta(\tau) \over p_-}$.}
    \eqn\formexp{
     ( \partial_\alpha y^+ \partial_ \alpha y^+ )^{ 3 \over 2} \delta(y^+) e^{ i k.y }
     }
     with $k$ obeying the condition \poleoper. This operator is formally a Virasoro primary but
    is not a proper local operator on the string worldsheet. Similar operators were shown
    to control  Regge physics in \psregge . Of course, our regime is closely connected with
     Regge physics so it is not a surprise that similar operators appear.
     The   operator \formexp , without the delta function, has spin $j=3$  in the $y^+,y^-$ plane, as we had in the field
     theory discussion. This is related to the factor of $p_-^2 $ that appears in \opepr . Notice that the complete operator \formexp , including the delta
     function, has total spin 2. It corresponds to the field theory operator ${ \cal U}_{j-1}$ with $j=3$.

      The operator \poleoper\ is the leading contribution in the worldsheet OPE \opepr .
      As we expand the exponentials at higher orders we pick up new operators which contain
      derivatives with respect to the transverse directions. Some of these
       operators can have
      transverse spin. These strings states have higher masses.
      They all have spin $j=3$ in the $y^\pm$ directions. Notice, however, that terms that come with odd powers of $\sigma$ in the $\ldots$ in
      \opepr\ vanish upon integration over $\sigma$. The reason is a $Z_2$ symmetry from interchanging $\sigma \rightarrow -\sigma$.
      This is completely analogous to the fact that terms that are not symmetric under the interchange of $z \rightarrow \overline{z}$ in the
      usual case where we integrate over the whole complex plane vanish upon integration. This is nothing else than the level matching
      condition. These terms are the same as the ones discussed above in connection with the singularities for odd powers of ${1 \over s}$. Now
      we understand why these poles are absent from \twopsf .

     In the end, the singularities arise from a fairly ordinary worldsheet operator
     product expansion in light cone gauge.
      On the other hand, we cannot associate the corresponding worldsheet operator to an
      ordinary closed string state. This is related to the fact that
       the operator product expansion of two energy flux operators in the
      field theory lead to   non-local operators in the $y^-$ direction. In the
      field theory we were not too disturbed by the appearance of operators that are non-local
      in the $y^-$ direction, so we should also not be surprised that  in string theory
      we also get string states that are non-local in the $y^-$ direction.  These string
      states are localized at $y^+=0$, carry zero $p_-$
       and are local in the transverse directions.

       The final conclusion of this discussion is that we should interpret the singularities in
       \twopsf\ as arising from the propagation, in the transverse space, of  non-local string
       states created by the operators like  \poleoper\ or \formexp .

%%% TO PUT FIGURES INSERT:
\ifig\stringope{ Small angle expansion of the energy correlation function. The expansion is
dominated by the propagation of a spin three non-local string state, denoted here by a thick red line.
  } {\epsfxsize2.5in\epsfbox{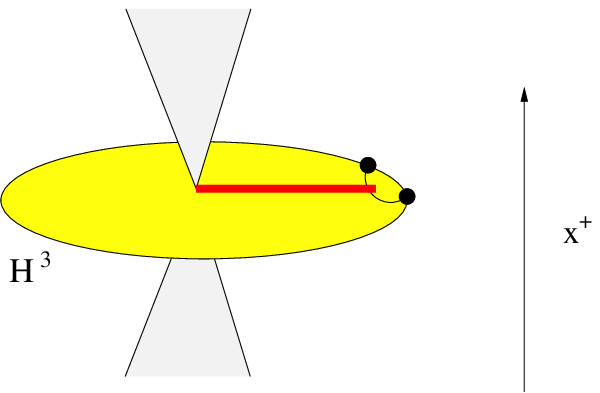}}

      We will now argue that the short distance singularity of the energy flux two point
      function is governed by the operator associated to the first
      singularity in \twopsf . Since we are working in the regime of large but fixed $\lambda$
      we might imagine that we could always expand the two point function as in the second line
      of \twopsf . This is
      not correct if the angle is very small. In that case the relevant relative momenta are
       of order $ t \sim {  1 \over |\theta_{12} |^2 }   { 1 \over R_{AdS}^2}$. Thus, at angles of
      order $\theta \sim \lambda^{-1/4}  $ we cannot use the approximations leading to \endres .
      However, we can use the interpretation given above to the poles in $t$ to write the
      flat space result in the first line of \twopsf\ as a sum over contributions of poles. Then each pole
      corresponds to the contribution of a physical (but non-local) string state that
     is localized in the $y^+$ direction but propagating in the transverse directions.
     We generalize this result to $AdS$ by replacing the transverse space by $H^3$.
     Now the non-local string states propagate on the
      $H_3$ subspace of $AdS_5$. These states propagate
     from the center of $H_3$, where the string state created by the localized operator
     insertion is concentrated,  to a region near the $H_3$ boundary, near the insertion of
     the two energy flux operators, see  \stringope .
     At large distances from the $H_3$ center we expect that the wavefunction of the
     non-local string state goes as  $1/|\vec W|^{\Delta}$, $|\vec W| \gg 1 $,
      with
     \eqn\dimeconr{
     \Delta \sim  m R_{AdS} \sim  \sqrt{2} \lambda^{ 1/4}
      + \cdots
     }
     where $m$ is given, to a good approximation, by  the mass of the
     flat space  state computed in \poleoper . Incidentally, we can calculate the conformal weight $\Delta$ of other (generally non local) operators
     with arbitrary spin in the same manner. They correspond to the string states
\eqn\formexpj{
     ( \partial_\alpha y^+ \partial_ \alpha y^+ )^{ j \over 2} \delta(y^+) e^{ i k.y }
     }
The mass of these states is given in flat space by $m^2 = - k^2 = {2 \over \alpha'} (j-2)$. Therefore
 \eqn\dimeconrj{\Delta(j) \sim \sqrt{2} \sqrt{j-2}
\lambda^{1/4} + \cdots}
This formula is expected to be a good approximation only for $j \ll \lambda^{1/2}$, since it was
derived assuming the flat space approximation. For very large values of $j$ we get a logarithmic 
behavior in $j$, see \gkpstring . Of course this is simply the analytic continuation of the leading 
Regge trajectory. 

      The dots in \dimeconr\ and \dimeconrj\ denote terms independent
     of $\lambda$ as well as higher order corrections.
     We then need to compute  the overlap of a wavefunction which decays like  $1/|\vec W|^{\Delta}$  for
     large $|\vec W|$ with the wavefunctions of the two gravitons associated to the energy flux
     insertions.  We find
     a behavior
     \eqn\smallanglbe{
     \langle { \cal E}( \theta_1) {\cal E}(\theta_2 ) \cdots \rangle \sim
      \theta_{12}^{ \Delta - 6  } \langle { \cal U}_{3-1}(\theta_2 ) \cdots  \rangle
     }
     where ${\cal U}_{3-1}$ is related to the lightest spin 3 non-local operator, with zero $p_-$,
      which at strong coupling
     has a large dimension \dimeconr . In string theory this expectation value is computed
     by inserting the operator \poleoper .

     In conclusion,  the structure of the OPE is precisely what we expected from general
     principles in any conformal field theory. At weak coupling the operator product expansion is
     dominated by operators of twist slightly bigger than two. This leads to correlation functions
     that are highly localized along certain jet directions. For any value of the
     coupling the  operator, or string state, that
     dominates has zero $p_-$ and spin $j=3$.
      At strong coupling, the operator
     acquires a large twist given in  \dimeconr . The fact that operators with
     spin $j > 2$ have large dimensions at strong coupling is seen to be intimately related with
     the fact that the energy distribution is uniform. Of course, this fact is also connected
     with the validity of the gravity approximation in the bulk.

    % Expectation values of light-ray operators \brodskylepage\ were considered
    % in the AdS/CFT context in
    % \brodsky .

\newsec{Summary, Conclussions and open problems}

Let us summarize some of our results.

We studied energy correlation functions in conformal field theories.
Energy correlation functions are an infrared finite quantity that is useful for
characterizing the states produced by localized operator insertions in  a field theory
\refs{\ellis,\ellissecond,\npcorrections}.

They can be computed for all values of the coupling since they involve the stress tensor
operator \tkachov\ and make no reference to a partonic description.
This is more manifest at strong coupling where the partons are difficult to see in
  the gravity or string
description.

After a conformal
transformation these energy correlation functions amount to measuring the
state along a null surface. More precisely, each ``calorimeter'' insertion corresponds to
an integral of the stress tensor along a lightlike line, $\int dy^- T_{--} $,  \exprg .

We have argued that the small angle
behavior of the energy correlation functions is controlled by an operator product expansion
which features non-local light-ray operators of definite spin. When two calorimeters come
close to each other we have a spin three operator
 $\langle {\cal E}(\theta_1) {\cal E}(\theta_2) \cdots \rangle \sim
 |\theta_{12}|^{ \tau_3 -4 }  \langle {\cal U}_{3-1} (\theta_2 ) \cdots \rangle $.
 These operators can be discussed
for any coupling.
 We recalled the   weak coupling expression for the twist \refs{\kloneloop} \threeitg ,
 and we also computed the twist at strong coupling  $\tau \sim \sqrt{2} \lambda^{1/4}$ \dimeconr ,
 after having identified the string
states that are dual to the operator ${\cal U}_{3-1}$. These are not ordinary closed string states. They are peculiar string states localized
along $x^+=0$ that have non-local vertex operators on the covariant worldsheet but do have a local description on the worldsheet in light-cone
gauge. Closely related string states appear in the Regge limit \psregge . Despite their unfamiliar features they control the short distance
singularities of energy correlation functions.

The light-ray operators that appear in the small angle behavior of the correlator are related to
the ones that control the moments of the parton distribution functions. In fact, one
can write a precise relation between the energy correlation functions on a special state
and a particular moment of the functions that govern the deep inelastic scattering amplitude \claimres .

We have  seen that energy flux one point functions in states created by currents or stress tensor insertions have an ``antenna''
pattern\foot{Remember that individual events do not present this pattern. This refers to the one point functions which consist of averages over
events.} which is determined by the three point functions in the conformal field theory \anglrebem \mostgengav . In the gravity description this
pattern is spherical but as we include higher order corrections to the gravity action we start seeing deviations from the spherical pattern
\valatwo \valuesfos . These deviations are sensitive to the spin of the operator that created the excitation in the conformal field theory.
 In the particular case of ${\cal N}=4$ super Yang Mills, the energy one point function   is
spherical for all values of the coupling. In more general  ${\cal N}=1$ superconformal theories
we find that the
antenna pattern,  \refnone \ttwotfour ,
depends on the parameters $a$ and $c$ that characterize the three point functions
in the current/stress tensor multiplet \freedmanc \onlyosborn .
 These results are exact expressions, valid for any coupling. They depend only on the two anomaly
 coefficients $a$ and $c$ defined in \freedmanc .
Demanding that the energy that calorimeters measure is
positive we get a constraint on $a$ and $c$, $ |a-c| \leq c/2$, which is saturated for free field
theories (of course, $c>0$).

We gave a general prescription for computing the energy correlation functions on the gravity side.
The operator insertion in the field theory produces a   string state that falls into the $AdS$ horizon.
Energy correlation functions depend on the wavefunction of this string state at the $AdS$ horizon.
The falling string is probed by particular shock waves associated to the insertion of each calorimeter.
This can be computed in a simple way by choosing a coordinate system in $AdS_5$ that is non-singular
at the horizon. We can view the computation of the energy correlation functions as taking a
snapshot of the falling string state as it crosses the horizon.
In the gravity approximation the result depends only on the momentum distribution of the
initial state and it is independent of the spin or any other property of
 the string state we consider. If the state carries a purely timelike momentum
 $q^\mu = (q^0,\vec 0)$,
 then the energy distribution on the detector is perfectly spherical with no fluctuations.
   As we include stringy corrections
we find small fluctuations that are inversely proportional to the square of the radius of
$AdS$ in string units (or $1/\sqrt{\lambda }$) \endres \correcn .
These fluctuations are small but they are not
gaussian \twoandthree . Since the shock waves we are considering are infinitely localized one
might worry that this leads to divergences. In fact, they would lead to divergent answers in
a field theory context (at least in some cases). The Regge behavior of string amplitudes at large
energies ensures that the results we obtain are finite.

It should be fairly straightforward to generalize this discussion to other dimensions.
%If the CFT is two dimensional, then we do not have any interesting angular dependence to explore,
%we could only see whether the energy goes to the left or the right.
The discussion in
2+1 dimensions might have some condensed matter applications, similar to  %\HerzogIJ
\ref\HerzogIJ{
  C.~P.~Herzog, P.~Kovtun, S.~Sachdev and D.~T.~Son,
  %``Quantum critical transport, duality, and M-theory,''
  Phys.\ Rev.\  D {\bf 75}, 085020 (2007)
  [arXiv:hep-th/0701036].
  %%CITATION = PHRVA,D75,085020;%%
}.

It would also be interesting to understand finite $N$ corrections.

There has been a great deal of progress in
 computing perturbative scattering amplitudes in ${\cal N}=4$ super Yang Mills, see
 \amplitudes\ for example.
From these scattering amplitudes one can compute the energy correlation functions. On the
other hand, since the energy correlation functions are already infrared finite, it would be
nice to see if any of the methods developed to compute amplitudes could be  extended
 to compute
the energy correlation functions directly, without having to compute the amplitudes first.

Another possible direction would be to consider the ``hadronization'' corrections for a non-conformal theory.
In particular one could imagine a non-conformal theory with a gravity dual.
 In a confining theory with a gravity dual we expect that
 these corrections will be large
 in the
large $N$ limit because the strings cannot break.
 One could  also  try to understand  situations
where the theory becomes free in the IR, such as $4+1$ dimensional super Yang-Mills,
or dimensionally regularized
${\cal N}=4$ super Yang Mills.

It would also be interesting to generalize this discussion to more complicated initial states such
as the one resulting from the collision of two closed string modes in the bulk. This would be
analogous to pp collisions.

Finally, this discussion  might have some implication for   black holes, since
energy correlations are a way of measuring the final state of Hawking radiation and its non-thermal
properties.

{\bf Acknowledgements}

We would like to thank  N. Arkani-Hamed,
Z. Bern, L. Dixon, G. Paz, N. Seiberg,  G. Sterman, M. Strassler,
 Y. Tachikawa, S. Thomas, T. Vachaspati, E. Witten  for discussions.

This work was supported in part by DOE grant \#DE-FG02-90ER40542.

\appendix{A}{ Positivity  of $ \int dy^- T_{--}$  }

Let us consider first free field theories.
The classical expression for the stress tensor for
the Maxwell  field, $T_{--} \sim \sum_{i=1,2}  F_{- i} F_{- i } $, is explicitly positive since it
is a sum of squares.
On the other hand, the quantum expectation value of $T_{--}$ can be negative. Let us recall why
this happens. Formally,
we also have the sum of squares of hermitian operators, so that we would also
expect a positive answer. However, when we normal order we subtract and infinite
constant. Then  the normal ordered expression is not a sum of squares of hermitian operators.
In fact, we have schematically $T_{--} \sim (a^\dagger )^2 + a^\dagger a +  a^2 $ where
we have separated the operator into terms with different numbers of creation and annihilation
operators. By considering  a state of the rough form $|\Psi \rangle = |0 \rangle + \epsilon a_1^\dagger
a_2^\dagger |0 \rangle $, and using that the vacuum expectation value of $T_{--}$ is zero we
find that $\langle \Psi | T_{--} |\Psi \rangle \sim Re[ c \epsilon ] + o(\epsilon^2 )  $
where $c $ is some number.
By taking $\epsilon $ to be a small complex number we see that we can make $T_{--}$ negative at
a point \twoparticle .

Let us now consider the integrated expression
${\cal E} = \int dy^- F_{- i} F_{ - i } $. This
expression has the schematic form
\eqn\expansf{
{\cal E} \sim  \int_0^\infty d p^+  p^+  ( a_{p^+}(\vec y) )^\dagger a_{p^+}(\vec y)
}
we thus see that we have the integral of products of operators and their adjoints. This is an
explicitly positive operator. We have used the variable $p^+ \sim - 2 p_-$ which is positive.
Notice that terms with two $a^\dagger $ or two $a$ operators have disappeared from \expansf\ due to
the following argument.
The integral over $y^-$ enforces that the total $p^+$ should be zero. However, creation
operators can only increase $p^+$, thus we do not obey the $p^+=0$ constraint with only
creation operators. Further discussion on the null energy condition for free fields
can be found in
\necflat .

Let us consider now an interacting field theory. If we choose the gauge $A_-=0$, then
the stress tensor $T_{--}$ continues to be quadratic in the fields and the above argument would
hold. Of course, this argument is not too convincing since we might be ignoring renormalization
subtleties or problems with the gauge choice. It would be nice to find a more general and solid
argument for an interacting field theory.

\appendix{B}{ Energy distributions in gravity for generic states}

\subsec{ Energy distributions for general states}

Here we show how to go between a discussion of $n$ point functions and
the computation of probabilities for seing various energy distributions on the detector.
Sometimes one might be interested in computing the probability functional
$\rho[ {\cal E}(\theta) ] $ for measuring a particular pattern of energy deposition on the
calorimeters. When one computes jet amplitudes one is computing probabilities of this kind,
where one integrates over certain regions, such as the low energy region between two jets, etc.

If we are given $\rho$ we can compute the $n$ point functions, as $\langle {\cal E}(\theta_1)
\cdots {\cal E}(\theta_n) \rangle = \int { \cal D} {\cal E} \rho[ {\cal E}]  {\cal E}(\theta_1)
\cdots {\cal E}(\theta_n) $.

Formally one can also go in the other direction by computing the generating functional for
energy correlation functions,
$\langle e^{ i  \int d^2 \theta \lambda(\theta)
\cal E(\theta) } \rangle $. This expression is a functional of $\lambda(\theta)$ and its expansion
in powers of $\lambda$ gives us the $n$ point functions.
Then $\rho$ is given by
\eqn\rhogiv{
\rho[ {\cal E}'  ] =  \int {\cal D} \lambda e^{ - i \int d^2 \theta \lambda(\theta) { \cal E}'(\theta)
}  \langle e^{ i  \int d^2 \theta \lambda(\theta)
\cal E(\theta) } \rangle
}

Just in order to see how this works, let us start with the $n$ point functions given in
\finalcorrint .
We can easily compute the expression
\eqn\lambdexp{
\langle e^{ i  \int d^2 \theta \lambda(\theta)
\cal E(\theta) } \rangle = \int d^4 q \rho(q) e^{ i \int d^2 \theta \lambda(\theta) {\cal E}_{q^\mu}(\theta) }
}
where ${\cal E}_{q^\mu}(\theta)$ is the function in \finarmpl\ and $\rho $ is defined in \finalcorrint .
% \finres .
After doing the functional integral in \rhogiv\ we get \eqn\densfore{ \rho[ {\cal E}  ] =  \int d^4 q \rho(q) \prod_{\theta} \delta[ {\cal
E}(\theta) - {\cal E}_q(\theta) ] } We see that we have a continuum of $\delta$ functions, one for each angle. But we are integrating only over
four variables. Thus, once we fix the energy at four points, the energy at all other points is also fixed.

The general, formal, string theory expression for the energy correlators has a similar form,
except that we have to integrate over the infinite number of
 variables specifying the string wavefunction. At finite $N$ we would also have many string states.

\subsec{ Bulk wavefunction for a localized state}

 Here we start with the wavefuntion $\phi_0(x) \sim e^{ - i q^0 t } e^{
  - { ( t^2 + \vec x^2 ) \over \sigma^2 } }$ that we mentioned in \opeffs .
Its Fourier transform is
 \eqn\waf{\eqalign{
 \tilde \phi_0(p)  = & \int d^4 x e^{ - i p x} \phi_0(x) = \int d^4 x e^{ i p^0 t - i \vec p \vec x }
 e^{ - i q^0 t - { ( t^2 + \vec x^2 ) \over \sigma^2 } }
 \cr
 & \sim \sigma^4 e^{ - { \sigma^2 \over 4 } [ ( p^0 - q^0 )^2 +  ( \vec p)^2 ] }
 }}
 The bulk wavefunction then has the form
 \eqn\wafn{
 \phi(W^+=0, W^-,W^\mu) \sim  (q^0)^{\Delta}
  \int_0^\infty d\tilde \lambda \tilde \lambda^{\Delta -1 } e^{ i \tilde \lambda W^- q^0/2}
   e^{ - { (\sigma q^0 )^2 \over 4 } [  ( \tilde \lambda W^0 - 1)^2 + | \tilde \lambda \vec W|^2 ] }
   }
   We are considering the case that $q^0 \sigma \gg 1$. We then see that
   as soon as $|\vec W| \gg  1/(\sigma q^0)$
   the answer is exponentially suppressed.
   In the region with large $|\vec W|\sim W^0 \gg 1 $  we can
   do the integral \wafn\ by saddle point approximation and we find
\eqn\approxw{
\phi(W)  \sim (q^0)^\Delta ( W^0)^{-\Delta}
 { 1 \over ( \sigma q^0 ) } e^{ i q^0 W^-/( 2 W^0) } e^{ - { (\sigma q^0)^2 \over 8 } }
}
We then insert this in \finalcorr\ to find an expression of the approximate form
\eqn\constrp{
\langle {\cal E}(\vec n'_1) {\cal E }(\vec n'_2) \rangle \sim
 q_0^2  e^{ - {(\sigma q^0)^2 \over 4 } }
 \int d \Sigma_3   { 1 \over (W^0)^{ \Delta + 2}} {  1 \over ( W^0 - \vec W . \vec n'_1)^3}
 { 1 \over ( W^0 - \vec W . \vec n'_2)^3}
 }
 where we used that $N^2$ in \finalcorr\ is not exponentially small, and in \constrp\ we have
 kept only the leading exponential behavior in $\sigma q^0$.  We have also approximated the
 integrand in the large $W^0$ region which we expect to dominate for the singular small angle
 behavior of the two point function.
 Finally we find  the singular small angle behavior
 \eqn\transin{
\langle {\cal E}(\vec n'_1) {\cal E }(\vec n'_2) \rangle \sim
 |\theta|^{ 2 \Delta -4 } e^{ -( \sigma q^0)^2/4}
 }
This is precisely the power we expect for the double trace contribution as was discussed in \jpms . This term is exponentially suppressed when
we consider a state with definite momentum. Thus, the term that gave the largest contribution in the deep inelastic scattering analysis in
\jpms\ does not contribute to energy correlators when we consider states created with definite momentum. They do contribute if the state does
not have definite momentum in $x$-space. In fact, we saw that a state with definite momentum in $y$ space is directly connected with the deep
inelastic scattering amplitudes \claimres .
 Such a state does not have definite momentum in
$x$-space and will not have the exponential suppression that we get in  \transin .

We should emphasize that the contribution to \transin\ is coming from the region where the
particle is crossing the horizon at a position that is close to where the calorimeters are
inserted.

\appendix{C}{ Computation of the energy and charge one point function in terms of the three point
functions of the CFT}

We denote by ${\cal O}$ any operator, which could be a scalar operator ${\cal S}$ or a
vector $\epsilon. j$ or a tensor $\epsilon_{ij} T_{ij}$.

Let us start by recalling some formulas for two point functions
\eqn\correcpr{\eqalign{
\langle 0| S(t,x) S(0,0) |0 \rangle =&
{ 1 \over \left[ - ( t - i \epsilon)^2 + |\vec x|^2 \right]^\Delta }
 \cr
\langle  0 | T(S(t,x) S(0,0)) |0 \rangle =& { 1 \over \left[   -t^2 +
|\vec x|^2 + i \epsilon \right]^\Delta } }}
 where the first is not time
ordered and  the second is time ordered. Of course the same prescription works for
vector or tensor operators.
The operator insertion with
a definite timelike momentum $q^\mu = ( q^0,\vec 0 )$ can be
written as
 \eqn\create{
 {\cal O}_q |0
\rangle  = \int dt e^{ - i q^0 t } { \cal O }(t) | 0 \rangle
}
and it creates a state with energy $E = q^0 > 0$. The fourier transform of the two point
function is
\eqn\fourtra{
\int d^4 x e^{- i q. x} { 1 \over \left[ - ( t - i \epsilon)^2 + |x|^2 \right]^\Delta }
= c(\Delta) \theta(q^0) (-q^2)^{\Delta - 2 }  ~,~~~~~~c(\Delta ) = { ( 2 \pi)^3 ( \Delta -1) \over
4^{\Delta -1} \Gamma(\Delta)^2 }
}
This is the norm of the state that \create\ creates.
This will also give us the total production cross section if the operator ${\cal O}$ couples to
the standard model.
As remarked in \keni\ the positive norm condition implies $\Delta \geq 1$.
% Just to recall harmonic oscillator formulas $x = a^\dagger e^{it} + a e^{- it } $.

We are interested in starting from the ordinary expressions for the correlation functions
in position space and extracting the limit that corresponds to the energy or charge correlators.
In doing so, it is important to order the operators appropriately.
For the non-time-ordered three point function the correct
prescription is
\eqn\threept{\eqalign{
\langle 0| & {\cal S}(x_2) {\cal S}(x_1) {\cal S}(x_3 ) |0  \rangle =
\cr
=&
{ 1 \over \left\{ [ - ( t_{23} - i \epsilon)^2 + (\vec x_{23})^2 ][-
(t_{13} - i \epsilon)^2 + (\vec x_{13})^2 ][-(t_{21} - i \epsilon)^2 + (\vec x_{21})^2 ] \right\}^{
\Delta/2} }
}}
If one considers tensor operators we get similar denominators and we choose the same
$i\epsilon$ prescription.
This $i\epsilon$ prescription is a simple way to enforce the right ordering of the
operators. Another way to say this is that an operator that is to the `left' of another
should have a more negative imaginary part in the time direction.
 When one does perturbation theory, it might
be convenient to use time ordering along a Keldysh contour. However, for our purposes this simple prescription suffices.

Let us first show how to extract the energy correlation for a state created from a scalar
operator with fixed momentum, or at least fairly well defined momentum, as in \opeffs .
 In this case, we know that the answer is independent of the angles
and that the overall coefficient is determined by energy conservation. Nevertheless it is instructive to discuss this case in detail since the computation is
the simplest and one can apply a similar method for other cases. Our method is  not too elegant and there is probably a more direct and elegant method than the
one we applied here.

We  extract the energy correlation by directly performing the limit and the integral in \energfl .
We use translation invariance to fix the position of the first operator at $x_3 =0$.
For simplicity we place the detector along the direction $z$, so that $x_1 = ( t, 0,0, r)$.
We will take the limit $r\to \infty$. If $t$ is generic, then the three point function will
decay as $1/r^8$ since it would be determined by the dimension of the stress tensor and
the operator product expansion. This would be decaying too rapidly in order to give a finite
large $r$ limit.
  However, there is a larger contribution from the region $ t \sim r $, the region
on the light-cone of the inserted operators. This is the region that will contribute.
Of course, this is precisely what we would  expect in a theory of massless particles.
It is convenient to define coordinates $x^\pm = t \pm r$. We will find that the region
with finite $x^-$ will contribute. In addition, only the $T_{--}$ component of the stress
tensor can contribute. The integral over $t$ can be traded for an integral over $x^-$.
So we first take the $r \to \infty$ limit. We then do the integral over $x^-$ and at the
end we do the integral over $x_2$.

Let us see what each of these steps gives us. In order to follow this appendix the reader
would need to have a copy of the paper by Osborn and Petkos \osborn , since we will make
frequent reference to it.
We start with the correlation function for
\eqn\correltens{
 \langle 0|
 {\cal S}(x_2) T_{--}(x_1) {\cal S}(x_3 ) |0 \rangle  \sim
  { 1 \over x_{23}^{2 \Delta -2} x_{12}^2 x_{13}^2 }
 \left( { x_{12}^+ \over x_{12}^2 } - { x_{13}^+ \over x_{13}^2 } \right)^2
 }
 from   equation (3.1) of \osborn . We are not going to keep track of overall numerical coefficients.
 After multiplying by $r^2$ and  taking
 the $r\to \infty$ limit we get
 \eqn\finalli{
 \lim_{r \to \infty} r^2 \langle 0|
 {\cal S}(x_2) T_{--}(x_1) {\cal S}(0 ) |0 \rangle  \sim
 { ( x_{2}^-  )^2\over ( x_2^2)^{\Delta -1} } {
 1 \over ( x^- - i \epsilon )^3 ( x^- + i \epsilon - x_2^-)^3 }
}
We now perform the integral over $x^-$.
Note that we can close the contour on either the upper or lower $x^-$ plane and pick one
of the two poles in \finalli. We then find
 \eqn\finaqtwo{
 \lim_{r \to \infty} r^2 \langle 0 |
 {\cal S}(x_2) \int dx^- T_{--}(x_1) {\cal S}(0 ) | 0 \rangle  \sim
 {1 \over ( x_2^2)^{\Delta -1}} {
 1 \over  ( x_2^- - 2 i \epsilon  )^3 }
}
We now integrate over the two transverse $x_2$ coordinates and use the wavefunction in
\create\ does not depend on them. We find
\eqn\closetof{\eqalign{
 \lim_{r \to \infty} r^2
 \int d^4 x_2 e^{ i q^0 t_2 } & \langle 0 |{\cal S}(x_2)  \int dx^-T_{--}(x_1) {\cal S}(0 ) | 0
  \rangle  \sim
  \cr
 \sim  &
\int dt_2 dz_2 e^{ i q^0 t_2 }{1 \over [ -(t_2 - i \epsilon )^2 + z_2^2 ]^{\Delta -2} }
 {
 1 \over  ( t_2 - z_2 - 2 i \epsilon  )^3 } \sim  \theta(q^0) ( q^0)^{2 \Delta - 3}
}}
where we have also done the remaining two integrals.
When we divide by the two point function \fourtra\ we get $\langle {\cal E} \rangle \sim q^0$.
The numerical coefficient can also be computed at each step. Of course, this gives the
right answer $\langle {\cal E} \rangle =  { g^0 \over 4 \pi } $ due to the Ward identity which
fixes the coefficient of the three point function \correltens\ in terms of the coefficient of
the two point function \correcpr , see eqns (6.15), (6.20) of \osborn .

This procedure can be repeated replacing the operator ${\cal S}$ by a current
$\epsilon . j$. The computations are identical but with more indices and we use a computer.
 In this
case the three point function of a stress tensor and two currents is fixed by conformal invariance, plus the Ward identity, up to one unknown coefficient.
Conformal invariance leaves two possible structures and the Ward identity fixes the coefficient of one of them. In this case we find, as expected, that the
energy correlation function depends on the angle with respect to the vector $\epsilon$. Here we simply quote the value of the parameter $a_2$, introduced in
\nepar , in terms of the parameters   $\hat e ,\hat c$   defined in (3.13) and (3.14) of   \osborn \foot{Here we added hats to the parameters $e$ and $c$ in
\osborn\ so that they are not confused with other parameters in the present paper. Note also that \osborn\ uses some of these letters with multiple meanings
through their paper.}. We find that \eqn\findre{ a_2 = { 3 ( 8 \hat e - \hat c) \over 2 ( \hat e  + \hat c )}  ~~ \to ~~~~ 3 { \sum_i (q_i^b)^2 - (q_i^{wf})^2
\over \sum_i (q_i^b)^2 + 2  (q_i^{wf})^2 } } where we indicated the value for a free theory with bosons and Weyl fermions of charges $q_i^b$ and $q_i^{wf}$.
  The combination $(\hat e + \hat c )$ is fixed
in terms of the coefficient of the two point function of two currents via 6.26 of \osborn .

We can do  this also for the correlation functions of the form
 $ \langle 0 | \epsilon^*_{ij} T_{ij}
 { \cal E} \epsilon_{ij} T_{ij}| 0  \rangle $. We can then compute the coefficients $t_2$, $t_4$
 introduced in \mostgengav
  \eqn\valuestt{\eqalign{
 t_2 = &    {30 (13 \hat a+4 \hat b-3 \hat c) \over 14 \hat a-2 \hat b-5 \hat c}  ~~ \to ~~~~
 { 15( - 4 n_v + n_{wf})
 \over (n_b + 12 n_v + 3 n_{wf} )}
 \cr
 t_4 = & - {15 (81 \hat a+32 \hat b-20 \hat c) \over 2 ( 14 \hat a-2 \hat b-5 \hat c)}  ~~ \to ~~~~
 { 15 ( n_b + 2 n_v - 2 n_{wf} )  \over 2 (  n_b + 12 n_v + 3 n_{wf} ) }
}} where $\hat a,\hat b,\hat c$ are defined in (3.19)-(3.21)  of \osborn . We have also indicated the result for  $n_v$, $n_b$ and $n_{wf}$ free  vectors, real
bosons, and Weyl fermions (one complex dirac fermion would give  $n_{wf} =2$)\foot{$\hat a, \hat b,\hat c$ here should not be confused with the $\hat e, \hat
c$ of the previous paragraph. In particular the two $\hat c$ are not the same \osborn . }.
 Again, the combination appearing
in the denominator is fixed in terms of the stress tensor two point function, see
(6.42) of \osborn .

Two combinations of
 these three coefficients are related to the values of $ a  $ and $ c $
defined
through the conformal anomaly\foot{Do not confuse these $a$ and $c$ with the parameters
$\hat a$ and $\hat c$ of the previous paragraph.}
\eqn\defconfan{
T^\mu_\mu = { c  \over  16 \pi^2 } W^2 - {  a  \over 16 \pi^2 }  E
}
where   $W$ is the Weyl tensor and $ E = R_{\mu \nu \delta \rho}R^{\mu \nu \delta \rho} - 4
R_{\mu \nu} R^{\mu \nu} + R^2 $ is the Euler density.
 $c$ also sets the   two point function of the stress
tensor. The coefficient $a$ can be expressed in terms of the three parameters in \valuestt\ as \eqn\resc{
 {a  \over c  } = {   (9 \hat a - 2 \hat b - 10 \hat c) \over 3 (14 \hat a - 2 \hat b - 5 \hat c )  }
 ~~ \to ~~~~ { 2 n_b + 124 n_v + 11 n_{wf} \over 6 n_b + 72 n_{v} + 18 n_{wf} }
 }
This follows from  (8.37) in \osborn . The values for a free theories were computed in \refs{\duffobs,\duff}.
 From the positivity conditions \constts\ it is possible to
get general bounds on this ratio.
%\foot{ We thank Y. Tachikawa for prompting us to compute this bound.}.
 We find
\eqn\boundar{
 { 31 \over 18}   \geq {  a  \over c }  \geq { 1 \over  3 }
 }
 where the lower bound is saturated by a free theory with only scalar bosons and the upper bound
 by a free theory with only vectors. Notice that this bound holds for any conformal theory while the more restrictive
 bound \implic\ holds for supersymmetric theories. We can also add that for a ${\cal N}=2$ supersymmetric theory we can find a similar bound by using as limiting cases free theories with only vector supermultiplets (${a \over c} = { 5 \over 4}$) and free theories with hypermultiplets (${a \over c} = { 1 \over 2}$). Therefore ${ 5 \over 4}   \geq {  a  \over c }  \geq { 1 \over  2 }$. This agrees with results from \takshap .

In an ${\cal N}=1$ supersymmetric theory there is a relation between the three parameters
$\hat a, \hat b,\hat c$ which is obtained by setting $t_4=0$ in \valuestt . In this case, the two coefficients
in \defconfan\ specify completely the three point functions of the stress tensor.
In a non-supersymmetric theory, we have one more parameter beyond the two in \defconfan .

Finally, we can repeat this exercise for the correlation function of three currents to
find that the coefficient introduced in \chargecor\
\eqn\charccof{
\tilde a_2 = { 3 \over 2 } { 5\hat a - 4 \hat b \over (\hat a + 4 \hat b )} ~~ \to ~~~~
3 { \sum_i C(r_i)_b - C(r_i)_{wf} \over  \sum_i C(r_i)_b + 2 C(r_i)_{wf}}
}
where $\hat a,\hat b$ are defined in eqn. (3.9) of \osborn , and are not the same as the ones in
the previous paragraphs. Here $r_i$ are the representations of the bosons and Weyl fermions.
And $C(r)$ is defined as $tr_r[ T^a T^b] = C(r) \delta^{ab}$.
Again, the combination $(\hat a+ 4 \hat b)$ sets the two point function of the current.
And $\tilde a_2 $ vanishes in a supersymmetric theory since there is only one structure
contributing for the supersymmetric case \onlyosborn\ and it vanishes for a free supersymmetric
theory.

One can also take similar limits of the parity odd part of the three point function of
three currents and one obtains the result in \chargfl .
Finally, starting from the correlation functions for two stress tensors and a current derived
in \gravanomaly\ one can derive the charge distribution function in \anomal . Both of these
results relate the corresponding anomaly to a charge asymmetry.

\appendix{D}{Energy one point functions in theories with a gravity dual}

In this appendix we present the
 calculation of energy one point functions for states created by current operators
 and the stress energy
operator.

\subsec{One point function of the energy with a current source}

We wish to compute the contributions to the energy
 one point function \nepar\ for a state created by a current operator at
  strong coupling. The
$AdS/CFT$ dictionary says we need to compute the bulk three point function between two bulk photons
and
the graviton. The bulk action \unicont\ contains two terms.
The first term contributes also to the current
two point function while the second term in \unicont\ does not. Thus, the first term  contributes
to the part of the energy one point function which is determined by the Ward identity and the
second one to the angular dependent term which is not fixed by the Ward identity. In principle,
the first term could also contribute to the angularly dependent part, but we have argued, based
on the results for ${\cal N}=4$ SYM that it contributes only to the constant part.
Thus in order to compute the coefficient $a_2$ in \nepar\ we need to compute the ratio of
the contribution of the second term in \unicont\ and the contribution of the first term in \unicont .

Since the graviton is localized in $W^+$ and the photon is localized in the other transverse
directions if the state has definite four dimensional momentum, we can approximate the computation
as a flat space computation. In this particular case, this approximation will be exact, but that
will not be the case when we discuss the three graviton vertex. Thus, we evaluate the vertex
expanding the flat space action, but we will insert the $AdS$ wavefunctions for the external
states.

In flat space our coordinates are $(x^+ , x^- , x^{1,2,3})$.
The metric is $ds^2 = - dx^+dx^- + dx^i dx^i$ (latin indices $i$ and $j$ go from 1
to 3).

We want to collect terms that are of  first order
 in the perturbation $h$. There are two such terms, one per factor of $g_{\mu\nu}$ in the action.
The determinant $g$ does not receive corrections and $\sqrt{g} =  {1 \over 2}$. Our perturbation is of the form $h = h_{++}(x^i,n^i) \delta(x^+)
(dx^+)^2$. We made explicit the fact that $h_{++}$ depends on the transverse coordinates and on a unit vector $n^i$ in the transverse space that
represents the position of our calorimeter.

Therefore, we want to calculate
\eqn\actb{
 S_1 = -{1 \over 4 g^2} \int {dx^+ dx^- d^3 x \over 2} 2 h_{++} F^{+ i} F^{+ j} g_{i j} = -{1 \over 4 g^2}\int dx^+ dx^- d^3 x h_{++}
F^{+ i} F^{+ j} g_{i j}
}
Notice that the contraction of $F$s is restricted to the 3 dimensional transverse space as the
 metric element $g_{--}$ is zero. We can do
the $x^+$ integral easily as $h_{++}$ is localized in this direction. Also, we will use the fact that the wave function of the photon is
localized in the transverse space \phipw. We represent this fact by writing
\eqn\ff{
F^{+ i} F^{+ j} g_{i j} = \alpha(x^+,x^-) \delta^3(x^i) \epsilon^i \epsilon^j g_{ij} =\alpha(x^+,x^-) \delta^3(x^i)
}
\noindent where $\epsilon^i$ represents the (normalized) polarization of the photon. Notice that we choose the polarization in the transverse
directions. Therefore $F^{+ i} \sim \partial^+ A^i$. Using these facts and performing the integrals we get
\eqn\actbb{
 S_1 = -{1 \over 4 g^2}\int dx^+ dx^- d^3 x h_{++} F^{+ i} F^{+ j} g_{i j} = -{1 \over 4 g^2}h_{++}(0,n^i) \int dx^- \alpha(0,x^-)
 }
In fact, $h_{++}$ evaluated at $x^i=0$ does
not depend on $n^i$, see \metrfl\ at $W^i =0$.
% as the only dependence on $n^i$ can come through $x^i n^j g_{ij}$.
 Therefore, this
term does not contribute to the angular dependence of the correlation function.

Let us  now  look at the other term in \unicont. We first need to compute the Weyl tensor. Let us start with the Riemann tensor. This tensor has
terms that go as ${1 \over 2} \partial^2 g$ and terms that go as $g\Gamma\Gamma$. Since we are in a flat space background only the first type of
term contributes. This yields
\eqn\rie{
 R_{+ i + j} = {1 \over 2} \partial_i \partial_j h_{++}
 }
All other terms are given by symmetry properties ( i.e. $R_{+ i + j}=-R_{i + + j}= -R_{+ i j +}= R_{i + j +}$)  or vanish. The Weyl tensor also
contain terms of the form ${1 \over 3} g_{\lambda \nu} R_{\mu \kappa}$. But,
\eqn\ricc{
 R_{\mu\nu} = g^{\lambda \rho} R_{\lambda \mu \rho \nu} \quad \longrightarrow \quad R_{++} = g^{i j} R_{+ i + j} = {1 \over 2} g^{i
j}\partial_i
\partial_j h_{++}
}
We see there is only one non vanishing term that is proportional to the laplacian inside the transverse space. There are also terms proportional
to the Ricci scalar inside the Weyl tensor, but we can see that these vanish in our case. The Weyl tensor is, then, given by
\eqn\weylc{
 C_{+ i + j} = {1 \over 2} \left(\partial_i \partial_j - {1 \over 3} g_{i j} \partial^k \partial_k\right) h_{++}
 }
The other components either vanish or are given by symmetry properties. There are four possible positions for the two plus signs, so we will
have four terms in the second term in \unicont\
(we are also using symmetry properties of $F^{+ i}$).
That is
\eqn\actions{ \eqalign{ S_2 = &  {\alpha \over
g^2 M^2_*} \int {dx^+ dx^- d^3 x \over 2} 4 C_{+ i + j} F^{+ i} F^{+ j} =
\cr = &
{\alpha \over g^2 M^2_*} \int dx^+ dx^- d^3 x  F^{+ i} F^{+ j}
\left(\partial_i \partial_j - {1 \over 3} g_{i j}
\partial^k
\partial_k\right) h_{++}
}}
Once again we can perform the integrals to obtain
\eqn\sss{
 S_2 = {\alpha \over g^2 M^2_*} \left(\partial_i \partial_j - {1 \over 3} g_{i j} \partial^k \partial_k\right)
h_{++}(x^i,n^i)\big|_{x^i=0}  \epsilon^i \epsilon^j \int dx^- \alpha(0,x^-)
}
We are interested in the quotient between the angularly dependent term \sss\ and the spherically
symmetric term \actbb
\eqn\aat{  -{4\alpha \over M^2_*}{ \left(\partial_i \partial_j - {1 \over 3} g_{i j} \partial^k
\partial_k\right) h_{++}(x^i,n^i)\big|_{x^i=0} \epsilon^i \epsilon^j \over h_{++}(0,n^i)}}

% Let us now consider the case in which the background is given by the $AdS$ metric.
%This can be done  by realizing that $C$ scales with a factor
%of a metric rescaling under Weyl transformations. Therefore, we only need to compute ${A \left(\partial_i
%\partial_j - {1 \over 3} g_{i j} \partial^k
%\partial_k\right) (y^5)^2 h_{++}(y^i,n^i)\big|_{y^5=1} \epsilon^i \epsilon^j \over h_{++}(y^5=1,n^i)}$ in y coordinates. Using the form of the
We use the explicit form of the perturbation \metrfl\ and
 we get the result
\eqn\aatt{
a^{AdS}_2 = -48 {\alpha \over M^2_*}
}
This gives the gravity result for the anisotropic part of the one point function \nepar\ of a state
produced by a current.

\subsec{One point function of the energy with a stress tensor source}

Now we want to repeat this calculation for the case where we have the stress tensor
 as a source. In this case we need to
consider 3 graviton interactions. There are 3 operators that contribute to this vertex. A natural parametrization is given by the action
\threeposver.

We will do first the calculation in a flat space
 background. As in the computation we did above we will get derivatives acting on the perturbation
 $h$. When we go to the $AdS$ background  we could get terms involving the background
 curvature. Such terms are isotropic and will not contribute to the terms that have maximal angular
 momentum. But they do give contributions to the terms that have smaller values of the angular momentum.
 The computations we do here give only  the leading contribution for $t_2$ and $t_4$ in
 \mostgengav .
 % When we transfer the result to an $AdS$ background we
 % In in this context, the contribution coming from each operator can be obtained exactly
% for the leading angular momentum term (the one with the highest power of $\cos \theta$). This is because we are interested in the term with the
%maximum number of uncontracted derivatives and this term does not depend on the background metric. Notice that this is not enough to calculate
%the one point function exactly, as there might be contributions from second (third) order terms to the angular momentum 0 (1) terms.
%Nevertheless, it will be sufficient if we are only interested in contributions that are leading order in ${1 \over M_{pl}^2 R^2_{AdS}}$. Notice
%also that it does not make any difference, in this approximation, whether we consider the Weyl tensor in \threeposver\ or the Riemann tensor, as
%their contributions to the term with the most uncontracted derivatives is the same.
We start from the action in \threeposver\ and we expand each term to cubic order.
%Each of these terms gives us schematically the following terms,
We focus on terms with highest angular momentum
in the transverse dimensions  (see
% We know from
\tseytlinhigher ).
% the expansions to cubic order and leading
%angular momentum of these coefficients.
We use the fact that we
need one of the metric perturbations to be $h_{++}$ while the other two only have purely transverse
indices. We find
\eqn\rmt{\eqalign{
R=&  - {1 \over 2} h_{++} h_{(1)}^{i j} (\partial^{+} )^2 h_{(2)}^{i j}, \quad
R_{\mu\nu\delta\sigma}R^{\mu\nu\delta\sigma}  = - 2 \partial_i \partial_j h_{++} h_{(1)}^{i k}
(\partial^{+})^2 h_{(2)}^{j k}, \quad
\cr
 R_{\mu\nu\delta\sigma} &R^{\delta\sigma \rho \gamma}  R_{\rho\gamma}^{~~ \mu\nu} =  - 6 \partial_i \partial_j \partial_k \partial_\ell h_{++} h_{(1)}^{i j} (\partial^+)^2
 h_{(2)}^{k \ell}
}}
\noindent Notice that expanding the determinant of the metric in the action does not contribute to the three point function. If we now use that
the wave function is going to be of the form
\eqn\rmtot{
h_{(1)}^{i j} \partial^{+2} h_{(2)}^{k \ell } =
\beta(x^+,x^-) \delta^3(\vec x) \epsilon^{i j}  \epsilon^{k \ell }
}
\noindent we can perform the integrals and calculate the quotients of the contribution to the three point function. After taking the derivatives
and evaluating at $\vec x = 0$ we obtain the ratios
\eqn\ratiob{t_2 = 48  {\gamma_1 \over R_{AdS}^2 M_{pl}^2}}
\eqn\ratiobb{t_4 = 4320 {\gamma_2 \over R_{AdS}^4 M_{pl}^4}}
Due to the issues we discussed above, there are terms contributing to $t_2$ which are of first
order in $\gamma_2/( R_{AdS} M_{pl})^4$, coming from the term with six derivatives in the action,
 which we neglected compared to the contribution of the four derivative terms. These formulas
 are also valid only to first other in the $\gamma_i$.

\listrefs

\bye